\documentclass[twocolumn, amsmath,amssymb,
apsfloats]{revtex4}
\usepackage{dcolumn}
\usepackage{bm}
\usepackage[pdftex]{graphicx}
\usepackage{graphicx}
\usepackage{epstopdf}
\usepackage{float}
\usepackage{url}
\usepackage{hyperref}
\usepackage{xstring}
\usepackage{tabularx}
\usepackage{xcolor}
\usepackage{array}
\usepackage[utf8]{inputenc} 
\usepackage[T1]{fontenc}     
\usepackage{textcomp}        
\usepackage{lmodern}        
\usepackage{multirow} 
\usepackage{array}

\def\KdS{Kerr--de~Sitter}

\def\be{\begin{equation}} 
\def\ee{\end{equation}}
\def\beq{\begin{equation}} 
\def\eeq{\end{equation}}
\def\bea{\begin{eqnarray}} 
\def\eea{\end{eqnarray}}
\def\din{\,\mathrm{d}} 
\def\dbe{\mathrm{d}}
\def\bet{\begin{tabular}} \def\ent{\end{tabular}}
\def\cale{{\cal E}} 
\def\calq{{\cal Q}}
\def\calphi{{\varPhi}}
\def\calpsi{{\varPsi}}
 
\def\calk{{\cal K}}

\def\cald{{\cal D}}
\def\oder#1#2{\frac{\dbe#1}{\dbe#2}}
\def\dgr{^\circ}

\def\spo{\!S\!P\!O}
\newcolumntype{Y}{>{\centering\arraybackslash}X}

\begin{document}	
	
\title{Shadows of naked singularities and superspinars related to the revisited Kerr--de~Sitter spacetimes}

\author{Daniel Charbul\'{a}k}
\email	{daniel.charbulak@physics.slu.cz}
\author{Zden\v{e}k Stuchl\'{i}k}
\email	{zdenek.stuchlik@physics.slu.cz}

\address{Research Centre for Theoretical Physics and Astrophysics, 
	Institute of Physics, 
	Silesian University in Opava, 
	Bezru\v{c}ovo n\'{a}m. 13, CZ-746 01 Opava, Czech Republic}

\date{\today}

\begin{abstract}
We construct shadows of superspinars described by the revisited Kerr--de Sitter (rKdS) naked singularity (NS) spacetimes and compare them with those of the standard KdSNS spacetimes. For all the classes of the rKdSNS spacetimes we determine local escape cones related to variety of fundamental frames: locally nonrotating frames (LNRFs), radially escaping frames, and circular geodesic frames related to marginally stable obits of the rKdSNS spacetimes. The local escape cones (and their complementary cones) are then applied to construct the shadow of the KdS superspinars related to the distant static observers represented by the LNRFs located near the so-called static radius where the spacetime is close to the asymptotically flat region of the Kerr spacetimes, or the superspinars radially approaching, due to the Universe's expansion, the cosmic horizon of the spacetime. Differences of the shadows in the rKdSNS and standard KdSNS spacetimes are established and demonstrated for sufficiently large values of the dimensionless cosmological constant. For the observationally given cosmological constant and masses of the largest objects in the Universe, the shadow differences are not observable using recent observational instruments. 
\end{abstract}

\maketitle

\section{Introduction}\label{intro}

The shadows of compact objects represent one of the most promising tools for testing
the physics of strong gravitational fields, both within classical general relativity
\cite{Bar:1973:BlaHol:,Cun-Bar:1973:ApJ:,Lum:1979:ASTRA:,Vie:1993:ASTRA:,FalckeMeliaAgol2000,BeckwithDone2005}
and beyond it.
In this context, black-hole and compact-object shadows provide a powerful observational
probe of possible deviations from general relativity encoded in the spacetime geometry
governing photon motion in the vicinity of extremely compact sources.

Such deviations naturally arise in a broad class of alternative and extended theories
of gravity, including modified $f(R)$ gravity
\cite{SotiriouFaraoni2010,DeFeliceTsujikawa2010,NojiriOdintsov2011},
braneworld scenarios inspired by extradimensional models
\cite{RandallSundrum1999,Dadhich2000,AlievGumrukcuoglu2005}, and
asymptotically safe gravity incorporating quantum corrections at high energies
\cite{Reuter1998,BonannoReuter2000,ReuterSaueressig2012,SaueressigEtAl2015,Eichhorn2023},
as well as higher-curvature theories such as Lovelock and Gauss-Bonnet gravity
\cite{Lovelock1971,BoulwareDeser1985,Abd-etal:2015:EPJC:}.
The impact of these theoretical extensions on photon dynamics and the optical properties
of compact objects has motivated a growing number of dedicated shadow studies in recent
years.

In addition to analytical studies, a substantial body of work has been devoted to the numerical modeling of compact-object shadows, providing a direct link between theoretical predictions and observable signatures. Early ray-tracing simulations in Kerr spacetimes established the basic appearance of black-hole shadows \cite{BroderickLoeb2009,Broderick2014}.

More importantly, numerical investigations have extensively explored the optical properties of exotic compact objects. These include naked singularities \cite{Nguyen2023,ShaikhJoshi2018}, scalar-hair configurations and boson stars \cite{Cunha2015,Vincent2016}, as well as other horizonless or superspinning objects, such as traversable wormholes \cite{Ohgami2015,Sokoliuk2022,Tangphati2023}. In parallel, parametrized deviations from the Kerr geometry and alternative gravity scenarios have been studied using numerical ray-tracing approaches \cite{Younsi2016} as well as more advanced general-relativistic magnetohydrodynamic and radiative-transfer simulations aimed at testing gravity with black-hole images \cite{Uniyal2025}. Such studies have demonstrated that these objects can produce shadow-like features with qualitative and quantitative differences from those of classical Kerr black holes, offering a potential observational avenue for testing strong-field gravity. The resulting diversity of shadow morphologies highlights the importance of numerical simulations in assessing the distinguishability of alternative compact-object models. 

Explicit constructions of black-hole shadows in alternative gravity frameworks have been presented, for example, in
Refs. \cite{NojiriOdintsov2025,RodriguezChagoyaOrtiz2024,Jafarzade:2023dak,Liu_2024,Luo2024}
and, in the context of braneworld models, in
Ref. \cite{Chur-Kon-Stu-Zhi:2021:JCAP:}.

Following the first direct astronomical images of the shadows of supermassive compact
objects obtained by the Event Horizon Telescope
\cite{Aki-etal:2019:ApJLa,Aki-etal:2019:ApJLb,Akiyama_2019:ApJLc,Akiyama_2019:ApJLd,Akiyama_2019:ApJLe,Akiyama:2019eap,Akiyama_2021:ApJLa,Akiyama_2021:ApJLb},
it has become clear that the theoretical interpretation of shadow observations crucially
depends on the assumed spacetime geometry and its ability to describe gravity in the
immediate vicinity of extremely compact sources.

In this context, considerable attention has been devoted to alternative configurations involving naked singularities and superspinars~--~objects exceeding the Kerr black-hole spin limit~--~as well as traversable wormholes. Such configurations may arise in extended or modified theories of gravity, including string-inspired, braneworld, and quantum-gravity-motivated models \cite{Horava2009Quantum,KehagiasSfetsos2009,GimonHorava2009,Dadhich2000,Aliev2005}, and they have also been considered in astrophysical contexts~--~for instance, in studies of superspinar evolution driven by accretion processes \cite{Stu-Hle-Tru:2011:CLAQG:}. The optical appearance and shadow-like features of these exotic compact objects can substantially differ from those of classical Kerr black holes, giving rise to potentially clear observational signatures \cite{Stu-Sche:2010:CQG:,Stu-Sche:2012:CQG:,Stu-Sche:2013:CLAQG:,Nedkova2013,Abdujabbarov:2016efm:,Abd-Ahm-Dad-Ata:2017:PHYSR4:,Wielgus_2020,Schee_2022}.

Observations of distant Type~Ia supernovae, relic radiation, and large-scale structure
indicate that the Universe is undergoing accelerated expansion, commonly attributed
to dark energy \cite{Riess1998,Perlmutter1999}, which in general relativity is naturally
modelled by a nonzero cosmological constant $\Lambda$ \cite{Kra-Tur:1995:GENRG2:,Kra:1998:ASTRJ2:}.
\footnote{Dark energy may also be described by alternative models, such as quintessence
	\cite{RatraPeebles1988,Caldwell1998,Far:2000:PHYSR4:,Far-Jens:2006:CQG:}
	or phantom energy \cite{Caldwell2002,Carroll2003}.
	While the cosmological constant represents the simplest realization, recent observations
	have raised tensions concerning its interpretation
	\cite{Riess2016,Verde2019,PhysRevD.103.L121302,2021APh...13102605D}.}
Beyond its cosmological role, a nonzero cosmological constant can also have nontrivial
astrophysical consequences
\cite{Stu-etal:2020:Universe:,Stu-Sla-Hle:2000:ASTRA:,Stu:2005:MODPLA:,Stu-Sche:2011:JCAP:,Stu-Nov-Hla:2025:}.
For this reason, Kerr--de Sitter (KdS) geometry, incorporating both rotation and a
cosmological constant, is frequently employed in theoretical studies of compact-object
optics and shadows
\cite{Stu-Char-Sche:2018:EPJC:,Stu-Char:2024:PHYSR4:,Char-Stu:2024:PHYSR4:}.

However, the classical KdS spacetime assumes a rigid cosmological constant that remains
unaffected even in regions of extremely strong gravity.
In contrast, the recently proposed revisited Kerr--de~Sitter (rKdS) geometry
\cite{Ovalle_2021} allows for a local deformation of vacuum energy in strong-field
regions, in closer accordance with expectations from quantum field theory in curved
spacetimes.
Such a modification can fundamentally alter the structure of geodesics and photon
orbits, and consequently the optical appearance of compact objects.
The aim of this work is therefore to determine how the structure and morphology of
superspinar shadows and naked singularities are modified within the rKdS framework.

The paper is structured as follows: In Sec. \ref{sec2}, we recall basic properties of the rKdS spacetimes, such as the event horizons, ergosphere and static radius, and compare the appropriate relations with those describing the standard KdS spacetimes. Most of the results can be found to varying degrees in Refs. \cite{Ovalle_2021,Omw-etal:2022:EPJC:,Omwoyo2023,Slany:2023:PHYSR4:}. Here, we compare the distribution of the parametric plane according to the properties of horizons and the erosphere in both rKdS and KdS geometries. In the remainder of the section, we compare the construction of the locally nonrotating reference frames (LNRF) and Carter's equations for the motion of test particles and photons and the shape of the black-hole shadow.

In Sec. \ref{sec3}, we discuss the properties of spherical photon orbits (SPOs) in rKdS spacetimes, as they are crucial for the construction of shadows. In particular, we consider their distribution in the equatorial plane, and thus the number of equatorial circular photon orbits (ECPOs). We also discuss the distribution of SPOs with respect to the ergosphere, the existence of polar SPOs and the related orientation of SPOs with respect to LNRF observers, and the stability of SPOs with respect to radial perturbations. We then classify rKdS spacetimes into seven classes according to the same criteria we used for KdS in Ref. \cite{Char-Stu:2017:EPJC:}~--~i.e., according to the number of horizons, the nature of the ergosphere as described in Sec. \ref{sec2}, and the properties of SPOs. We compare this classification between the two types of geometries. 

In Sec. \ref{sec4}, we construct the shadows of rKdS and KdS superspinars for an LNRF observer located at the static radius of the rKdS spacetime, and we introduce observable quantities. We compare these quantities as calculated for the rKdS and KdS geometries. Furthermore, to assess the influence of cosmic repulsion in rKdS and KdS geometries, we analyze the image of a radially escaping superspinar for the currently estimated value of the cosmological constant $\Lambda=1.1\times 10^{-52}\mathrm{m}^{-2}$ \cite{Riess1998,Planck2018}. We consider two characteristic observation distances: the first corresponds to the scale of the local cosmic void (Local Void), and the second to the distance of the most massive known black hole, TON 618 \cite{ShemmerNetzerMaiolino2004,Ziolkowski:2008aq}. Subsequently, in order to emphasize the role of the cosmological constant, we construct shadows of the rKdS superspinars also for significantly increased values of $\Lambda$, when the object is located near the corresponding cosmological horizon. We also discuss the critical value of $\Lambda$, above which it is possible, due to the current observational capabilities, to clearly distinguish whether the observed image originates from rKdS or KdS geometry. In the last part of this section, we construct the light-escape cones (LECs) for a source orbiting around the rKdS superspinar on marginally stable circular orbits of test particles that define the radial dimensions of the Keplerian disk. Such a study of LECs is relevant for the investigation of optical effects such as self-illumination, self-eclipse, or self-heating related to accretion disks orbiting the superspinars.

Section \ref{sec5} is devoted to the conclusions of this work.

\section{Revisited \KdS\, superspinar (NS) geometry} \label{sec2}

As in our previous works, we denote here by $M$ the mass of the central gravitating object (the naked singularity or superspinar), by $J$ its internal angular momentum, which we replace with a dimensionless spin parameter  $a=J/M^2$, and by $\Lambda$ the cosmological constant, which we conveniently replace with the dimensionless cosmological parameter $y = \frac{1}{3}\Lambda  M^2$. Furthermore, we use the Boyer-Lindquist spheroidal coordinates $t,r,\theta,\phi$ and the geometric system of units in which $c = G = 1$. It is convenient to set $s/M \rightarrow s$, $t/M \rightarrow t, r/M \rightarrow r$~--~or, equivalently, to set $M=1$.

As a reminder and for the comparison of both rKdS and KdS geometries, we adopt the style from Ref. \cite{Slany:2023:PHYSR4:} and write down the line element for both of these geometries collectively~--~terms that are present only in the standard KdS metric and do not occur in the rKdS case are marked with a box.

The rKdS/KdS geometry is then described by a line element of the form  
\bea
\dbe s^2 &=&
\frac{a^2\sin^2\theta \boxed{\Delta_{\theta}}- \Delta}{\boxed{I^2}\rho^2}\dbe t^2 +\frac{\rho^2}{\Delta_{\boxed{r}}}\din r^2 + \frac{\rho^2}{\boxed{\Delta_{\theta}}}\din \theta^2 \label{ds2} \\
&+&\frac{2a\sin^2\theta}{ \rho^2}\left[\Delta_{\boxed{r}}-\left(r^2 +   a^2\right) \right]\din t \din \phi+\frac{A \sin^2\theta}{\boxed{I^2}\rho^2}\din \phi^2, \nonumber 
\eea
where
\bea
\Delta_{\boxed{r}}&=& r^2-2r+a^2-yr^4 -\boxed{a^2yr^2}, \label{Delta}\\
\rho^2&=&r^2 + a^2\cos^2\theta, \label{rho2}
\eea
and
\be A=(r^2+a^2)^2\boxed{\Delta_{\theta}}-a^2\Delta \sin^2\theta. \label{A} \ee 

As follows from the previous note, by omitting the terms in boxes, we obtain the expression for the new rKdS geometry. 

Terms that occur exclusively in the KdS case are
\bea
\Delta_\theta&=&1 + a^2y\cos^2\theta \\
I&=&1 + a^2 y.
\eea

\subsection{Event horizons} \label{ssec_event_hor}

The event horizons are given by the roots of the equation
\be
\Delta =0. \label{Delta=0}
\ee
The roots of Eq. (\ref{Delta=0}) were given in the original paper \cite{Ovalle_2021} in terms of the cosmological constant $\Lambda$, and discussion of the number and character of these roots in terms of the cosmological parameter $y$ was presented in Ref. \cite{Slany:2023:PHYSR4:}. In this paper, we present a new partition of the parameter plane $(a^2,y)$ into regions corresponding to black-hole spacetimes and NS spacetimes. First, we recall that the positive roots $0<r_{i}<r_{o}<r_{c}$ of the quartic Eq. (\ref{Delta=0}), if they exist, are interpreted as the inner BH (Cauchy) horizon, the outer BH horizon and the cosmological horizon, respectively, while the negative root $r^{-}_{c}<0$ is interpreted as the dual of the cosmological horizon \cite{Omw-etal:2022:EPJC:}. From the properties of Eq. (\ref{Delta=0}) it can be deduced that the partitioning of the parameter plane $(a^2\text{-}y)$ into individual regions is given by the functions $y_{min(h)}(a^2)$ and $y_{max(h)}(a^2)$, which are defined as the local minima and maxima, respectively, of the function
\be
y_{h}(r;a^2)\equiv \frac{r^2-2r+a^2}{r^4}, \label{yh(r,a2)}
\ee
where the equation $y=y_{h}(r;a^2)$ represents the solution of Eq. (\ref{Delta=0}) with respect to the variable $y$. Although the plot of the function $y_{h}(r;a^2)$ has already been presented in Ref. \cite{Slany:2023:PHYSR4:} for significant values of the parameter $a^2$, it is reproduced here in Fig. \ref{fig_yh(r,a2)} for later reference.

\begin{figure}[htbp]
	\includegraphics[width=\linewidth]{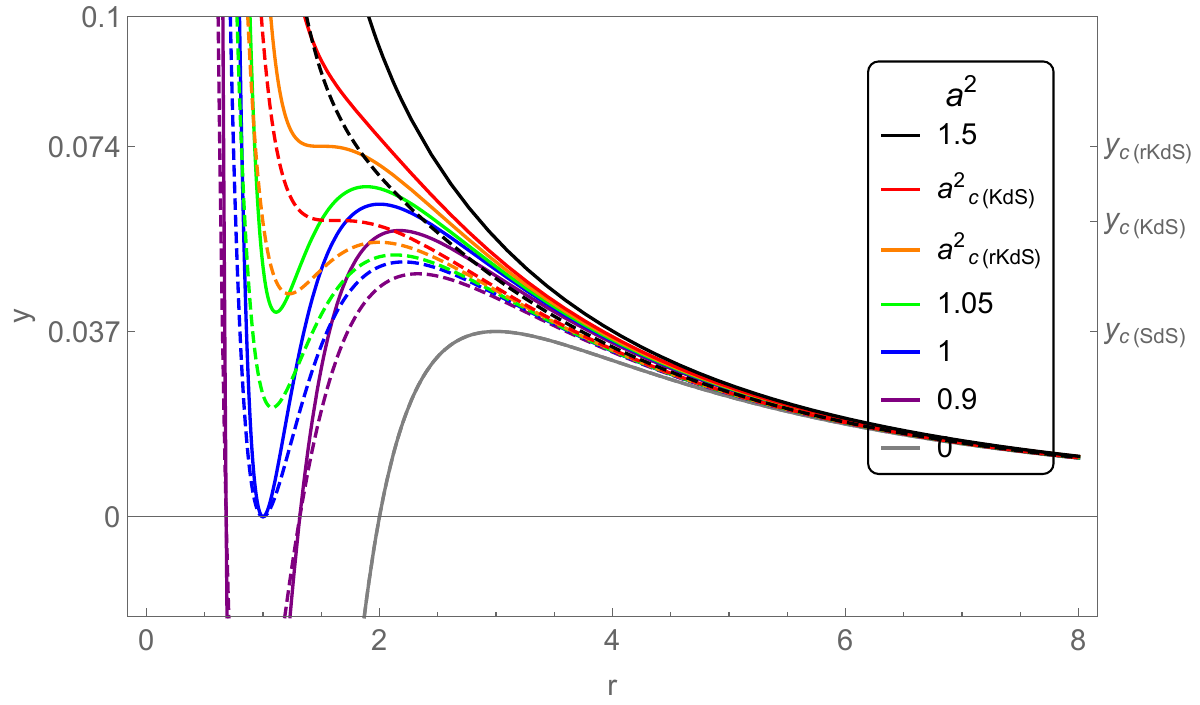}
	\caption{Functions $y_{h}(r;a^2)$ shown for significant values of the spin parameter $a^2$. The solid lines indicate functions corresponding to rKdS spacetimes, while the dashed lines indicate functions corresponding to KdS spacetimes. The static region $\Delta_{\boxed{r}} >0$ of a family of spacetimes with given spin parameter $a^2_{0}$, corresponds to the area under the curve $y=y_{h}(r;a^2_{0})$. The intersections of the line $y=y_{0}$ with the curve $y=y_{h}(r;a^2_{0})$ represent the horizons of the chosen spacetime with parameters $(a^2_{0},y_{0})$.}\label{fig_yh(r,a2)}
\end{figure}

\begin{figure}[htbp]
	\centering
	\begin{tabular}{c}
		\includegraphics[width=\linewidth]{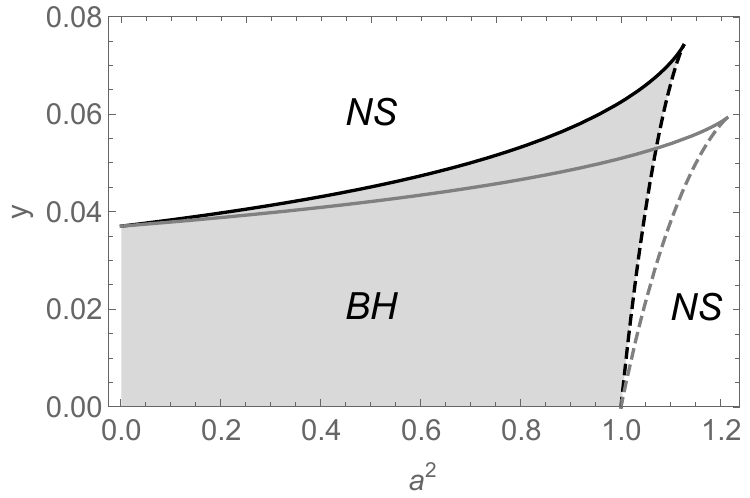}\\
		(a)\\
		\includegraphics[width=\linewidth]{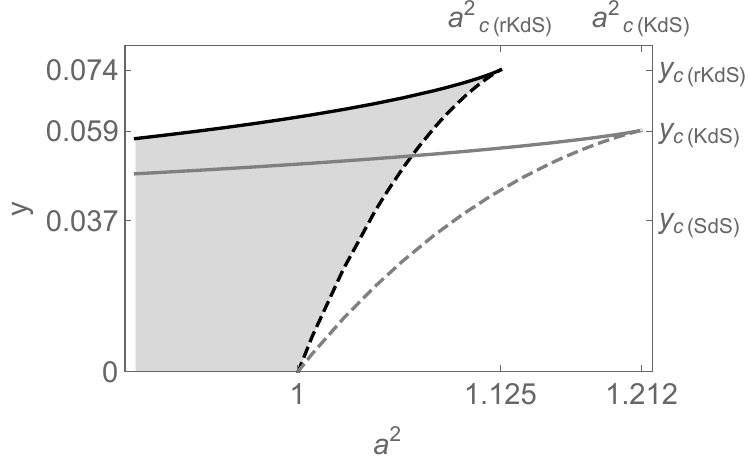}\\
		(b)		
	\end{tabular}
	\caption{(a) Behavior of the functions $y_{max(h)}(a^2)$ (full black curve) and $y_{min(h)}(a^2)$ (dashed black curve) dividing the parameter plane into the regions corresponding to the BH and NS spacetimes of the rKdS geometry. The corresponding curves for KdS spacetimes are also shown in gray for comparison. (b) Detail of the figure above with the critical spacetime parameters displayed.}\label{fig_ymaxminh(a2)}
\end{figure}

In the case of rKdS geometry, it can be shown that the functions $y_{min(h)}(a^2)$ and $y_{max(h)}(a^2)$ can be expressed in the form 
\be
y_{max/min(h)}(a^2)\equiv \frac{-8a^4+36a^227\pm (9-8a^2)^{3/2}}{32a^6}. \label{ymaxminh(a2)}
\ee
The behavior of the functions defined by Eq. (\ref{ymaxminh(a2)}) is shown in Fig. \ref{fig_ymaxminh(a2)} in comparison with the functions $y_{max/min(h)}(a^2)$ corresponding to the KdS spacetimes. These functions, which we introduced in our previous article \cite{Char-Stu:2024:PHYSR4:}, are defined by much more complex expressions, so we will not repeat them here. The meaning of the functions $y_{min(h)}(a^2)$ and $y_{max(h)}(a^2)$ is as follows: \footnote{Here and in the following we restrict our discussion to positive values of radii $r>0$.}

For $0<a^2<1$ and $0<y<y_{max(h)}(a^2)$, or $1< a^2<a^2_{c(rKdS)}\equiv 9/8=1.125$ and $y_{min(h)}(a^2)<y<y_{max(h)}(a^2)$, the geometry describes the rKdS BH spacetimes with four horizons: $r^{-}_{c}<0<r_{i}<r_{o}<r_{c}$. Therefore, the functions $y_{max/min(h)}(a^2)$ give for a fixed parametr $a^2$ the upper/lower limits for the cosmological parameter $y$, for which the rKdS BH spacetimes can exist. If for a given $0<a^2<a^2_{c(rKdS)}$, $y=y_{max(h)}(a^2)$, then $r_{o}=r_{c}$, i.e., the outer and the cosmological horizons coalesce and the geometry describes marginal NS spacetime. Especially for $a^2=0$ and $y=y_{c(SdS)}\equiv 1/27=0.037= \lim_{a^2 \to 0^{+}}(y_{max(h)}(a^2))$ the geometry describes marginal Schwarzschild~de--Sitter NS spacetime. If for a given $a^2$, $y=y_{min(h)}(a^2)$, then $r_{i}=r_{o}$-- i.e., the inner and the outer BH horizons coalesce, and the geometry corresponds to extreme rKdS BH spacetime. In the extreme case $a^2=a^2_{c(rKdS)}$ and $y=y_{c(rKdS)}\equiv 2/27=0.074$, $r_{i}=r_{o}=r_{c}$ and the geometry describes the ultraextreme rKdS NS spacetime with all horizons in the region of $r>0$ merging (cf. Ref. \cite{Char-Stu:2024:PHYSR4:}). In all other cases, the geometry corresponds to the rKdS NS spacetime \cite{Slany:2023:PHYSR4:}.     

 \subsection{Ergosphere} \label{ssec_ergos}
Phenomena such as the Penrose process - the possibility of extracting energy from the rotational energy of a black hole; or superradiance, the amplification of a wave reflected from the ergosphere under certain conditions; and the various variations of these that can be found in the literature, lead us to examine the ergosphere. A detailed discussion of the ergosphere for the case of classical KdS spacetime has been carried out in Ref. \cite{Stu-Char:2024:PHYSR4:}. It is therefore inspiring to see what modifications, if any, correspond to the rKdS metric.

The region of the ergosphere is determined by the condition 
\be
g_{tt}\equiv \frac{a^2 \sin^2\theta - \Delta}{\rho^2}\geq0,
\ee
with equality defining its boundary, the so-called ergosurface. In the equatorial plane, where the rotational effects are strongest, this leads to the quartic equation
\be
r (yr^3-r+2)=0, \label{cub. eq. ergos}
\ee
which has one trivial solution $r=0$, and for $y\leq 1/27$ three other real solutions
\be
r^{(-)}_{erg}=-\frac{2}{\sqrt{3y}}\cos[\frac{1}{3}\arccos \sqrt{27y}], \label{r(-)erg}
\ee
\be
r^{\pm}_{erg}= \frac{2}{\sqrt{3y}}\cos[\frac{\pi}{3}\pm \frac{1}{3} \arccos \sqrt{27y}].\label{rpmerg}
\ee
It holds that $r^{(-)}_{erg}<0<r^{+}_{erg}<r^{-}_{erg}<r_{c}$. As can be seen, similarly to the pure Kerr metric, and contrary to the standard KdS spacetime, where the $a^2y$ term appears in the result, in the rKdS case this result does not depend on the value of the spin parameter $a^2$. However, for $a^2y\ll1$, which is practically fulfilled in the case of astrophysically relevant values of the cosmological parameter $y$, the rKdS and KdS results are indistinguishable (cf. Ref. \cite{Stu-Char:2024:PHYSR4:}). In the limit of $y\to 0$, $r^{+}_{erg}=2$, which is the result known for the case of the Kerr metric. 

In the case of rKdS geometry, from Eq. (\ref{rpmerg}), one can derive the critical value of the cosmological parameter $y$ for which the inner and cosmological ergospheres meet each other in the equatorial plane at radius $r_{erg}$, i.e., for which $r^{+}_{erg}=r^{-}_{erg}\equiv r_{erg}$. This value is $y= 1/27=y_{c(SdS)}$. Substituting this into Eq. (\ref{rpmerg}), we get $r_{erg}=3$, which corresponds to the so-called static radius $r_{s}$, which generally depends on the cosmological parameter by the relation $r_{s}=y^{-1/3}$ (see below). 

In the standard KdS spacetime, $r_{erg}=r_{s}$, but in this case, the critical value of the cosmological parameter is dependent the on spin parameter via the relation (see Ref. \cite{Char-Stu:2024:PHYSR4:})
\be
\frac{27y}{(1-a^2y)^3}=1,
\ee
which can be solved with respect to $y$ to obtain 
\be
y_{erg-s}(a^2) \equiv \frac{1}{a^2}+\frac{3\cdot 2^{1/3}}{2a^{3}} \frac{2^{1/3}(\sqrt{4+a^2}-a)^{2/3}-2}{(\sqrt{4+a^2}-a)^{1/3}}. \label{yerg}
\ee

In the case of rKdS BH/NS spacetimes, the ergosphere, which is determined by the condition $g_{tt}<0$, extends in the equatorial plane in the region of positive radii to radii $r_{o}<r<r^{+}_{erg}$/$r^{-}_{erg}<r<r_{c}$ for $y<1/27$, and $0<r<r_{c}$ for $y>1/27$. Therefore, the radius $r^{+}_{erg}$ corresponds to the boundary of the so-called inner--or in the case of BH spacetimes, the so-called BH--ergosphere, while $r^{-}_{erg}$ corresponds to the boundary of the cosmological ergosphere (see Ref. \cite{Char-Stu:2024:PHYSR4:}). The radius $r^{(-)}_{erg}$ corresponds to the ergosphere in the negative-radius region, which is not the focus of this paper.

In general, any point on the boundary of the ergosphere can be determined by the dependence of its latitudinal coordinate $\theta_{erg}$ on the radius in the form $\theta=\theta_{erg}(r)$. For the rKdS case,
\be
\theta_{erg} = \arcsin \sqrt{\frac{\Delta}{a^2}}. \label{theta_erg}
\ee
In the case of KdS geometry, the dependence on the latitudinal coordinate is complicated by the presence of the term $\Delta_{\theta}$, which leads to \cite{Stu-Char:2024:PHYSR4:}
\be
\theta^{(KdS)}_{erg}=\arcsin \sqrt{\frac{I-\sqrt{I^2-4y\Delta_{r}}}{2a^2y}}. \label{theta_erg_KdS}
\ee

\begin{figure*}[h!]
	\centering
	\begin{tabular}{ccc}
		\includegraphics[width=0.3\textwidth]{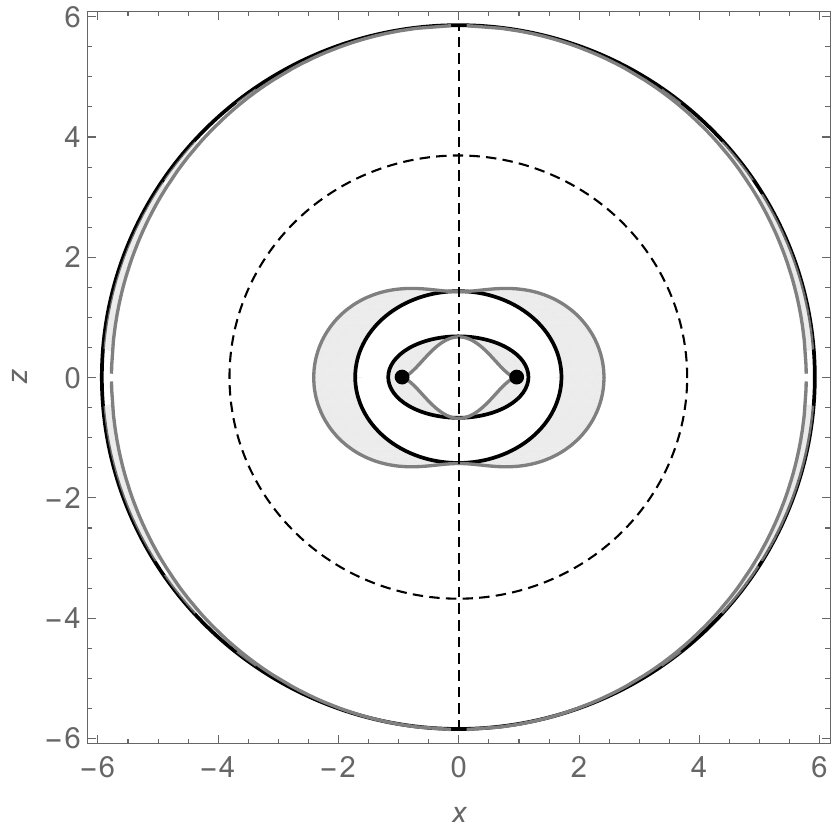}&\includegraphics[width=0.3\textwidth]{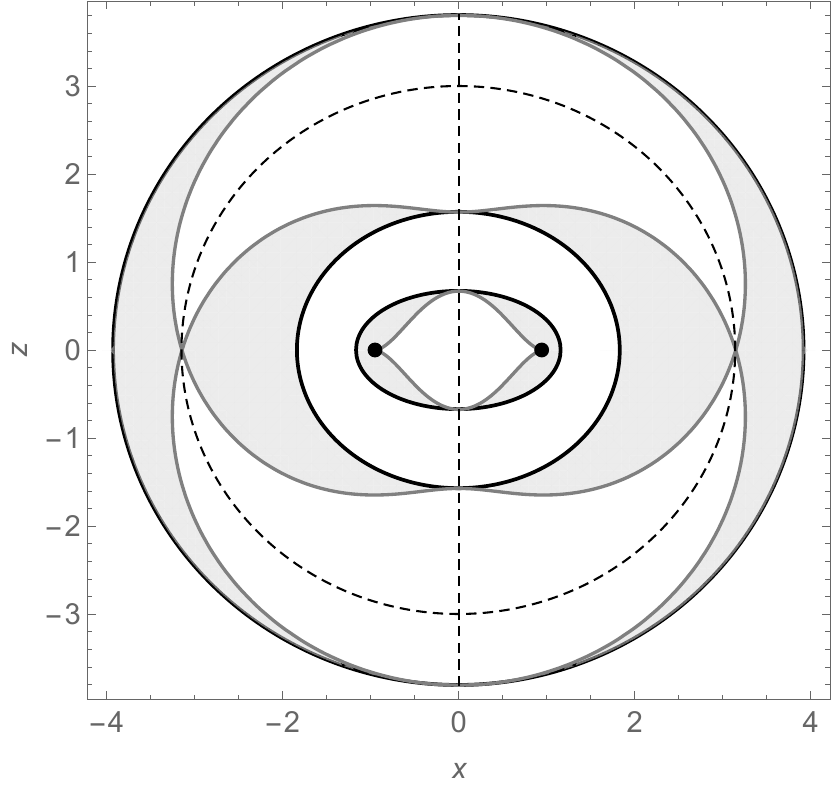}&\includegraphics[width=0.3\textwidth]{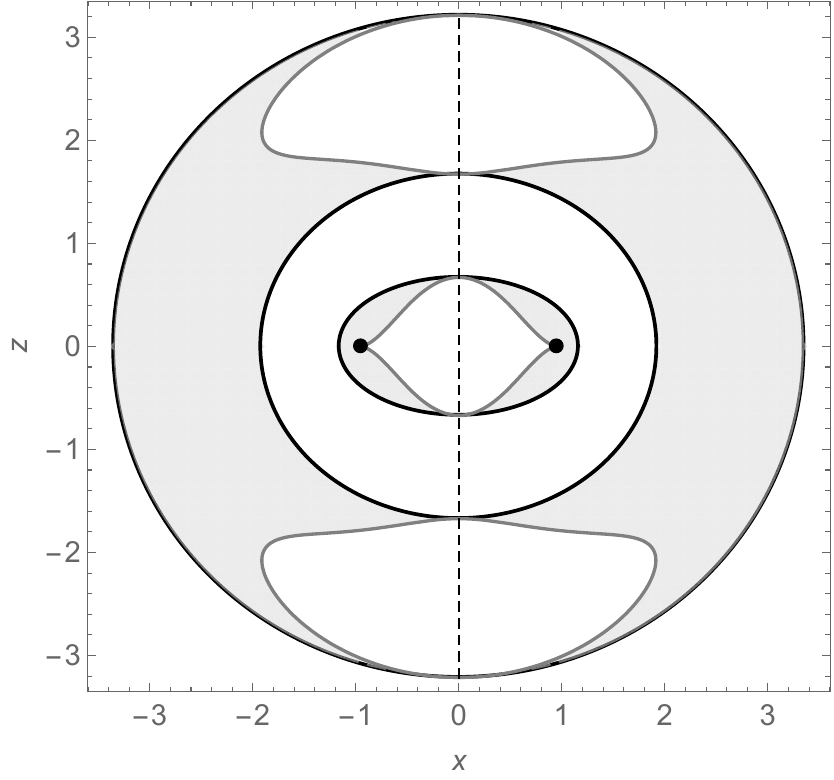}\\
		(a) $a^2=0.9, y=0.02$& (b) $a^2=0.9, y=1/27=0.037$& (c) $a^2=0.9, y=0.045$\\
		\includegraphics[width=0.3\textwidth]{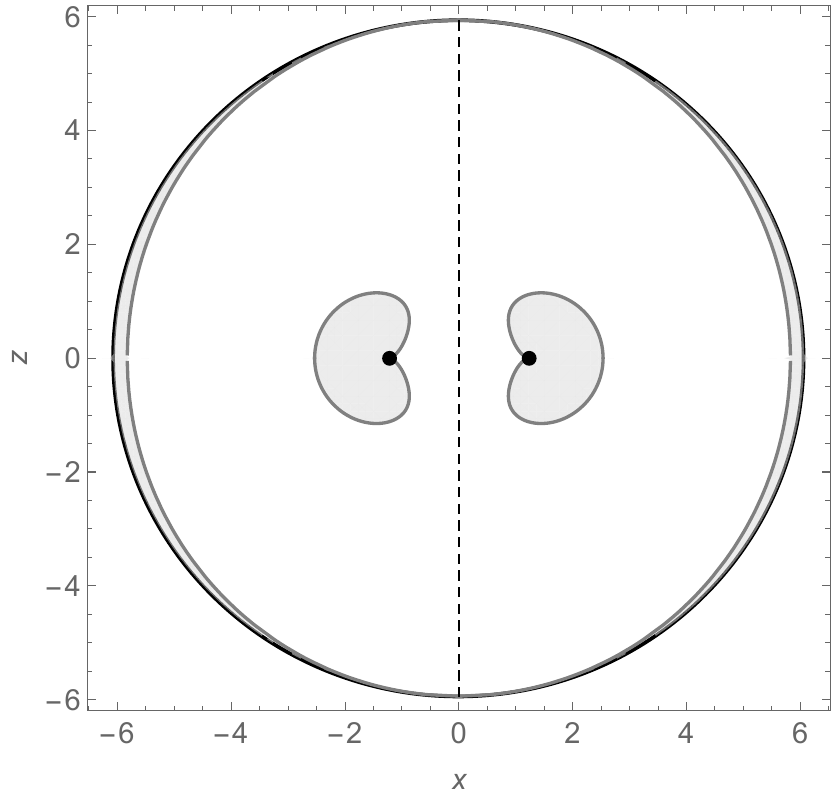}&\includegraphics[width=0.3\textwidth]{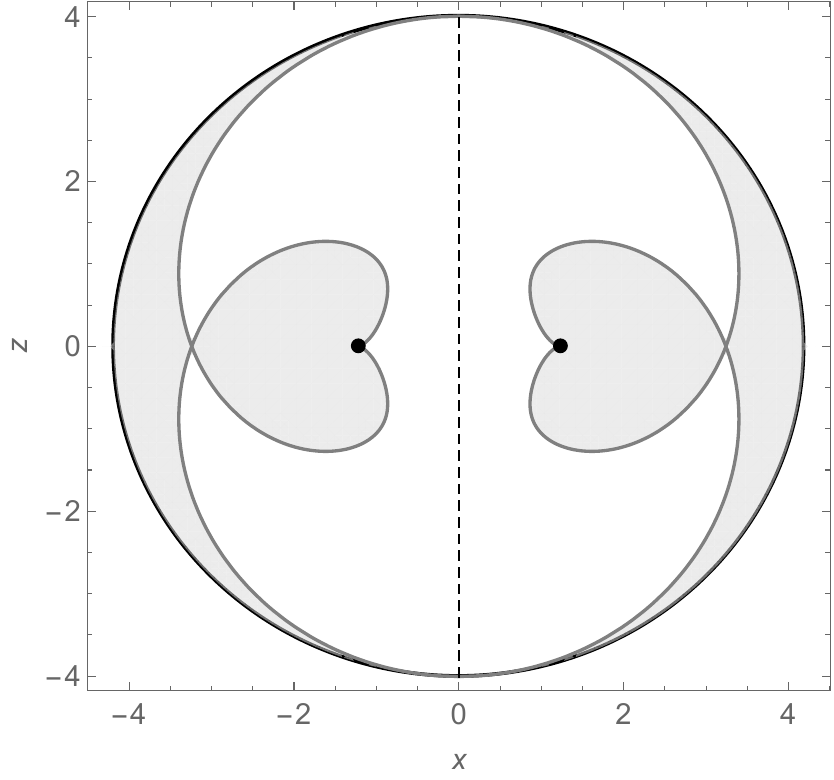}&\includegraphics[width=0.3\textwidth]{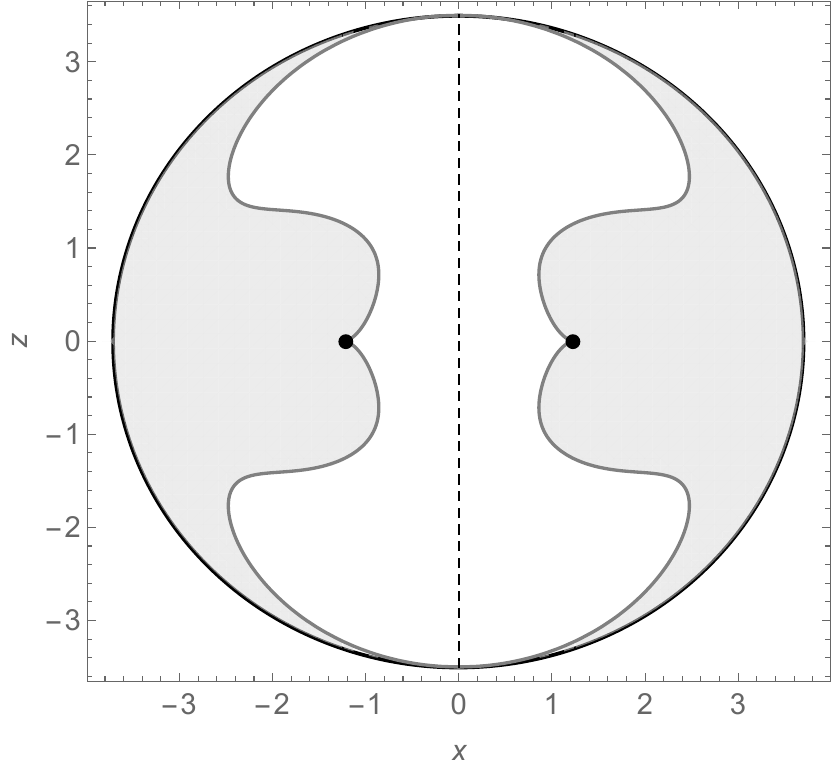}\\
		(d) $a^2=1.5, y=0.02$& (e) $a^2=1.5, y=1/27=0.037$&(f) $a^2=1.5, y=0.045$\\
	\end{tabular}	
\caption{Basic types of ergosphere in rKdS spacetimes compared to horizons and possible static radii shown as sections on the $xz$ Kerr-Schild coordinate plane. Ergosurfaces are shown as gray curves, BH and cosmological horizons are shown as black curves, black dots represent cuts with ring singularity, dotted ellipses represent static radii, and dotted vertical lines represent the spin axis. The ergosphere is shown as a gray region. The top row of figures shows BH spacetimes; the bottom shows NS spacetimes. The left column corresponds to the cases $y<1/27$, the middle column to the cases $y=1/27$, and the right column to the cases $y>1/27$. }     
\label{fig_ergospheres}	
\end{figure*}

It can be seen that, as in the Kerr and KdS cases, the ergosphere is widest in the equatorial plane and touches the horizons at the intersections with the axis of rotation. Further, in rKdS spacetimes, the ergosphere never interferes with the $r=0$ disk, contrary to the standard KdS case where this occurs for the $a^2y>1$ case. Otherwise, it can be concluded that the behavior of the ergosphere is qualitatively the same in both rKdS and KdS spacetimes. Its individual types in rKdS spacetimes are shown in Fig. \ref{fig_ergospheres}.

\clearpage
\subsection{Static radius} \label{stac rad}
The static radius $r_{s}$ is defined as the radial distance at which the gravitational attraction of the central mass is balanced by cosmological repulsion. It is independent of the spin parameter and for both rKdS and KdS spacetimes, $r_{s}=y^{-1/3}$ \cite{Slany:2023:PHYSR4:}. 
  
\subsection{LNRFs}
The four-velocity of the locally nonrotating observer is given by the standard formula
\be
u_{LNRF}=\gamma (\frac{\partial}{\partial t}+\Omega_{LNRF}\frac{\partial}{\partial \phi}), \label{uLNRF}
\ee
where the normalization condition $u^{\mu}u_{\mu}=-1$ implies
\be
\gamma=\frac{1}{\sqrt{-g_{tt}+g_{t \phi}^{2}/g_{\phi \phi}}}=\sqrt{\frac{A}{\rho^2 \Delta}}, \label{gamma}
\ee
and 
\be
\Omega_{LNRF}=-\frac{g_{t \phi}}{g_{\phi \phi}}=a\frac{(r^2+a^2)-\Delta}{A} \label{OmegaLNRF}
\ee
is the angular velocity of the LNRF observer.
The LNRF tetrad of basis vectors can then be expressed using metric coefficients of the form
\bea e^{\mu}_{(t)}&=&u^{\mu}_{LNRF}=\gamma (\delta^{\mu}_{t}+\Omega_{LNRF}\delta^{\mu}_{\phi})\nonumber \\
&=&\sqrt{\frac{A}{\rho^2 \Delta}}(\delta^{\mu}_{t}+\Omega_{LNRF}\delta^{\mu}_{\phi}),\\
e^{\mu}_{(r)}&=&\frac{1}{\sqrt{g_{rr}}} \delta ^{\mu}_{r}=\sqrt{\frac{\Delta}{\rho^2}} \delta ^{\mu}_{r},\\
e^{\mu}_{(\theta)}&=&\frac{1}{\sqrt{g_{\theta \theta}}}\delta ^{\mu}_{\theta}=\frac{1}{\sqrt{\rho^2}}\delta ^{\mu}_{\theta},\\ 
e^{\mu}_{(\phi)}&=&\frac{1}{\sqrt{g_{\phi \phi}}}\delta ^{\mu}_{\phi}=\sqrt{\frac{\rho^2}{A\sin^2\theta}}\delta ^{\mu}_{\phi},\label{e^mu_(phi)} 
\eea
and the corresponding dual 1-forms 
\bea
\omega ^{(t)}_{\mu}&=&\gamma^{-1}\delta^{t}_{\mu}=\sqrt{\frac{\rho^2 \Delta}{A}} \delta^{t}_{\mu},\\
\omega ^{(r)}_{\mu}&=&\sqrt{g_{rr}}\delta^{r}_{\mu}=\sqrt{\frac{\rho^2}{\Delta}}\delta^{r}_{\mu},\\
\omega ^{(\theta)}_{\mu}&=&\sqrt{g_{\theta \theta}}\delta^{\theta}_{\mu}=\sqrt{\rho^2}\delta^{\theta}_{\mu},\\
\omega ^{(\phi)}_{\mu}&=&\sqrt{g_{\phi \phi}}(\delta^{\phi}_{\mu}-\Omega_{LNRF}\delta ^t_{\mu}),\nonumber\\
&=&\sqrt{\frac{A\sin^2\theta}{\rho^2}} (\delta^{\phi}_{\mu}-\Omega_{LNRF}\delta ^t_{\mu}).
\eea
The locally measured contravariant components $p^{(a)}$ and the locally measured covariant components $p_{(a)}$ of the four-momentum $p^{\mu}$ are given by the projections
\bea
p^{(a)}&=&\omega^{(a)}_{\mu}p^{\mu}, \label{p(a)contra} \\
p_{(a)}&=&e^{\nu}_{(a)}p_{\nu}. \label{p(a)cov}
\eea
The orthonormality of the tetrad $e_{(a)}e_{(b)}=\eta_{(a)(b)}$, where $\eta_{(a)(b)}$ is the local Minkowskian metric, and the duality condition $\omega^{(a)}_{\mu}e^{\mu}_{(b)}=\delta^{(a)}_{(b)}$ then ensures that the locally measured quantities are related by the special relativistic way 
\be
p^{(a)}=\eta^{(a)(b)}p_{(b)},\quad p_{(a)}=\eta_{(a)(b)}p^{(b)}. \label{p(a)cov vs p(a)contra}
\ee

\subsection{Equation of geodetic motion of the test particle }
The rKdS spacetimes possesses the same symmetries as the standard KdS spacetimes, and therefore the same associated Killing vector fields, i.e., the time Killing vector field  $\xi_{(t)}=\partial/\partial t$ and the axial Killing vector field $\xi_{(\phi)}=\partial/\partial \phi$. That is, the projections $\cale \equiv -\xi^{\mu}_{(t)}p_{\mu}$ and $\calphi \equiv \xi^{\nu}_{(\phi)}p_{\nu}$ of the four-momentum $p^{\mu}=\din x^{\mu}/\dbe \lambda$ of the test particle are the constants of motion, called "energy" and "angular momentum," respectively. Here, $\lambda$ denotes the affine parameter which is related to the proper time $\tau$ of the particle by the relation $\tau=m \lambda$. Another, trivial, constant of motion is the rest mass of the particle $m=\sqrt{-p^{\mu}p_{\mu}}$. It turns out that, as in the case of the standard KdS spacetimes, also in the rKdS case, the Hamilton-Jacobi equation, describing motion of test particles, is completely separable. One consequence is the existence of the so-called fourth Carter constant $\calk$, connected with hidden symmetry of the rKdS spacetime. Since the metric (\ref{ds2}) has formally the same form as the Kerr metric, where the only difference is the presence of the cosmological term $-yr^4$ in Eq. (\ref{Delta}) \cite{Slany:2023:PHYSR4:}, the same properties of the separation constant $\calk$ as in the Kerr case can be derived - it is non-negative $\calk \geq 0$, while the equality $\calk=0$ occurs only for photons moving along the spin axis $\theta=0,\pi$ \cite{Mis-Tho-Whe:1973:Gravitation:}, or the so-called principal null congruence (PNC) photons moving along trajectories of constant latitude \cite{Bic-Stu:1976:BULAI:}. Another consequence is that the Carter equations, governing the geodesic motion of test particles, can be written in the following separated form:
\bea
\rho^2 p^{r}&=&\pm \sqrt{R}, \label{CarterR}\\
\rho^2 p^{\theta}&=&\pm \sqrt{W}, \label{CarterW}\\
\rho^2 p^{\phi} &=& \frac{a P_{r}}{\Delta}-\frac{P_{\theta}}{\sin^2\theta }\label{CarterPhi} \\
\rho^2 p^{t} &=&\frac{(r^2+a^2)P_{r}}{\Delta}-a P_{\theta}.\label{CarterT} 
\eea

Here
\be
R(r) = P_{r}^2 - \Delta (m^2r^2+\calk), \label{R} 
\ee

\be
W(\theta) = \calk-a^2m^2\cos^2\theta- P^{2}_{\theta}\csc^{2} \theta, \label{W}
\ee
\be P_{r}=\cale (r^2+a^2)-a\calphi,\ee
\be P_{\theta}=a\cale \sin^2\theta-\calphi. \ee 

Note that the Carter equations (\ref{CarterR})-(\ref{CarterT}) for the rKdS metric can be formally obtained from the equations in the KdS case by simply replacing $\Delta_{r}\to \Delta$ and setting $\Delta_\theta=1$, $I=1$, or $a^2y=0$ for short \cite{Slany:2023:PHYSR4:}. \footnote{Of course, the parameters $a^2, y$ of the spacetimes described by the rKdS metric can give arbitrary values $a^2y>0$.} 

In Eqs. (\ref{R}) and (\ref{W}), the constant $\calk$ is often replaced by the constant of motion $\calq$, obtained by combining $\cale$, $\calphi$ in the form
\be
\calq=\calk-(\calphi-a\cale)^2, \label{calQ}
\ee
which is in fact the original Carter-derived separation constant of that name. While $\calk$ is always non-negative, $Q$ can take negative values, where zero corresponds to motion confined to the equatorial plane.

In analogy with our work \cite{Stu-Char-Sche:2018:EPJC:}, we introduce rescaled motion constants called specific energy $E=\cale/m$, specific angular momentum $\Phi=\calphi/m$ and rescaled Carter constants $K=\calk/m^2$ and $Q=\calq/m^2$. Given Eq. (\ref{calQ}), it is obvious that
\be
Q=K-(\Phi-aE)^2.\label{Q}
\ee

\section{Classification of the  {\lowercase{r}}KdS spacetimes according to character of photon motion} \label{sec3}
In this section we compare the properties of photon motion in rKdS and KdS spacetimes, and propose a classification of rKdS spacetimes in analogy with our previous work \cite{Char-Stu:2017:EPJC:}.

\subsection{Carter equations for photon motion}
First, for $\cale \neq 0$, we introduce the constants of photon motion $\ell \equiv \Phi/\cale$: the impact parameter, which is commonly replaced by the rescaled impact parameter $X\equiv \ell - a$, and the rescaled fourth Carter motion constant $q\equiv \calq/\cale^2=\calk/\cale^2-X^2$. Using these constants of motion, the Carter equations for photon motion can then be written in the form
\bea
\frac{\rho^2}{\cale} k^{r}&=&\pm \sqrt{\bar{R}}, \label{CarterRphot}\\
\frac{\rho^2}{\cale} k^{\theta}&=&\pm \sqrt{\bar{W}}, \label{CarterWphot}\\
\frac{\rho^2}{\cale} k^{\phi} &=&\frac{a}{\Delta}(r^2-aX)+\frac{X+a\cos^2\theta}{\sin^2\theta} \label{CarterPhiphot} \\
\frac{\rho^2}{\cale} k^{t} &=&\frac{(r^2+a^2)(r^2-aX)}{\Delta}+a(X+a\cos^2\theta),\label{CarterTphot} 
\eea
where
 \bea
\bar{R}(r)&=&(r^2 - aX)^2 -\Delta(X^2 + q),\label{barR}\\
\bar{W}(\theta)&=&(X^2+q)-\frac{(a\cos^2\theta+X)^2}{\sin^2\theta}.\label{barW}
\eea
Here, $k^{\mu}=\din x^{\mu}/\dbe \lambda^\prime$ is the wave 4-vector of the photon, and $\lambda^\prime$ is the affine parameter, such that $k^{\mu}k_{\mu}=0$. 

In the case of rKdS spacetime, the function $\bar{W}(\theta)$ describing the latitudinal motion is identical to the corresponding function in the pure Kerr case; therefore, the character of photon motion in rKdS and Kerr spacetimes is the same. In the Kerr case, this motion has been studied in detail--e.g., in Ref. \cite{Bic-Stu:1976:BULAI:}--the results of which we will not repeat here. For the radial function $\bar{R}(r)$, it can be seen that it is formally identical for both types of rKdS and KdS spacetimes, and that the only difference is in the definition of the term $\Delta_{\boxed{r}}$. The question then arises whether and how this modification is reflected in the nature of the radial motion of photons.

The radial motion of photons in the case of the KdS spacetimes has been studied extensively in Ref. \cite{Char-Stu:2017:EPJC:} by analyzing the effective potentials $X_{\pm}(r)$ that determine the radial turning points. They are given by the condition $\bar{R}(r)=0$, from which it follows that
\be
X=X_{\pm}(r)\equiv \frac{ar^2\pm \sqrt{\Delta [r^4+q(a^2-\Delta)]}}{a^2-\Delta}. \label{eff.pot.}
\ee
In Ref. \cite{Char-Stu:2017:EPJC:}, we have particularly focused on the study of the properties of SPOs. Following this work and in order to compare the results for the KdS and rKdS spacetimes, we also present here a classification of the rKdS spacetimes in terms of the properties and structure of the SPOs. For the sake of brevity, however, we will not repeat the whole analysis here. Instead, we only outline the basic steps here, referring the reader to the aforementioned work for details of the procedure used.

\subsection{SPOs} \label{ssec_SPOs}
\begin{figure}[htbp]
	\centering
	\begin{tabular}{c}
		\includegraphics[width=\linewidth]{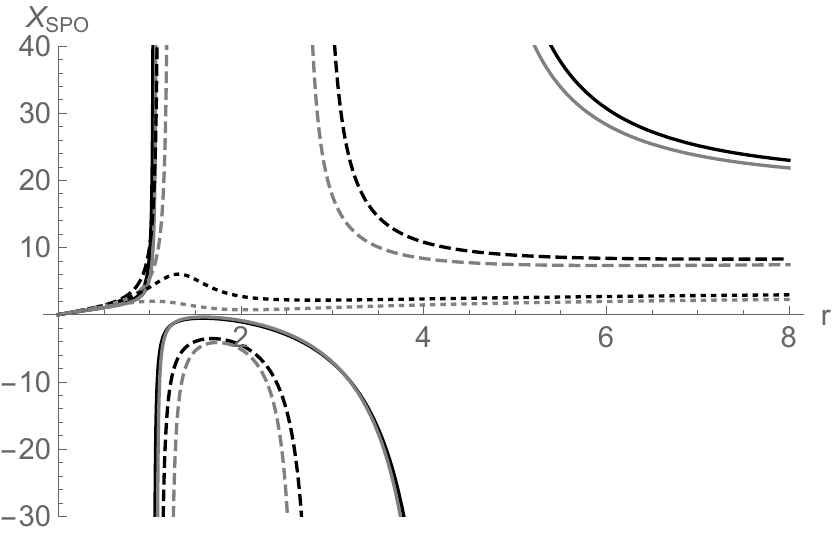}\\
		(a)\\
		\includegraphics[width=\linewidth]{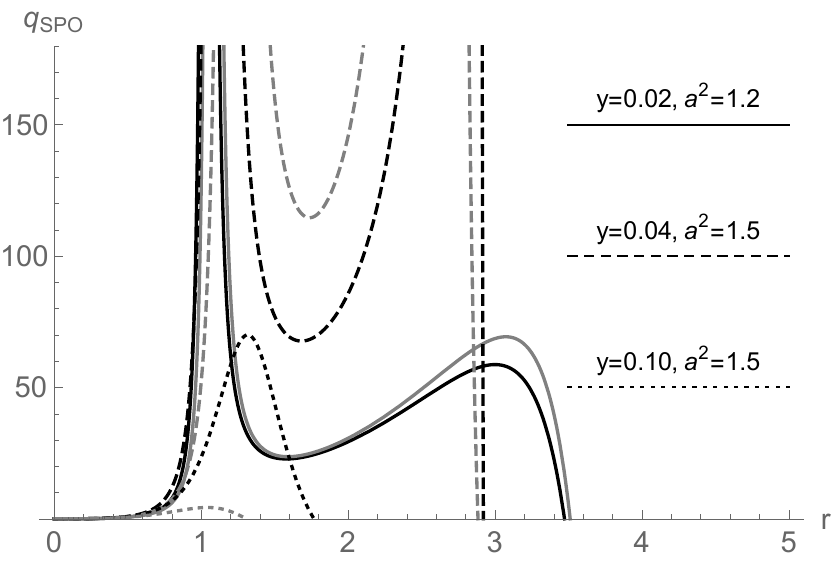}\\
		(b)		
	\end{tabular}
	\caption{Typical behavior of the functions (a) $X_{\spo}(r)$ and  (b) $q_{\spo}(r)$. Solid curves qualitatively describe the behavior corresponding to spacetimes IV and V, dashed curves correspond to spacetimes VI, and dotted curves to spacetimes VII, according to the classification introduced in Ref. \cite{Char-Stu:2017:EPJC:}, which is recalled below. For comparison, we show the cases corresponding to rKdS (black) and KdS (gray) spacetime.}\label{fig_Xspo_qspo}
\end{figure}

The SPOs are based on the conditions 
\be
\bar{R}(r)=0,\quad \oder{\bar{R}}{r}=0,\label{cond_spo}
\ee
which imply that the photon constants of motion in these orbits can be written in the form
\be
X=X_{\spo}(r) \equiv \frac{r(r^2+2a^2-3r)}{a(2yr^3-r+1)}, \label{Xspo}
\ee
or, alternatively,
\be
\ell=\ell_{\spo}(r) \equiv \frac{(1+2a^2y)r^3-3r^2+a^2r+a^2}{a(2yr^3-r+1)}, \label{lspo}
\ee
and 
\bea
q&=&q_{\spo}(r) \\
&\equiv& -\frac{r^3}{a^2}\; \frac{r(r-3)^2+4a^2(yr^3-1)}{(2yr^3-r+1)^2}.\nonumber \label{qspo}
\eea
There is another solution $X=r^2/a$, $q=-r^4/a^2$ to the system of Eq. \ref{cond_spo}, but it does not satisfy the latitudinal motion reality condition $\bar{W}(\theta)\geq0$, and therefore we do not consider it. The plot of the functions $X_{\spo}(r)$ [$\ell_{\spo}(r)$] and $q_{\spo}(r)$ is shown in Fig. \ref{fig_Xspo_qspo}. It can be seen that their behavior is qualitatively the same as in the standard KdS case (cf. Refs. \cite{Char-Stu:2017:EPJC:,Stu-Char:2024:PHYSR4:}).

For later reference we present the divergence points of the functions $X_{\spo}(r)$ ($\ell_{\spo}(r)$) and $q_{\spo}(r)$ at positive radii, which read
\be
r^{\pm}_{d(\spo)}=\frac{2}{\sqrt{6y}}\cos\left[ \frac{\pi}{3}\pm \frac{1}{3}\arccos \sqrt{\frac{27y}{2}}\right] , \label{rdexpm}
\ee 
where $0<r^{+}_{d(\spo)}<r^{-}_{d(\spo)}$ for both rKdS and KdS geometries. For rKdS case, they merge at $r=1.5$ for $y=2/27=y_{c(rKdS)}$, such that for $y>2/27$, there are no divergences.

For the KdS case (see Ref. \cite{Char-Stu:2024:PHYSR4:}), the value of the cosmological parameter, for which the points of divergence coalesce, depends on the spin parameter implicitly according to the relation
\be
\frac{27y}{2(1-a^2y)^3}=1, \label{rdex_coal}
\ee
which can be solved with respect to $y$ in the form 
\be
y=y^{(KdS)}_{d(\spo)}(a^2) \equiv \frac{1}{a^2}-\frac{6}{\sqrt{2}a^3}\sinh \left[1/3 \sinh^{-1} \frac{\sqrt{2}a}{2}\right] . \label{yd(spo)}
\ee

Stable/ marginally stable/ unstable SPOs exist on radii where the condition $\mathrm{d}^2R/\mathrm{d}r^2\gtreqless0 $ is satisfied, which in region of positive radii $r>0$ implies the inequality 
\be
3yr^4-(1+4a^2y)r^3+3r^2-3r+a^2\gtreqless0. \label{stab_cond}
\ee

\subsection{Significant types of the SPOs}
There are two important types of SPOs that play a key role in the classification of rKdS spacetimes, namely, equatorial circular photon orbits (ECPOs) and polar SPOs that intersect the rotation axis.
\subsubsection{ECPOs}
\begin{figure}
	\includegraphics[width=\linewidth]{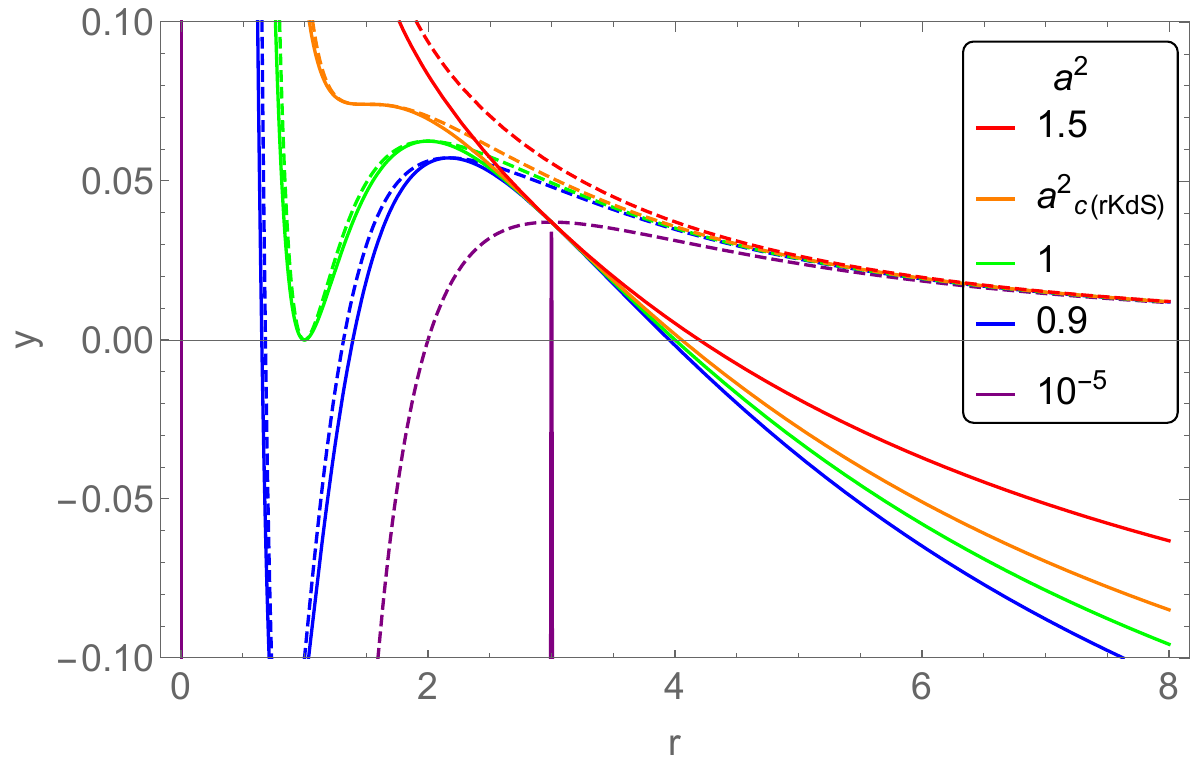}
	\caption{Functions $y_{ECPO}(r;a^2)$ (solid curves) and $y_{h}(r;a^2)$ (dashed curves) in rKdS geometry shown for some representative values of the cosmological parameter $a^2$.  } \label{fig_yecpo}
\end{figure}

The ECPOs are determined by the equation
\be
q_{\spo}(r)=0. \label{qspo=0}
\ee
It is convenient to solve Eq. (\ref{qspo=0}) with respect to the variable $y$, which in the rKdS case has the form
\be
y=y_{ECPO}(r;a^2)\equiv \frac{4a^2-r(r-3)^2}{4a^2r^3}, \label{yECPO(r,a2)}
\ee
so that we can compare the positions of the ECPOs with the horizons by comparing the behavior of the functions $y_{ECPO}(r;a^2)$, $y_{h}(r;a^2)$ in one graph. This is done in Fig. \ref{fig_yecpo}, where for given spacetime parameters $(a^2_{0}, y_{0})$, the number and distribution of ECPOs relative to the horizons can then be interpreted as the intersections of the line $y=y_{0}$ with the curves $y=y_{ECPO}(r;a^2_{0})$ and $y=y_{h}(r;a^2_{0})$. It can be seen that these functions have common local extrema. Restricting ourselves to the stationary region $\Delta>0$, it is clear that in the BH spacetimes there are two ECPOs, an inner one with radius $r^{+}_{ph}$ and an outer one with radius $r^{-}_{ph}$, $r^{+}_{ph}\leq r\leq r^{-}_{ph}$, which interval defines the extension of the SPOs in the equatorial plane, while in NS spacetimes there is only one ECPO with radius $r^{-}_{ph}$, with the SPOs extending in the range $0<r\leq r^{-}_{ph}$. It can be shown that the circular orbits with radii $r^{+}_{ph}/r^{-}_{ph}$ correspond to photons with a positive/negative impact parameter $\ell_{\spo}(r^{+}_{ph})>0/\ell_{\spo}(r^{-}_{ph})<0$, which we call corotating/counterrotating (see bellow). In all cases, the ECPOs are unstable.

\begin{figure}
	\includegraphics[width=\linewidth]{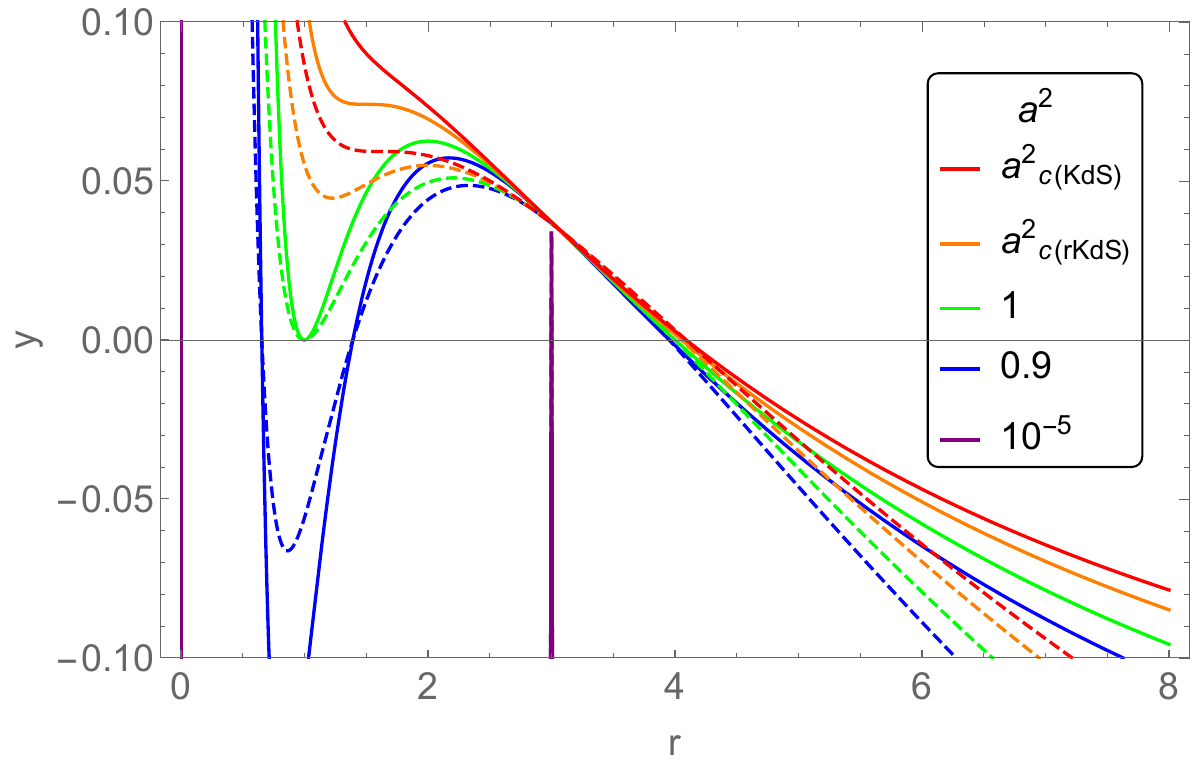}
	\caption{Comparison of functions $y_{ECPO}(r;a^2)$ for both rKdS and KdS spacetimes, shown for significant values of the spin parameter $a^2$. The solid lines indicate functions corresponding to rKdS spacetimes, while the dashed lines indicate functions corresponding to KdS spacetimes.} \label{fig_yecpo_comp}
\end{figure}

The corresponding function in the KdS case has a significantly different form, so we list it separately again:
\be
y=y^{(KdS)}_{ECPO}(r;a^2)\equiv \frac{2\sqrt{r(3r^2+a^2)}-r(3+r)}{r^2a^2}. \label{yECPO(r,a2)_KdS}
\ee
A comparison of the functions defined in Eqs. (\ref{yECPO(r,a2)}) and (\ref{yECPO(r,a2)_KdS}) is shown in Fig. \ref{fig_yecpo_comp} for selected parameters $a^2$, including significant values in both rKdS and KdS geometries.

The radii $r^{\pm}_{ph}$ for rKdS spacetimes are given by the equations   
\be
r^{+}_{ph}=\frac{2}{1+4a^2y}[1+\sqrt{1-12a^2y}\cos\frac{4\pi+\Psi}{3}], \label{rphplus}
\ee
and
\be
r^{-}_{ph}=\frac{2}{1+4a^2y}[1+\sqrt{1-12a^2y}\cos\frac{\Psi}{3}], \label{rphminus}
\ee
where
\be
\cos \Psi=\frac{16a^4y(1+2a^2y)-2a^2(18y-1)-1}{(1-12a^2y)^{3/2}}. \label{cos(Psi)}
\ee
In the case of naked singularities, the expression for $r^{+}_{ph}$ has no physical meaning, since it is imaginary. Moreover, if $a^2y>a^2_{c(rKdS)}\times y_{c(rKdS)}=1/12=0.083$, expression (\ref{rphminus}) must be replaced by
\be
r^{-}_{ph}=\frac{2}{1+4a^2y}[1+\sqrt{12a^2y-1} \sinh\frac{\bar{\Psi}}{3}], \label{rphminus_new}
\ee
where
\be
\sinh \bar{\Psi}=\frac{16a^4y(1+2a^2y)-2a^2(18y-1)-1}{(12a^2y-1)^{3/2}}. \label{sinhbar(Psi)}
\ee

For comparison with the KdS spacetimes, the radii of ECPOs in that case are given by the expressions
\be
r^{(KdS)+}_{ph}=\frac{2}{I^2}(1-a^2y+\sqrt{1-14a^2y+a^4y^2} \cos\frac{4\pi+\Xi}{3}), \label{rphplus_KdS}
\ee
and
\be
r^{(KdS)-}_{ph}=\frac{2}{I^2}(1-a^2y+\sqrt{1-14a^2y+a^4y^2} \cos\frac{\Xi}{3}), \label{rphminus_KdS}
\ee
where
\be
\cos \Xi=\frac{2a^2I^4-33a^2y(1-a^2y)+a^6y^3}{(1-14a^2y+a^4y^2)^{3/2}}. \label{cos(Xi)}
\ee

In the case $a^2y>a^2_{c(KdS)}\times y_{c(KdS)}=7-4\sqrt{3}=0.072$, the formula (\ref{rphminus_KdS}) must be replaced by
\be
r^{(KdS)-}_{ph}=\frac{2}{I^2}(1-a^2y+\sqrt{14a^2y-a^4y^2-1} \sinh\frac{\bar{\Xi}}{3}), \label{rphminus_KdS_new}
\ee
where
\be
\sinh \bar{\Xi}=\frac{2a^2I^4-33a^2y(1-a^2y)+a^6y^3}{(14a^2y-a^4y^2-1)^{3/2}}. \label{sinhbar(Xi)}
\ee

\subsubsection{Polar SPOs} \label{ssec_polar_SPOs}
\begin{figure}
	\includegraphics[width=\linewidth]{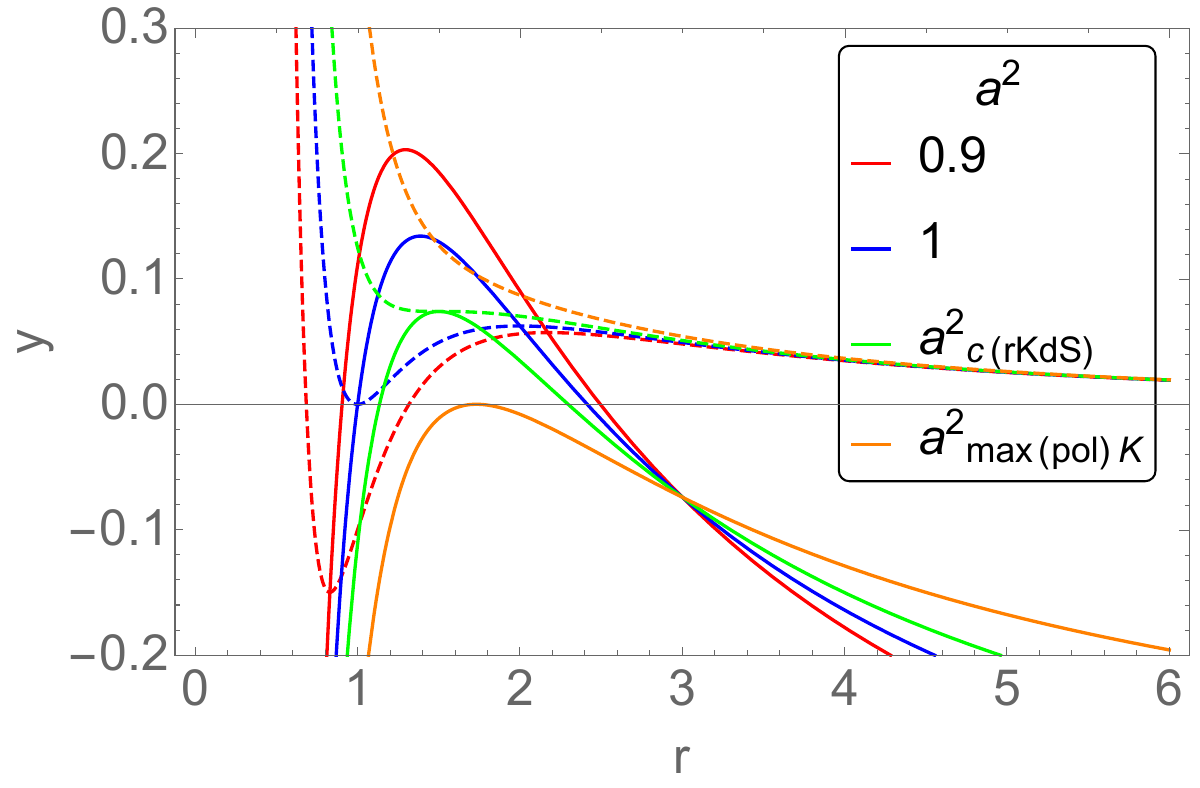}
	\caption{Behavior of the function $y_{pol}(r;a^2)$ (full curves) compared to the function $y_{h}(r;a^2)$ (dashed curves) for some representative values of the cosmological parameter $a^2$. For a given parameter $a^2$, both curves have the same color. } \label{fig_ypol}
\end{figure}

The latter type of SPOs is given by the condition
\be
l_{\spo}(r)=0. \label{lspo=0}
\ee
Solving Eq. (\ref{lspo=0}) with respect to the variable $y$ yields
\be
y=y_{pol}(r;a^2)\equiv \frac{r^2(3-r)-a^2(1+r)}{2a^2r^3}. \label{ypol(r,a2)}
\ee
The number and distribution of the radii of polar SPOs relative to the horizons can then be interpreted in the same manner as in the case of the ECPOs, with Fig. \ref{fig_ypol} displaying the behavior of the function $y_{pol}(r;a^2)$ compared to the function $y_{h}(r;a^2)$. This figure shows, and the calculation confirms, that the graph of the function $y_{pol}(r;a^2)$ intersects the graph of the function $y_{h}(r;a^2)$ at its local extrema, if any. It also can be shown that the raising/descending parts of the curves $y=y_{pol}(r;a^2)$ correspond to the stable/unstable polar SPOs. The local maxima $y_{max(pol)}(a^2)$ of the function $y_{pol}(r;a^2)$ correspond to the coalescence of these orbits, so that for $y<y_{max(pol)}(a^2)$ there are two polar orbits, while for $y>y_{max(pol)}(a^2)$ there are no polar orbits.

We also compare the behaviour of the functions $y_{pol}(r;a^2)$ for selected values of the parameter $a^2$ for both the rKdS and KdS geometries (see Fig. 8).

\begin{figure}
	\includegraphics[width=\linewidth]{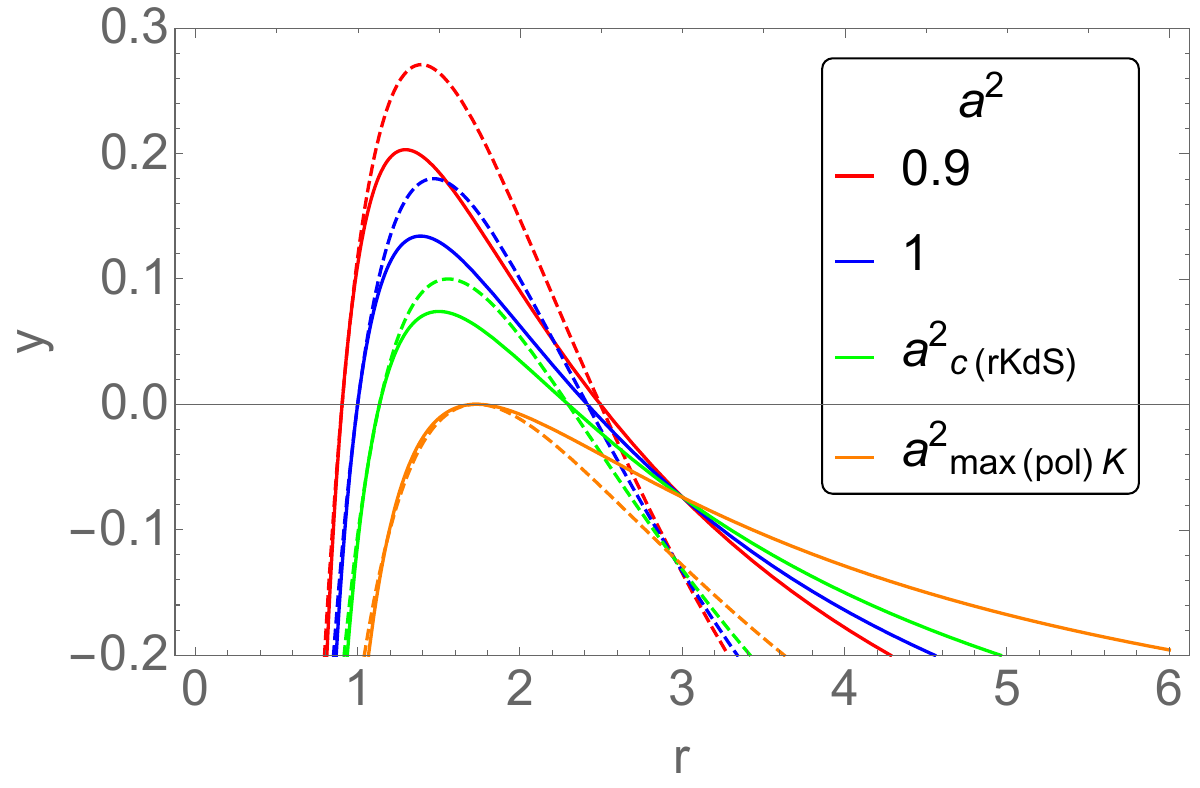}
	\caption{Comparison of functions $y_{pol}(r;a^2)$ for rKdS (full curves) and KdS (dashed curves) spacetimes for the same parameters as in Fig. \ref{fig_ypol}. } \label{fig_ypol_comp}
\end{figure}
One can derive that for rKdS spacetime,
\be
y_{max(pol)}(a^2)\equiv \frac{\sqrt{(9+a^2)^3}-a^3-27a}{27a^3}. \label{ymaxpol(a2)}
\ee
Apparently 
\be
y_{max(pol)}(a^2_{c(rKdS)})=y_{c(rKdS)}. \label{ymaxpol(a2crit)=ycrit}
\ee
The function $y^{(rKdS)}_{max(pol)}(a^2)$ is decreasing for $a^2>0$, while $\lim_{a^2 \to 0^{+}}y^{(rKdS)}_{max(pol)}(a^2)=\infty$, and its zero, which corresponds to the maximum value of the spin parameter enabling the existence of the polar orbits in the Kerr case, is given by 
\be
a^2=a^2_{max(pol)K}\equiv 9+6\sqrt{3}=1.3923. \label{a2max(pol)K}
\ee
This is the same result as for KdS spacetime in our work \cite{Char-Stu:2024:PHYSR4:}, \footnote{It is clear that the results obtained in this work for $y\to 0$ must be the same as in the case of the standard KdS spacetime, since in this limiting case both KdS and rKdS geometries describe the same, i.e., the Kerr spacetime.} where the function analogous to (\ref{ymaxpol(a2)}) is 
\be
y^{(KdS)}_{max(pol)}(a^2)\equiv a^{-2}(\sqrt{\frac{a^2_{max(pol)K}}{a^2}}-1). \label{ypolmax}
\ee

\begin{figure}
	\includegraphics[width=\linewidth]{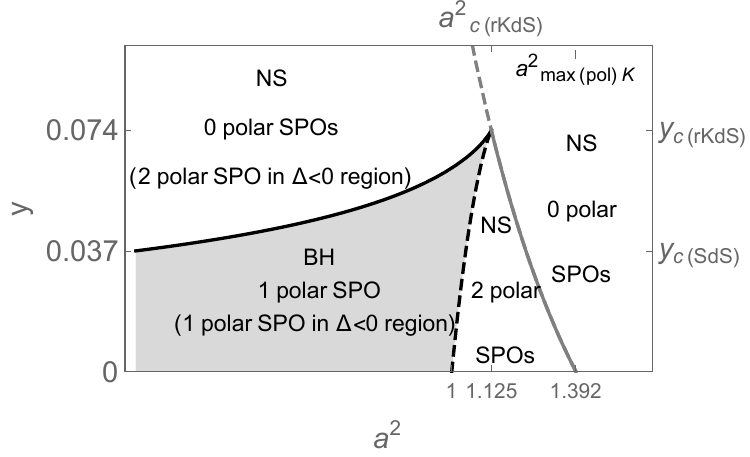}
	\caption{Behavior of the function $y_{max(pol)}(a^2)$ (gray curve) compared to the functions $y_{max(h)}(a^2)$ (black solid curve) and $y_{min(h)}(a^2)$ (black dashed curve) and the induced partition of the parameter plane $(a^2\text{-}y)$ into regions corresponding to BH and NS spacetimes endowed with the indicated number of polar SPOs in the stationary region $\Delta >0$. The dashed part of the curve $y_{max(pol)}(a^2)$ is physically irrelevant, since it separates spacetimes with two polar orbits beyond the cosmological horizon from spacetimes without polar orbits, so effectively there is no difference between them.  
	} \label{fig_ymaxpol}
\end{figure}

It follows from the above that in the rKdS BH spacetimes, there is just one polar SPO in the stationary region between the outer and cosmological horizons, which is unstable. In the rKdS NS spacetimes with spacetime parameters, $[1<a^2<a^2_{c(rKdS)},y<y_{min(h)}(a^2)]$ or $[a^2_{c(rKdS)}<a^2<a^2_{max(pol)K},y<y_{max(pol)}(a^2)]$, there are two polar SPOs in the stationary region $0<r<r_{c}$, with the inner one stable and the outer one unstable.  In the rKdS NS spacetimes with spacetime parameters $[a^2<a^2_{c(rKdS)},y>y_{max(h)}(a^2)]$, both polar SPOs are in the dynamical region $r>r_{c}$; therefore, parts of the curves $y_{pol}(r; a^2)$ with values $y_{h}(r;a^2)<y<y_{max(pol)}(a^2)$ and also the parts of the curve $y_{max(pol)}(a^2)$ with values $y>y_{c(rKdS)}$ are physically irrelevant. The behavior of the function $y_{max(pol)}(a^2)$ in relation to the functions $y_{max(h)}(a^2)$ and $y_{min(h)}(a^2)$ is shown in Fig. \ref{fig_ymaxpol}.

The radii of the polar SPOs in the stationary region of the rKdS spacetimes are
\be
r^{\pm}_{pol}=\frac{1}{1+2a^2y}[1+2\sqrt{1-\frac{a^2(1+2a^2y)}{3}}\cos\frac{\pi \pm \Phi}{3}], \label{rpolpm}
\ee
where
\be
\cos \Phi= \frac{a^2I(1+2a^2y)-1}{(1-\frac{a^2I(1+2a^2y)}{3})^{3/2}}. \label{cosPhi}
\ee

If we compare the expressions (\ref{rpolpm}) and (\ref{cosPhi}) with the corresponding relations for polar SPOs in the KdS case, we see that they have a similar structure \cite{Stu-Char:2024:PHYSR4:}:
\be
r^{(KdS)\pm}_{pol}=\frac{1}{I}[1+2\sqrt{1-\frac{a^2I^2}{3}}\cos(\frac{\pi\pm\psi}{3})], \label{rpolpm_KdS}
\ee

\be
\cos\psi=\frac{a^2I^2-1}{(1-\frac{a^2I^2}{3})^{3/2}}.\label{cosvarphi}
\ee
\subsection{rKdS spacetime classification criteria} \label{ssec_class_criteria}
In Ref. \cite{Char-Stu:2017:EPJC:}, we have proposed a classification of the KdS spacetimes into eight classes denoted by Roman numerals I--VIII. Here, we classify them according to the same criteria, which are the number of horizons as discussed in Sec. \ref{ssec_event_hor}, the character of the ergosphere as addressed in Sec. \ref{ssec_ergos}, and the properties of the SPOs. For these, we then look at their distribution in the equatorial plane, and thus the number of ECPOs, and relative to the ergosphere, the existence of polar SPOs and the associated orientation of the SPOs, and the stability of the SPOs with respect to radial perturbations.

\subsubsection{BH and NS spacetimes}
The main criterion for the classification of rKdS spacetimes is the number of horizons, i.e. into BH and NS spacetimes. In the $(a^2\text{-}y)$ plane, this corresponds to the region bounded by the $y_{min(h)}(a^2)$ and $y_{max(h)}(a^2)$ curves, and the outer region, as discussed in Sec. \ref{ssec_event_hor}. However, according to other criteria, the family of BH spacetimes is further subdivided into classes I--III, and the family of NS spacetimes into classes IV--VII, see below.

\subsubsection{Relative position of SPOs and ergosphere}
The reader can easily conclude that for the spacetime parameters $a^2, y=y_{erg\text{-}ph}(a^2)$, where
\be
y_{erg\text{-}ph}(a^2)\equiv \frac{4(\sqrt{9-16a^2}-1)}{(3+\sqrt{9-16a^2})^3}, \label{yerg-ph}
\ee
the radii $r^{+}_{erg}$, $r^{+}_{ph}$ given by Eqs. (\ref{rpmerg}) and (\ref{rphplus}) are coincident. Thus, according to our classification proposed, the curve $y=y_{erg\text{-}ph}(a^2)$ separates in the parameter plane $(a^2\text{-}y)$ the regions corresponding to the spacetimes of class I, where the SPOs in the stationary region $\Delta>0$ do not interfere with the inner ergosphere, from those where they do. It holds $y_{erg\text{-}ph}(0)=1/27=y_{c(SdS)}$, $y_{erg\text{-}ph}(0.5)=0$, $y_{erg\text{-}ph}(a^2)>0$ for $0\leq a^2 <0.5$, which are the same results as in the KdS case. 

On the other hand, the merger of $r^{-}_{erg}$ and $r^{-}_{ph}$ occurs simultaneously with the merger of $r^{+}_{erg}$, when $y=y_{c(SdS)}$. Thus, the SPOs cannot partially interfere with the cosmological ergosphere; they can either partially interfere with the inner ergosphere, which is the case for class II, corresponding to the BH spacetimes, and for classes IV, V, which correspond to the NS spacetimes,  or the entire SPO region is overlapped with the ergosphere. This is true for spacetimes with $y>1/27$, where the ergosphere covers the entire stationary region in the equatorial plane  (see Sec. \ref{ssec_ergos}). According to our proposed classification, this corresponds to class III in the case of BH spacetimes, or to one of the classes VI, VII in the case of NS spacetimes, whose distinction takes into account other classification criteria (see below). In summary, the line $y=y_{c(SdS)}$ separates, in the parameter plane $(a^2\text{-}y)$ according to our classification, class II from class III, which both correspond to BH spacetimes, and classes IV, V from classes VI, VII, which all correspond to NS spacetimes.

The analogy of the function $y_{erg\text{-}ph}(a^2)$ defined in Eq.~(\ref{yerg-ph}) for the case of KdS spacetime, which similarly determines the boundary between classes~I and II of KdS spacetimes, is parametrically derived in Ref.~\cite{Char-Stu:2017:EPJC:}. 
 
\subsubsection{Orientation and energy of SPOs}
The orientation of the SPOs is standardly defined by the sign of the ratio $k^{(\phi)}/k^{(t)}$ of the locally measured azimuthal and time components of the photon's four-momentum, which defines a directional angle $\calpsi$ such that $\sin \calpsi=k^{(\phi)}/k^{(t)}$. If $\sin \calpsi>0$, we call the appropriate SPO \textit{prograde}, for $\sin \calpsi<0$, we name it \textit{retrograde}. Using relations (\ref{p(a)contra})--(\ref{p(a)cov vs p(a)contra}), where we substitute from (\ref{CarterPhiphot}) and (\ref{CarterTphot}), and the definition of the impact parameter $\ell$, thereby repeating the procedure in \cite{Char-Stu:2017:EPJC:}, we arrive at 
\be
\sin\Psi = \frac{\rho^2\sqrt{\Delta}}{A \sin\theta}\frac{\ell}{1-\Omega_{LNRF}\ell}. \label{sinpsi(ell)}
\ee 
It can be seen that the sign of the impact parameter $\ell$ determines the orientation of the SPOs; therefore, the polar orbits separate the regions of retrograde SPOs (with $\ell<0$) and prograde SPOs (with $0<\ell<1/\Omega_{LNRF}$). However, it can be shown in the same way as we have done in Ref. \cite{Char-Stu:2017:EPJC:} that there exist SPOs with positive impact parameter $\ell>1/\Omega_{LNRF}>0$ that appear to be locally retrograde. The assumption of standard future-oriented photons $k^{(t)}>0$ implies that $\Phi<0$ and $E<0$ simultaneously. A more detailed analysis shows that the inequality $\ell>1/\Omega_{LNRF}$, together with the reality condition $\bar{R}\geq 0$, can be satisfied only in the ergosphere for photons with impact parameter $X\geq X_{+}(r)>0$, which is the conclusion we also reached in Reg. \cite{Char-Stu:2017:EPJC:} for the KdS spacetimes.

\begin{figure}
	\includegraphics[width=\linewidth]{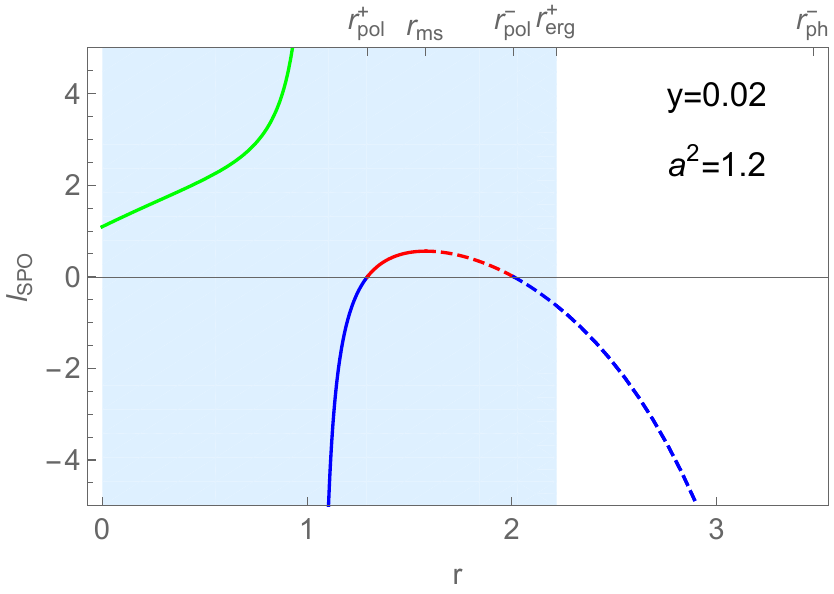}
	\caption{Function $\ell_{\spo}(r)$ plotted for cosmological parameters corresponding to rKdS NS spacetime with two polar orbits. The polar SPOs are given by zeros of $\ell_{\spo}(r)$. Red/blue parts of the curve correspond to prograde/retrograde SPOs, while the green part corresponds to retrograde SPOs with negative covariant energy. Solid/dashed lines indicate stable/unstable orbits, which are separated by the marginally stable orbit $r_{ms}$ (see below). Orbits with negative energy are stable. The gray area indicates the ergosphere.    
	} \label{fig_lspo_1.2_0.02}
\end{figure}

In the rKdS BH spacetimes, there is just one polar SPO $r^{-}_{pol}$, such that for $r^{+}_{ph}<r<r^{-}_{pol}$ the SPOs are prograde, while for $r^{-}_{pol}<r<r^{-}_{ph}$ they are retrograde. In the rKdS NS spacetimes with two polar SPOs, the orbits are prograde in the range $r^{+}_{pol}<r<r^{-}_{pol}$, and retrogade in the range $0<r<r^{+}_{pol}$ and $r^{-}_{pol}<r<r^{-}_{ph}$; in the NS spacetime with no polar orbits, there are only retrograde SPOs (see Sec. \ref{ssec_polar_SPOs}). The distribution of differently oriented orbits in NS spacetime with two polar orbits is shown in Fig. \ref{fig_lspo_1.2_0.02}.

The curve $y_{max(pol)}(a^2)$ forms additional separation in the parameter plane $(a^2\text{-}y)$ dividing the spacetimes of classes VI and IV into the subclasses IVa, IVb and VIa, VIb, where the label "a" indicates a NS spacetime with two polar SPOs, while "b" denotes a NS spacetime with no polar SPOs.

The function $y^{(KdS)}_{max(pol)}(a^2)$ induces an analogous division of KdS spacetime in the appropriate parameter plane. 

\subsubsection{Stability and energy of SPOs}
The condition of the stability or instability of the SPOs with respect to radial perturbations is given by the additional relation
\be
\frac{\mathrm{d^2}\bar{R}}{\mathrm{d}r^2}\lessgtr 0, \label{stability_conditions}
\ee
which must be satisfied simultaneously with Eq. (\ref{cond_spo}). The boundary between stable and unstable orbits is given by the $r_{ms}$ radius of marginally stable orbits, which are the solution to the equation
\be
\frac{\mathrm{d^2}\bar{R}}{\mathrm{d}r^2}= 0. \label{rms_conditions}
\ee  
In the (r)KdS BH spacetimes, there are unstable SPOs only. In the case of the (r)KdS NS spacetime, the distribution of stable and unstable orbits, as well as other characteristics, can also be related to the behavior of the characteristic functions $q_{\spo}(r)$, $q_{r}(r)$, where the function $q_{r}(r)$ is defined by the reality condition of the effective potentials $X_{\pm}(r)$. For $q\geq 0$ \footnote{For $q<0$, the conditions for reality of the latitudinal motion imply a constraint $-a^2<q<0$, while the constraints imposed on the impact parameter $X$ in this case are stronger than those arising from the conditions for the reality of radial motion. That is, if the 'latitudinal conditions' are satisfied, the 'radial conditions' are automatically satisfied.}, the reality condition of the $X_{\pm}(r)$ functions, as follows from Eq. (\ref{eff.pot.}), is that
\be
q\leq q_{r}(r)\equiv \frac{r^4}{\Delta -a^2}. \label{qr}
\ee
The following discussion in this subsection concerns rKdS spacetime; details for the KdS case can be found in Ref. \cite{Char-Stu:2017:EPJC:}.  

The function $q_{r}(r)$ is obviously positive for $r$ outside the ergosphere on whose boundary it diverges, while it is negative for $r$ inside the ergosphere, in which case it is irrelevant. It has a local minimum $\frac{27}{1-27y}$ at $r=3$. Note that for $y\lessgtr1/27=y_{c(SdS)}$, the minimum is positive/negative.

\begin{figure*}[h!]
	\centering
	\begin{tabular}{cc}
		\includegraphics[width=0.45\textwidth]{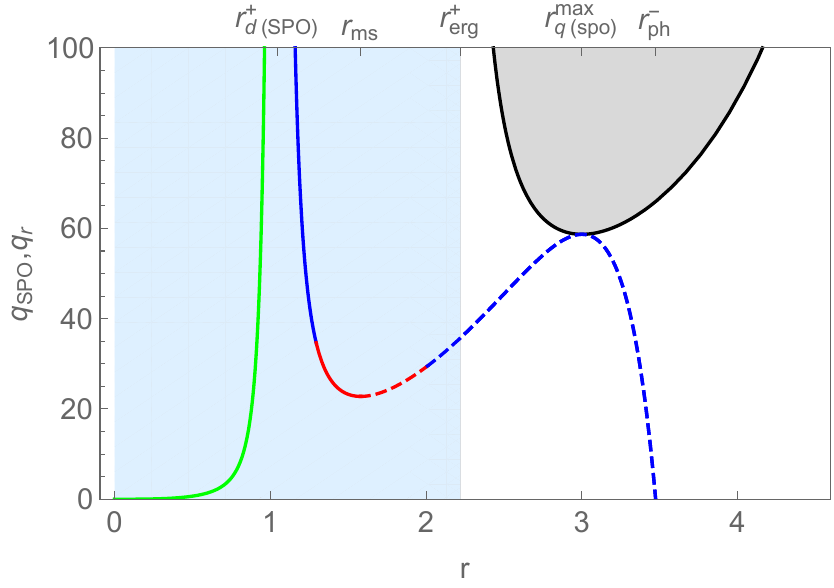}&\includegraphics[width=0.45\textwidth]{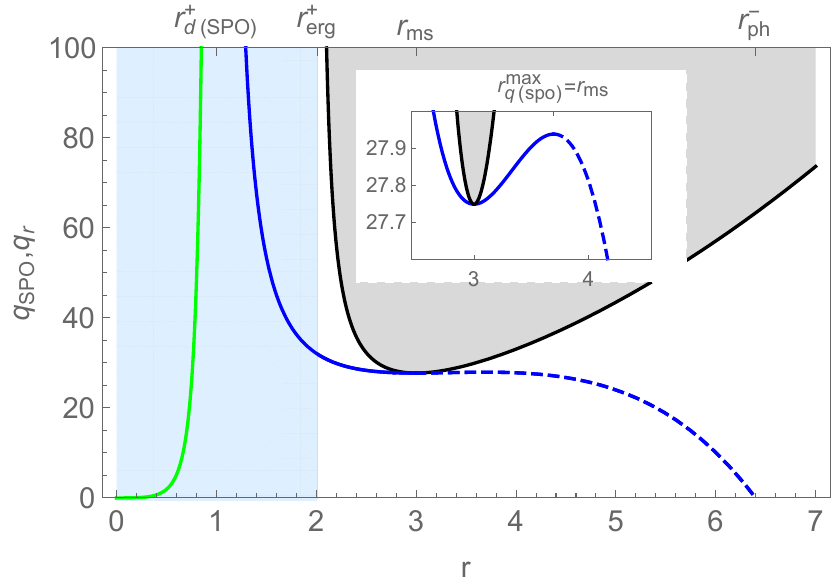}\\
		(a) Class IV: $a^2=1.2, y=0.02$& (b) Class V: $a^2=25, y=0.001$\\
		\includegraphics[width=0.45\textwidth]{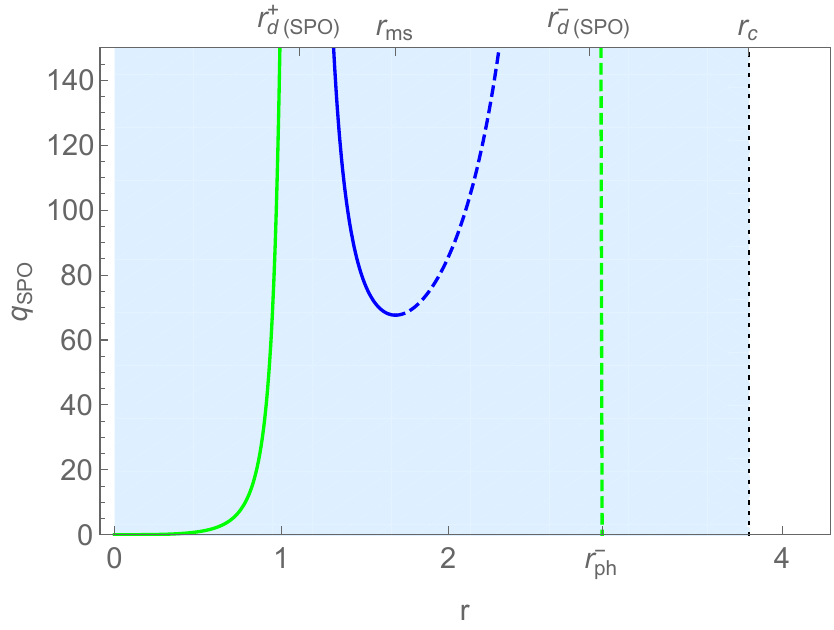}&\includegraphics[width=0.45\textwidth]{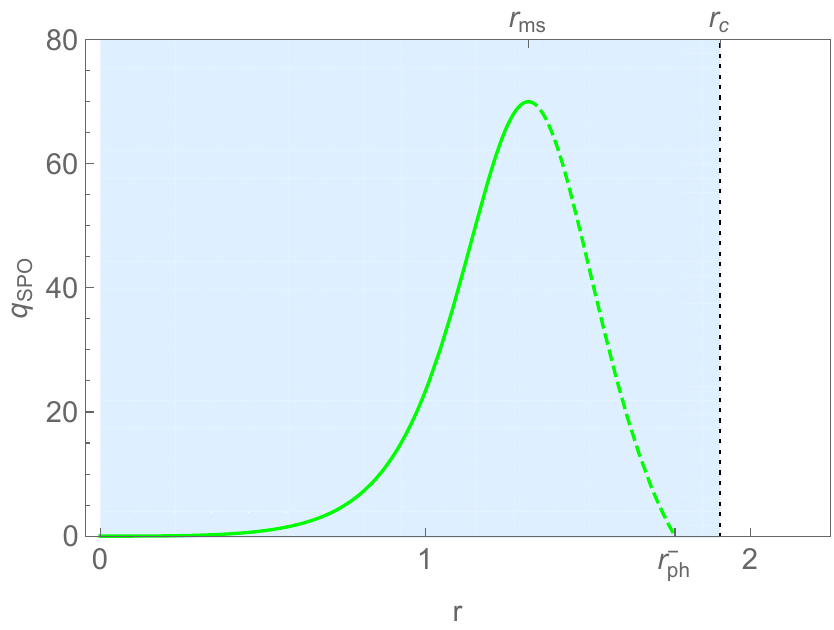}\\
		(c) Class VI: $a^2=1.5, y=0.04$& (d) Class VII: $a^2=1.5, y=0.10$\\
	\end{tabular}	
	\caption{Behavior of the function $q_{\spo}(r)$ plotted in different styles corresponding to specific characteristics of the SPOs. Full/dashed curves correspond to stable/unstable SPOs. Green lines correspond to negative-energy SPOs, which are retrograde, while red/blue lines correspond to progade/retrograde SPOs with $E>0$. The function $q_{r}(r)$ is shown in black. The blue regions indicate the ergosphere, while gray shading indicates a forbidden region, where the reality condition (\ref{qr}) is not met. The cosmological horizon $r_c$ lies outside the plot in cases (a) and (b). The function $q_{\spo}(r)$ is negative in cases (c) and (d).  }     
	\label{fig_qspo_qr}	
\end{figure*}

The function $q_{\spo}(r)$ has a local extreme at the radius $r_{ms}$ corresponding to the marginally stable SPO, such that for $0<r<r_{ms}$/$r_{ms}<r\leq r^{-}_{ph}$ the SPOs are stable/unstable. 
The character of this extreme, and the other stationary point at $r=3$, which can be a local minimum, maximum, or saddle point, determines our proposed further classification,  as shown in Fig. \ref{fig_qspo_qr}, where the functions $q_{\spo}(r)$, $q_{r}(r)$ are both plotted.

For rKdS spacetime parameters corresponding to class IV [see Fig. \ref{fig_qspo_qr}(a)], the function $q_{\spo}(r)$ has two divergence points given by Eq. (\ref{rdexpm}), such that $q_{\spo}(r)\to \pm \infty$ as $r\to r^{\pm}_{d(\spo)}$. The radius $r^{+}_{d(\spo)}$ represents the boundary between orbits with positive and negative energy, $E\lessgtr 0$ for $0<r<r^{+}_{d(\spo)}$/$r^{+}_{d(\spo)}<r\leq r^{-}_{ph}$; regarding the radius $r^{-}_{d(\spo)}$, $r^{-}_{ph}<r^{-}_{d(\spo)}$, and therefore is irrelevant. In this case, the radius $r_{ms}<3$ of the marginally stable orbit corresponds to the local minimum of the function $q_{\spo}(r)$, while at $r=3$ a local maximum occurs, which merges with the local minimum of the function $q_{r}(r)$.

In the case of rKdS spacetime of class V, the extremum of $q_{\spo}(r)$ at $r=3$ is a local minimum that coalesces with the local minimum of $q_{r}(r)$.  The radius $r_{ms}>3$ of the marginally stable SPO then corresponds to the local maximum of $q_{\spo}(r)$ [see Fig. \ref{fig_qspo_qr}(b)]. Therefore, as in the class V of the standard KdS spacetimes, there exists a small region $3<r<r_{ms}$ of unusual bound orbits, which are retrograde and have $E>0$. For more details and consequences for constructing the light-escape cones, see Ref. \cite{Char-Stu:2024:PHYSR4:}. It can be shown that this situation occurs if $a^2>9$, such as in standard KdS spacetimes, and for $y<y_{classV}(a^2)$, where
\be
y_{classV}(a^2)\equiv \frac{9-a^2}{27(9-4a^2)}. \label{yclassV}
\ee
The function $y_{classV}(a^2)$ gives higher values than the corresponding function in the standard KdS case, which definition is given parametrically in Ref. \cite{Char-Stu:2017:EPJC:}. Again, for the sake of brevity, we will not repeat the construction of this definition here. The other characteristics of rKdS spacetimes of class V are qualitatively the same as in rKdS spacetimes of class IV, described below.

The rKdS spacetimes of class~VI correspond to the NS case with $y_{c(SdS)}<y<y_{c(rKdS)}$. Since the ergosphere covers the whole stationary region in the equatorial plane, the divergence points of $q_{r}(r)$ have merged, and this function is negative at $r>0$. It holds that $q_{\spo}(r)\to + \infty$ as $r\to r^{\pm}_{d(\spo)}$; both radii $r^{\pm}_{d(\spo)}$ represent the boundary between positive and negative energy SPOs, such that $E<0$ for $0<r<r^{+}_{d(\spo)}$ and/or $r^{-}_{d(\spo)}<r\leq r^{-}_{ph}$, while $E>0$ for $r^{+}_{d(\spo)}<r<r^{-}_{d(\spo)}$ [see Fig. \ref{fig_qspo_qr}(c)]. The radius $r_{ms}<3$ of the marginally stable SPO corresponds to the local minimum of $q_{\spo}(r)$.

The rKdS spacetimes of class VII correspond to the cases $y>y_{max(h)}(a^2)$ for $0\leq a^2<a^2_{c(rKdS)}$, or $y>y_{c(rKdS)}$ for $a^2_{c(rKdS)}<a^2$. The divergences $r^{\pm}_{d(\spo)}$ have merged (see Sec. \ref{ssec_SPOs}), and the whole region of SPOs is nested in the ergosphere, and all orbits have only negative energies $E<0$ [see Fig. \ref{fig_qspo_qr}(d)]. The radius $r_{ms}$ corresponds to the only local maximum of the function $q_{\spo}(r)$ at $r>0$. 

Analogous to how the parametric plane is divided by the curve $y_{c(rKdS)}$ in the rKdS case, in the KdS case it is divided by the curve $y_{d(\spo)}(a^2)$, defined by Eq.~(\ref{yd(spo)}), into areas corresponding to classes~VI and VII of KdS spacetimes~\cite{Char-Stu:2017:EPJC:}.

\clearpage

\subsection{Classification of rKdS spacetimes}
\begin{figure}[htbp]
 	\centering
 	\begin{tabular}{c}
 		\includegraphics[width=\linewidth]{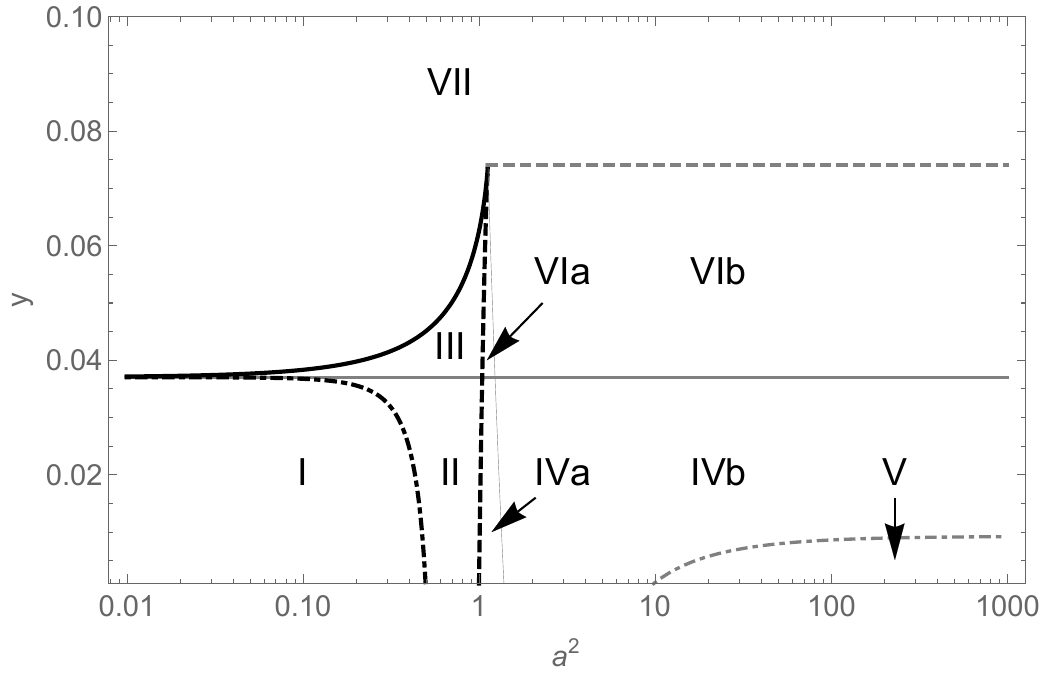}\\
 		(a)\\
 		\includegraphics[width=\linewidth]{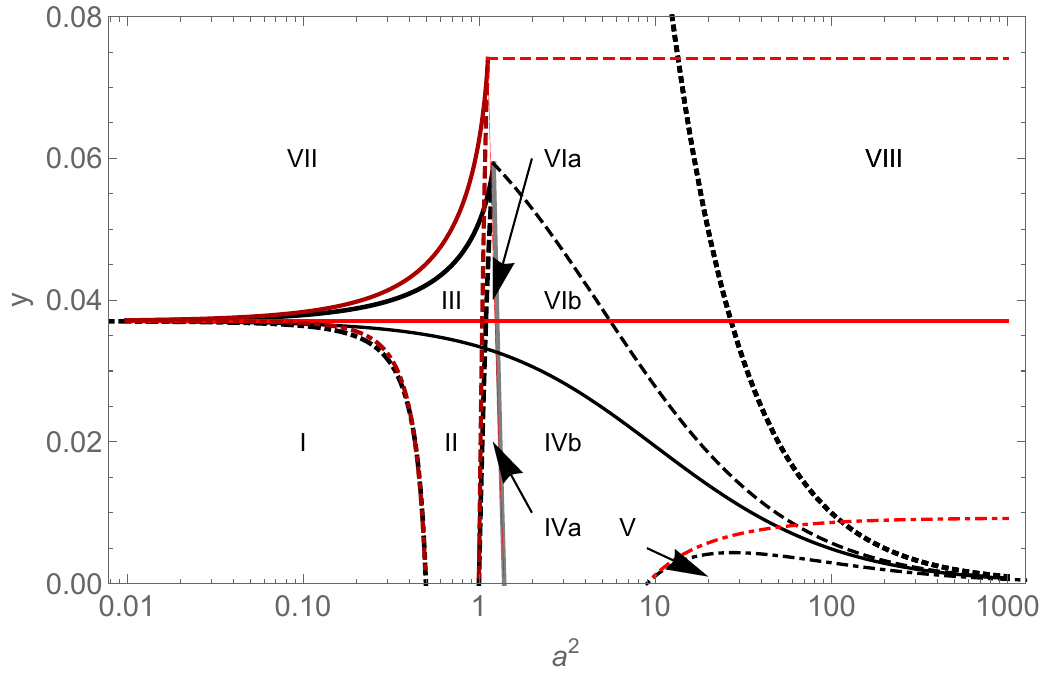}\\
 		(b)		
 	\end{tabular}
 	\caption{(a) Classification of rKdS spacetime into individual classes I--VII and the boundary curves $y_{max(h)}(a^2)$ (full black curve), $y_{min(h)}(a^2)$ (black dashed curve), $y_{erg\text{-}ph}(a^2)$ (black dash-dotted curve), $y_{max(pol)}(a^2)$ (thin gray curve), $y_{classV}(a^2)$ (gray dash-dotted curve), $y_{c(SdS)}$ (full gray line), $y_{c(rKdS)}$ (gray dashed line). (b) Comparison of the classification of KdS and rKdS spacetimes. Roman numerals denote classes of KdS spacetimes and  the appropriate boundary curves are shown in the same styling as the corresponding boundary curves of the rKdS classes in panel (a). The black dotted curve separating classes~VII and VIII has no equivalent in panel (a), because there are only seven rKdS classes of spacetime. The boundary curves rKdS of spacetime are distinguished here by corresponding shades of red. }\label{fig_(y-a2)_plane}
 \end{figure}

 All curves $y_{min(h)(a^2)}$, $y_{max(h)}(a^2)$, $y_{erg\text{-}ph}(a^2)$, $y_{max(pol)}(a^2)$, $y_{classV}(a^2)$, $y_{c(SdS)}$, $y_{c(rKdS)}$  dividing the parameter plane $(y\text{-}a^2)$ into individual classes I--VII are shown in Fig.~\ref{fig_(y-a2)_plane}.

\section{Shadows and LEC{\lowercase{s}} in local reference frames of {\lowercase{r}}KdS spacetimes} \label{sec4}

\subsection{Shadows in LNRFs}
Our main purpose in this paper is to compare some aspects of the shadows of superspinars observed by observers located in the LNRF on the static radius $r_{s}$ of the rKdS spacetimes with the corresponding properties of shadows constructed on the static radius of the KdS spacetimes in our previous paper \cite{Char-Stu:2024:PHYSR4:}. The reason for this focus is that at and around the static radius, the KdS and rKdS spacetimes are closest to the asymptotically flat region of Kerr spacetimes, and the LNRFs are close to the systems of static observers. It is also advantageous that the static radius is identical for both KdS and rKdS geometries for a given cosmological parameter $y$. 

\subsubsection{Directional angles of photons in the LNRFs}
The locally measured directional angles ($\alpha, \beta$) of a photon are defined using the locally measured frame components $k^{(a)}$ of the photon four-momentum by the standard relations  \cite{Stu-Sche:2010:CQG:}
\bea
\cos \alpha&=&k^{(r)}/k^{(t)},\label{cosalpha}\\
\sin \alpha \cos \beta&=&k^{(\theta)}/k^{(t)},\label{sinalphacosbeta}\\
\sin \alpha \sin \beta&=&k^{(\phi)}/k^{(t)}. \label{sinalphasinbeta}
\eea
Here, $k^{(a)}$ are defined by relations analogous to Eq. (\ref{p(a)contra}):
\bea
k^{(a)}&=&\omega^{(a)}_{\mu}k^{\mu}, \label{k(a)contra} \\
k_{(a)}&=&e^{\nu}_{(a)}k_{\nu}. \label{k(a)cov}
\eea
Without loss of generality, we can, as usual, insert $k^{(t)}=1$ into the relations (\ref{cosalpha})-(\ref{sinalphasinbeta}).\\

\subsubsection{Celestial coordinates}
In order to compare with the results obtained for the KdS metric, we define, as in Ref. \cite{Char-Stu:2024:PHYSR4:}, the celestial coordinates $\tilde{\alpha}$, $\tilde{\beta}$ for the distant observer, which are related to the directional angles $\alpha$ and $\beta$ by the relations
\bea
\tilde{\alpha}&=&- k^{(\phi)}/k^{(t)}=-\sin \alpha \sin \beta  \nonumber \\
\tilde{\beta}&=& k^{(\theta)}/k^{(t)}=\sin \alpha \cos \beta . \label{stereograph}
\eea

\subsubsection{Astronomical observables}
\begin{figure*}[h!]
	\centering
	\begin{tabular}{cc}
		\includegraphics[width=0.45\textwidth]{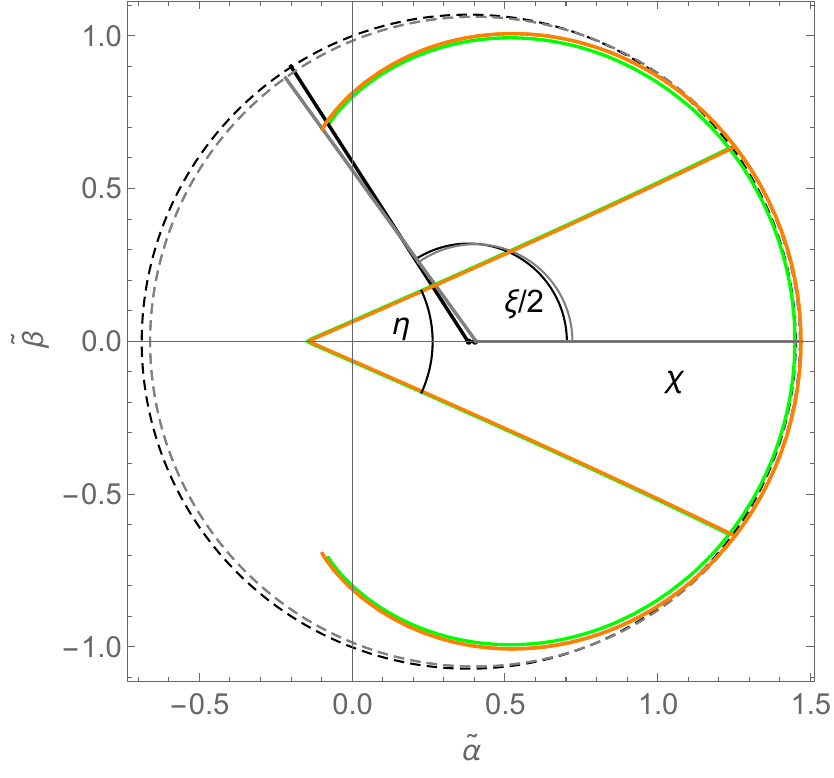}&\includegraphics[width=0.45\textwidth]{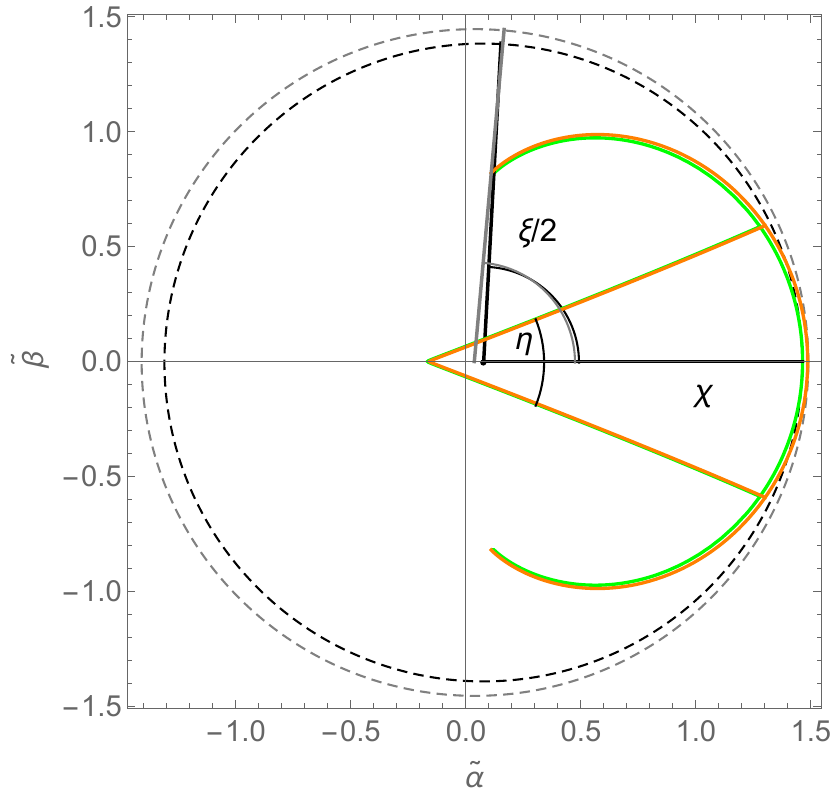}\\
		(a) Class IVa: $a^2=1.2, y=0.02$& (b) Class IVb: $a^2=1.5, y=0.02$\\
	
	\end{tabular}	
	\caption{Comparison of the shadows of rKdS (green) and KdS (orange) superspinars observed by an LNRF observer located in the equatorial plane at the static radius for the cases of (a) spacetime with polar SPOs and (b) spacetime without polar SPOs. The C-shaped arc corresponds to photons with constants of motion close to $q_{\spo}$ and $X_{\spo}$, while the wedge shape corresponds to the shadow of the surface of the superspinar with radius $R=0.1$, which acts as an absorbing surface (see Ref. \cite{Char-Stu:2024:PHYSR4:}). The ratio of the respective arcs of the osculating circles (dashed circles) and their radii $\chi$ defines the magnitude of the central angle $\xi$ in radians.The angle $\eta$, characterizing the angular size of the shadow of the superspinar, is given by the deviation of the straight lines formed by connecting the top of the wedge with its end points.}     
	\label{fig_shads_LNRF}	
\end{figure*}

We define the astronomical observables $\xi$, $\eta$, $\chi$ in Fig.~\ref{fig_shads_LNRF}. Then, we give their behavior and special values for both rKdS and KdS geometries. In order to compare these geometries, we choose a LNRF observer at a static radius that is identical for both geometries, which is convenient for this purpose. Figure~\ref{fig_shads_LNRF} corresponds to an observation in the equatorial plane.

\clearpage

In our previous paper \cite{Char-Stu:2024:PHYSR4:}, we found that the (non) presence of polar SPOs significantly affects the behavior of some observables. In what follows, we will therefore examine these quantities and their special values separately for the two types of spacetime in each subsection.  We then pay special attention to observations from the equatorial plane and on the rotation axis.

\subsubsection{Appearance of observables corresponding to spacetimes with polar SPOs}

\begin{figure*}[h!]
	\centering
	\begin{tabular}{cccc}
		\includegraphics[width=0.24\textwidth]{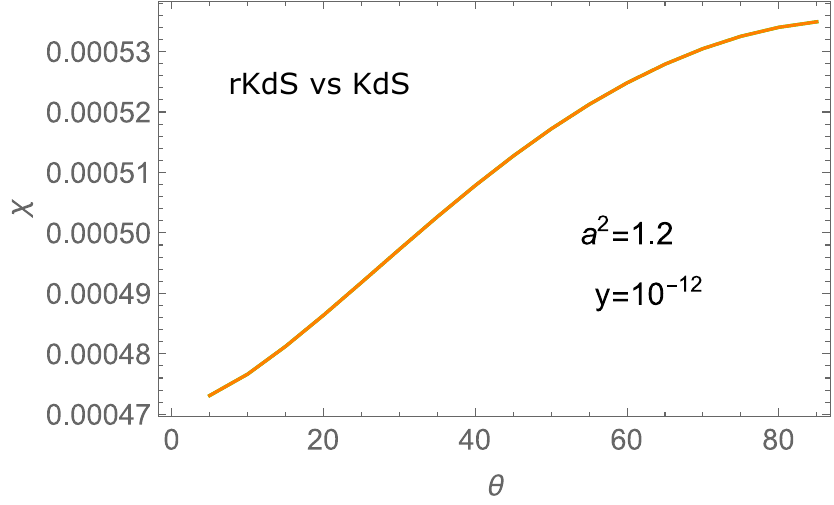}&\includegraphics[width=0.24\textwidth]{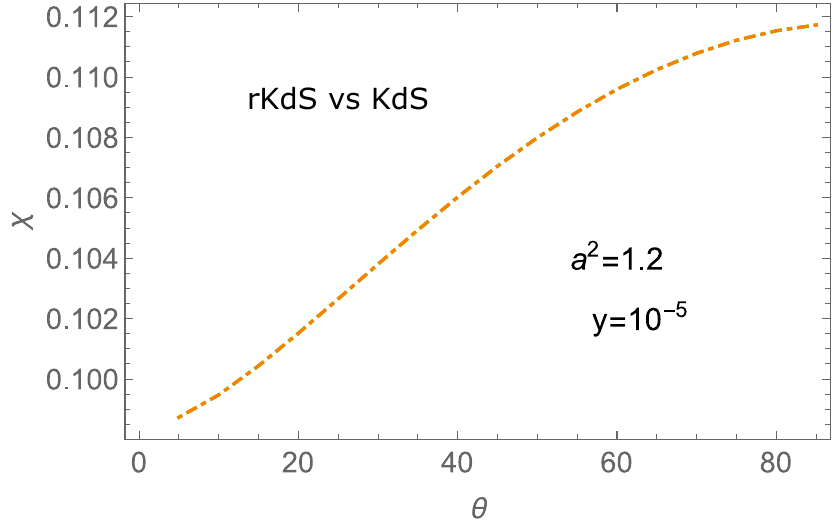}&\includegraphics[width=0.24\textwidth]{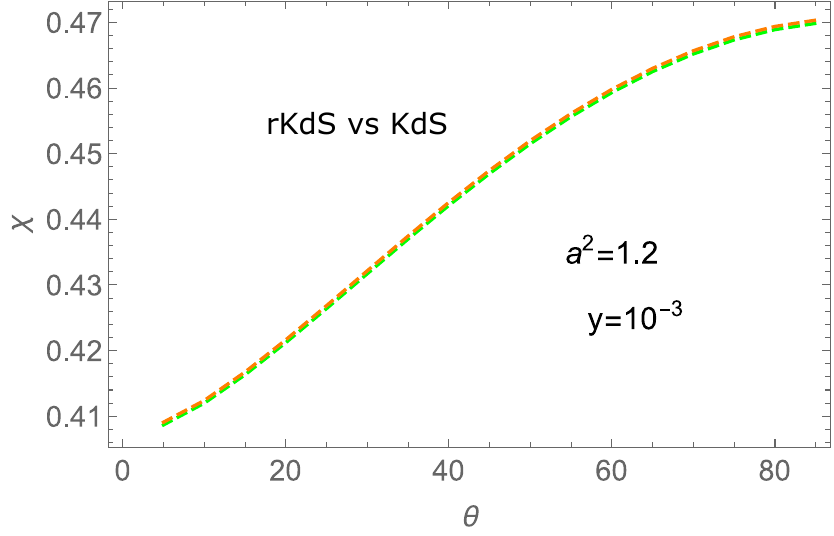}&\includegraphics[width=0.24\textwidth]{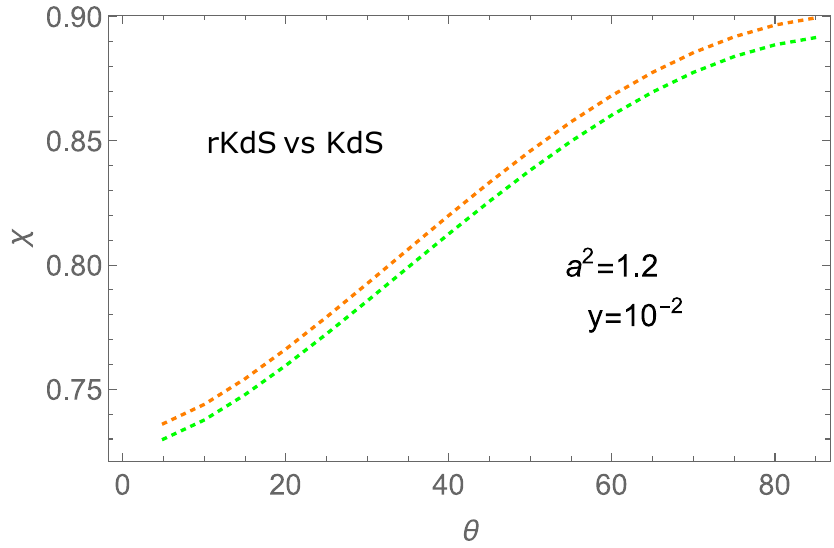}	
	\end{tabular}	
	\caption{Comparison of the dependence of the magnitude of the angular radius $\chi$, characterizing the angular size of the shadow of the superspinar in the observer's sky, on the latitudinal coordinate $\theta_{0}$ of the observer for rKdS and KdS geometries for the case of spacetimes with polar SPOs. A slightly magnifying effect is seen with increasing observer latitude, and with increasing cosmological parameter $y$, being slightly more pronounced for KdS spacetimes.}     
	\label{fig_chi_theta_1.2}	
\end{figure*}

First, to get an idea of the angular size of the shadow of the superspinar on the observer's sky, we present in Fig. \ref{fig_chi_theta_1.2} the angular radius $\chi$ of the osculating circle, which closely surrounds the observed shadow, as a function of the observer's latitude $\theta_{0}$ for a selected representative value of the spin parameter $a^2$ and some values of the cosmological parameter $y$.

\begin{figure*}[h!]
	\centering
	\begin{tabular}{ccc}
		\includegraphics[width=0.33\textwidth]{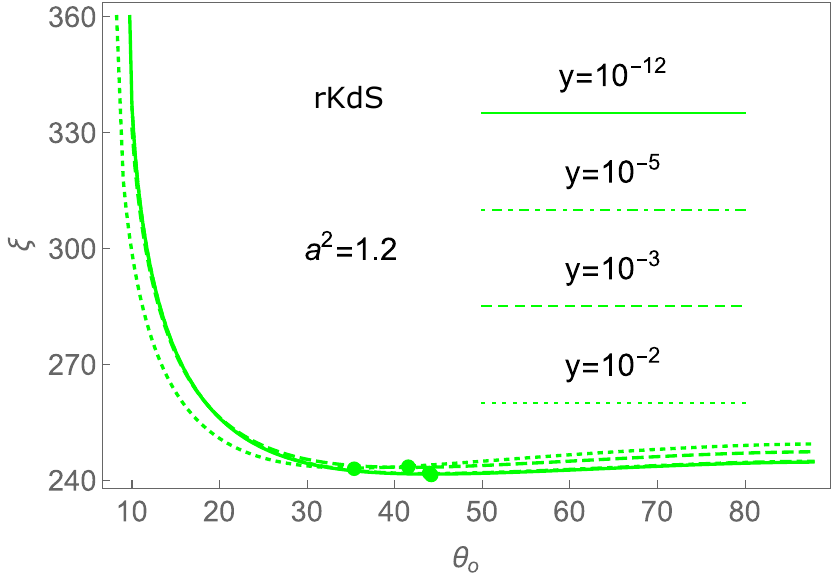}&\includegraphics[width=0.33\textwidth]{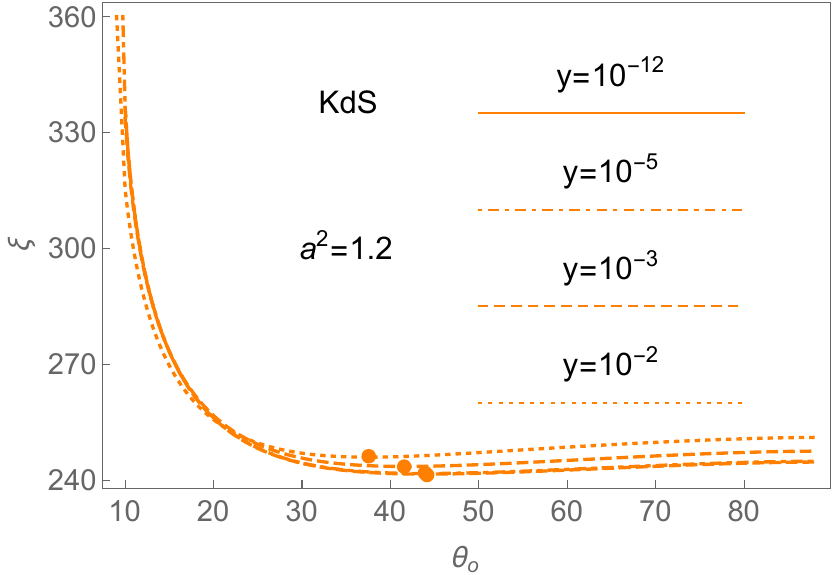}&\includegraphics[width=0.33\textwidth]{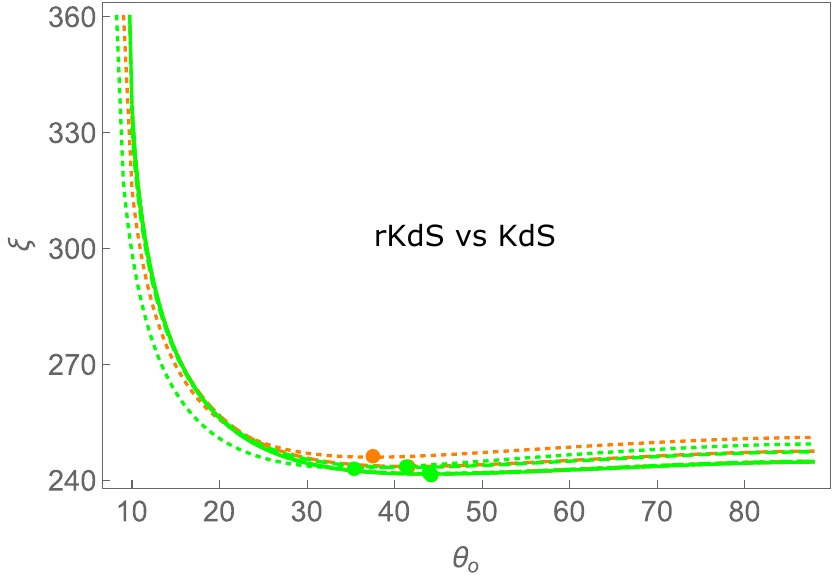}		
	\end{tabular}	
	\caption{Comparison of the dependence of the magnitude of the central angle $\xi$ on the latitudinal coordinate $\theta_{0}$ of the observer for rKdS and KdS geometries for the case of spacetimes with polar SPOs. The local minima $\xi_{min}$ are highlighted with dots. }     
	\label{fig_ksi_theta_1.2}	
\end{figure*}

\begin{figure*}[h!]
	\centering
	\begin{tabular}{ccc}
		\includegraphics[width=0.33\textwidth]{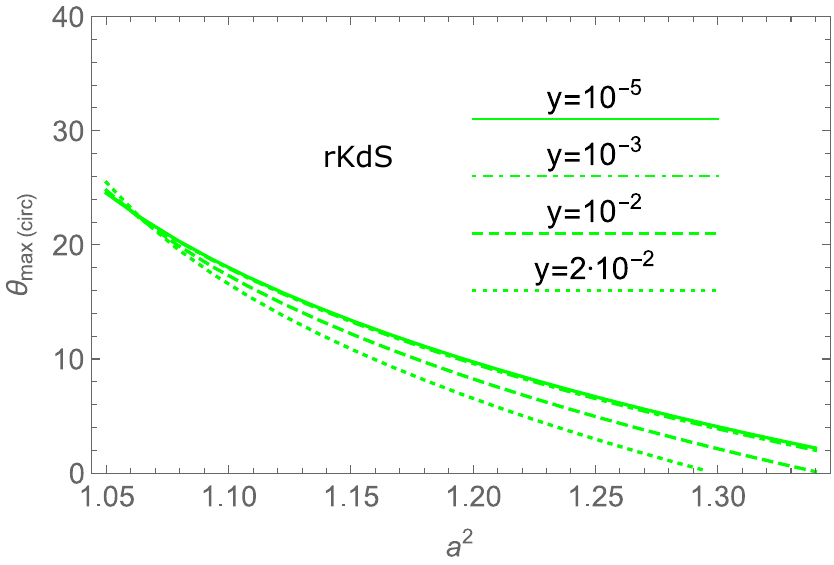}&\includegraphics[width=0.33\textwidth]{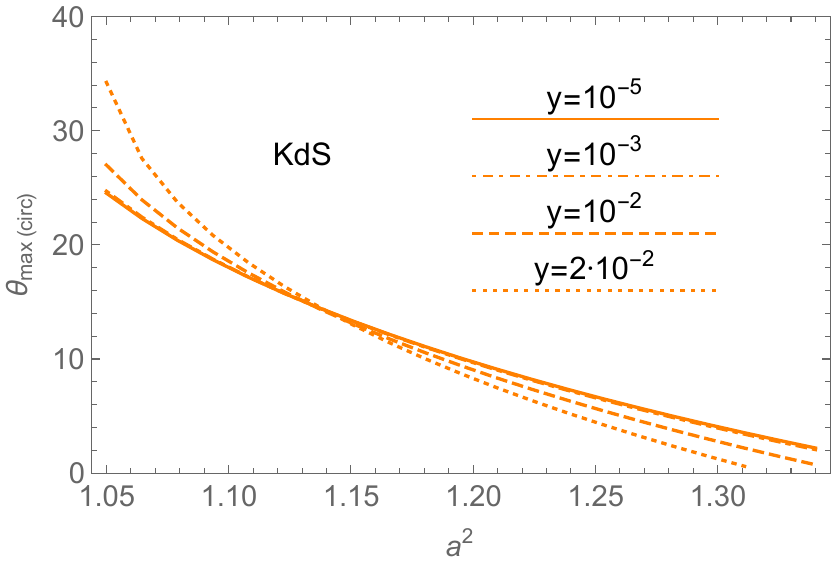}&\includegraphics[width=0.33\textwidth]{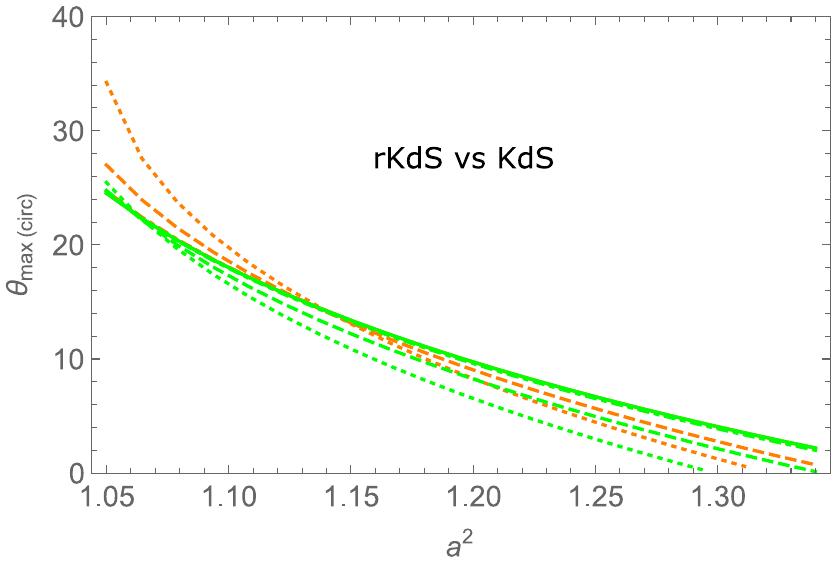}		
	\end{tabular}	
	\caption{Comparison of the size dependence of the maximum latitude coordinate $\theta_{max(circ)}$ of the observer for the rKdS (green) and KdS (orange) spacetimes with polar SPOs, for which the observer sees a circular light fringe around the shadow of the superspinar, such that for $\theta_{0}>\theta_{max(circ)}$ an arc appears. The magnitudes of the angles on both axes are expressed in degrees. It can be seen that as the cosmological parameter $y$ decreases, the results become indistinguishable both within a given geometry and between these geometries.}     
	\label{fig_thetamaxcirc_a2}	
\end{figure*}

Figure \ref{fig_ksi_theta_1.2} shows the central angle $\xi$ versus the latitudinal coordinate $\theta_{o}$ of the observer for some selected values of the spacetime parameters corresponding to spacetimes with polar SPOs. It can be seen that, as in KdS geometry, there is a minimum angle $\theta_{max(circ)}$ such that for $0\dgr \leq \theta_{o}<\theta_{max(circ)}$ the observer sees a circle of light, flanking the shadow of the superspinar, which for $\theta_{max(circ)}<\theta_{o}\leq 90 \dgr$ becomes an arc. \footnote{Here and in the following discussion we restrict ourselves to the 'northern hemisphere' $0\dgr \leq \theta \leq 90\dgr$ for reasons of symmetry.} The comparison of the angle $\theta_{max(circ)}$ for both geometries is shown in Fig. \ref{fig_thetamaxcirc_a2}.

Figure \ref{fig_ksi_theta_1.2} further shows that with increasing observer latitude $\theta_{o}$ the magnitude of the observed angle $\xi$ first decreases sharply to reach a certain minimum $\xi_{min}$ at an appropriate point $\theta_{\xi min}$, after which it slowly increases again to the value $\xi_{eq}$ corresponding to the observation in the equatorial plane. The comparisons of the angles $\xi_{min}$, $\theta_{\xi min}$ and $\xi_{eq}$ for both geometries are shown in Figs. \ref{fig_ksimin_a2}, \ref{fig_thetaksimin_a2} and \ref{fig_ksieq_a2pol}, respectively. In all figures below, the sizes of all angles are expressed in degrees.

\begin{figure*}[h!]
	\centering
	\begin{tabular}{ccc}
		\includegraphics[width=0.33\textwidth]{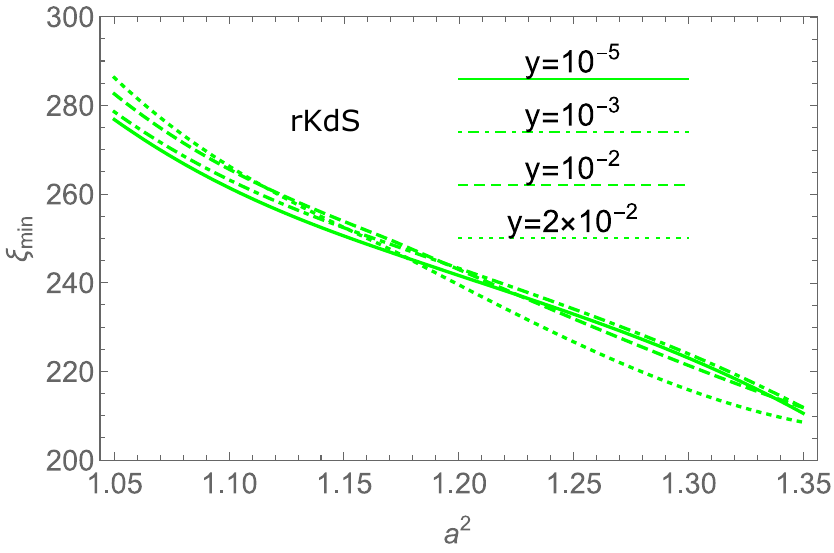}&\includegraphics[width=0.33\textwidth]{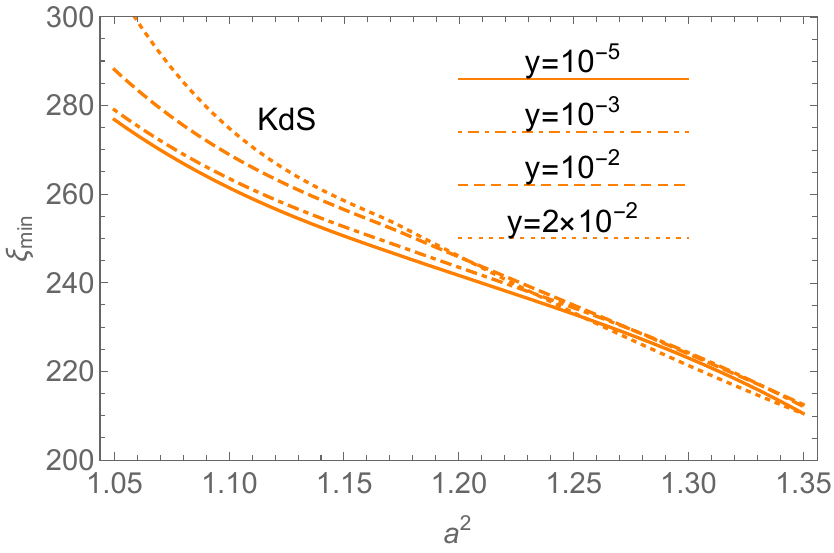}&\includegraphics[width=0.33\textwidth]{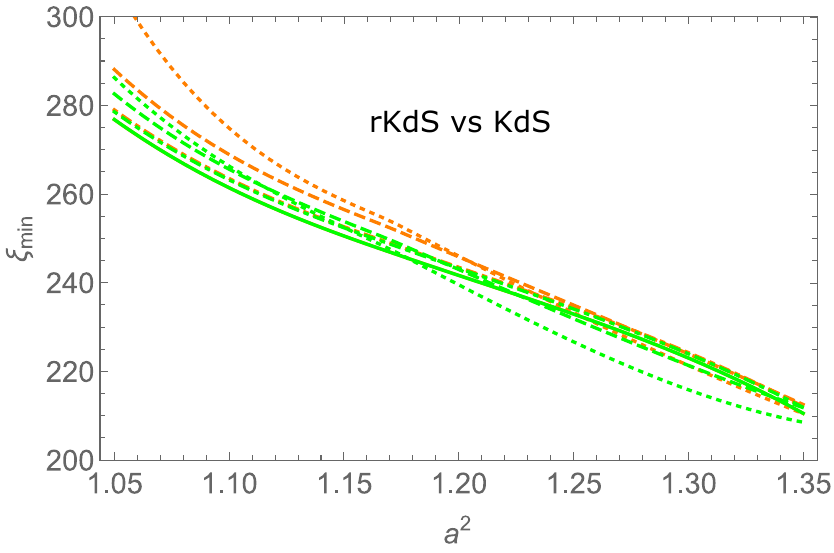}		
	\end{tabular}	
	\caption{Comparison of the dependence of the minimum of the central angle $\xi_{min}$ on the parameters $a^2,y$ corresponding to spacetimes with polar SPOs for rKdS and KdS geometries. For $y$ values of order less than $10^{-5}$, the graphs become indistinguishable.}     
	\label{fig_ksimin_a2}	
\end{figure*}

\begin{figure*}[h!]
	\centering
	\begin{tabular}{ccc}
		\includegraphics[width=0.33\textwidth]{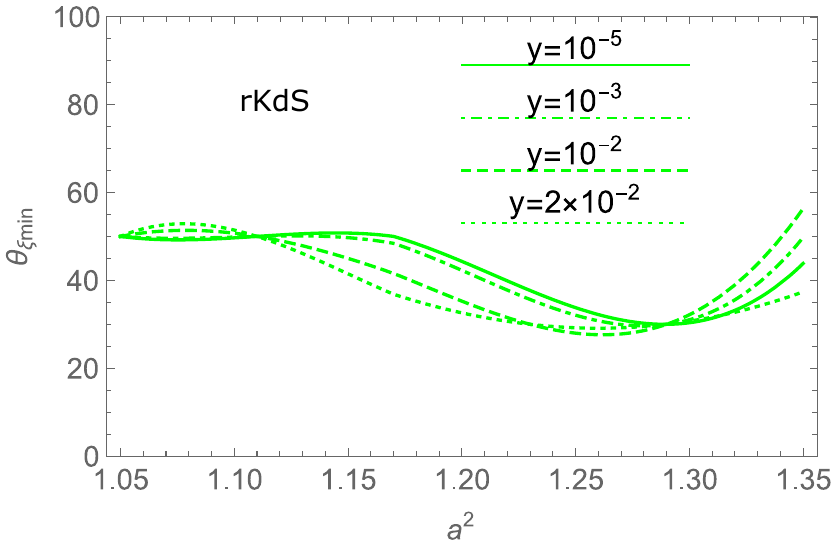}&\includegraphics[width=0.33\textwidth]{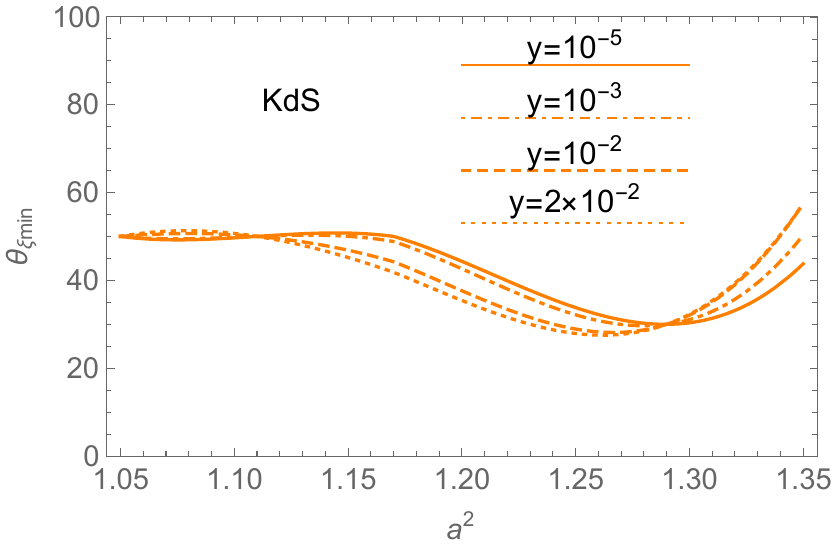}&\includegraphics[width=0.33\textwidth]{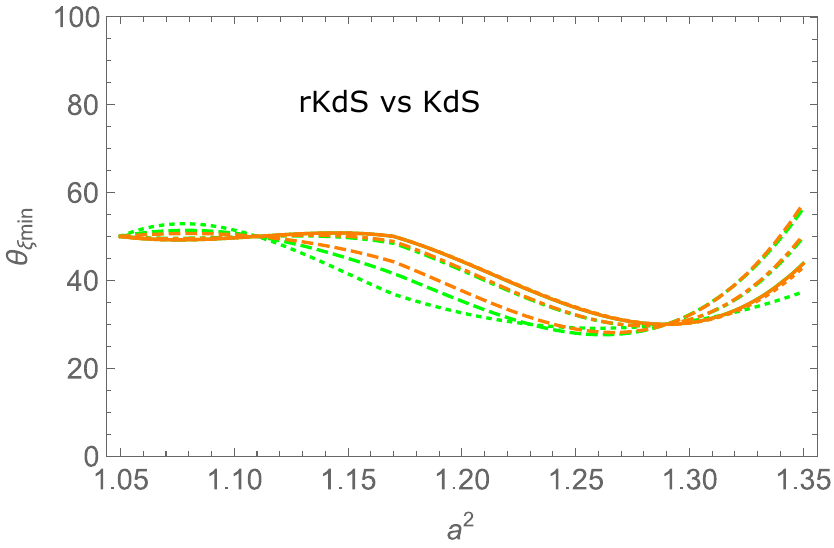}		
	\end{tabular}	
	\caption{Comparison of the latitudinal coordinate $\theta_{\xi min}$ corresponding to minimum central angle $\xi_{min}$ for rKdS and KdS geometries that appears in spacetimes with polar SPOs. For $y$ values of order less than $10^{-5}$, the graphs become indistinguishable.}     
	\label{fig_thetaksimin_a2}	
\end{figure*}

\begin{figure*}[h!]
	\centering
	\begin{tabular}{ccc}
		\includegraphics[width=0.33\textwidth]{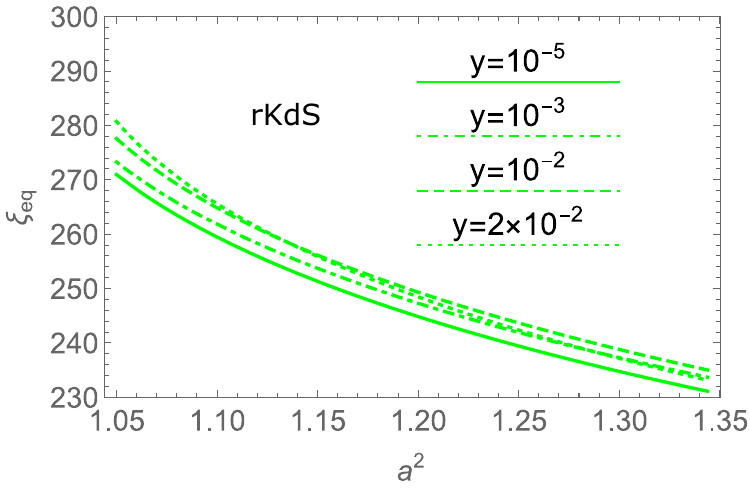}&\includegraphics[width=0.33\textwidth]{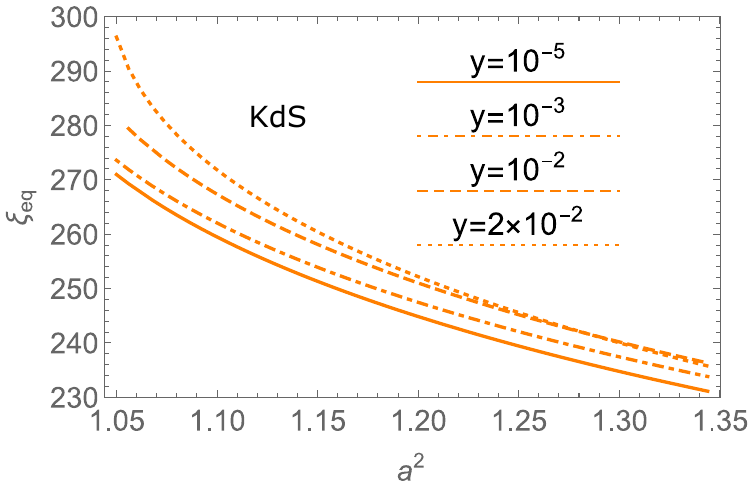}&\includegraphics[width=0.33\textwidth]{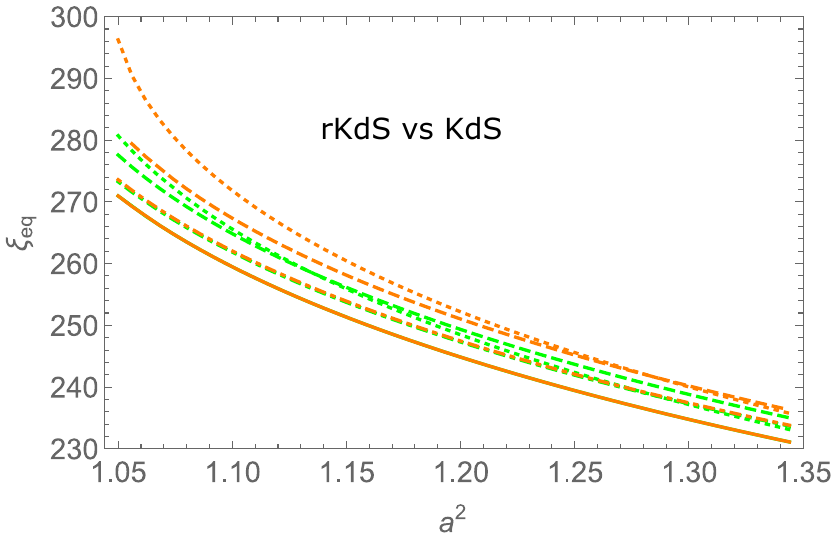}		
	\end{tabular}	
	\caption{Comparison of the central angle $\xi_{eq}$ corresponding to observation in the equatorial plane $\theta_{0}=90\dgr$. The $\xi_{eq}$ angle values are expressed in degrees.}     
	\label{fig_ksieq_a2pol}	
\end{figure*}

\clearpage
\subsubsection{Appearance of observables corresponding to spacetimes with no polar SPOs}
\begin{figure*}[h!]
	\centering
	\begin{tabular}{cccc}
		\includegraphics[width=0.24\textwidth]{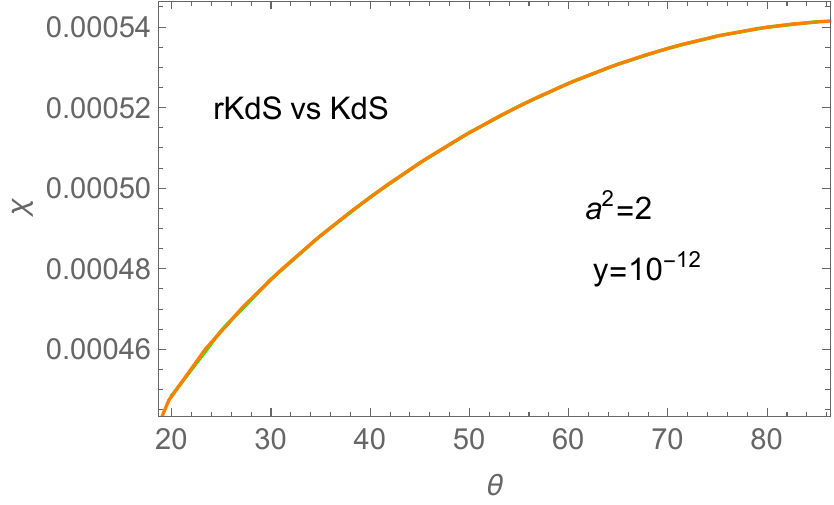}&\includegraphics[width=0.24\textwidth]{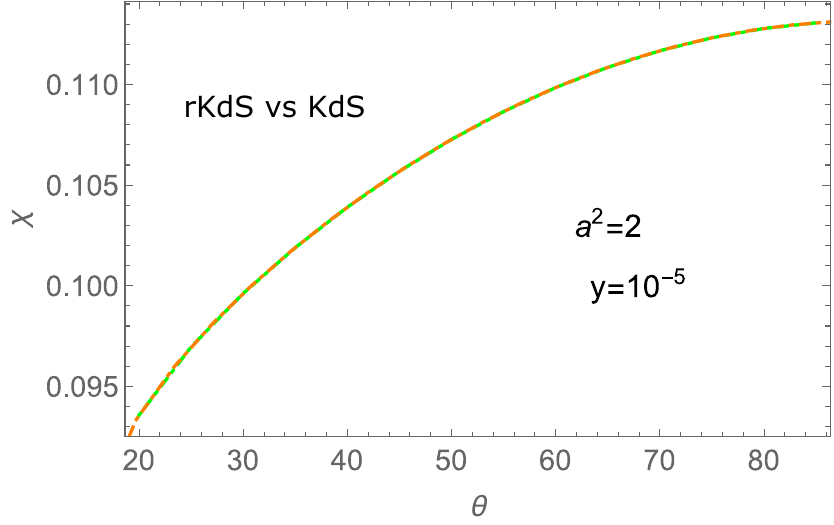}&\includegraphics[width=0.24\textwidth]{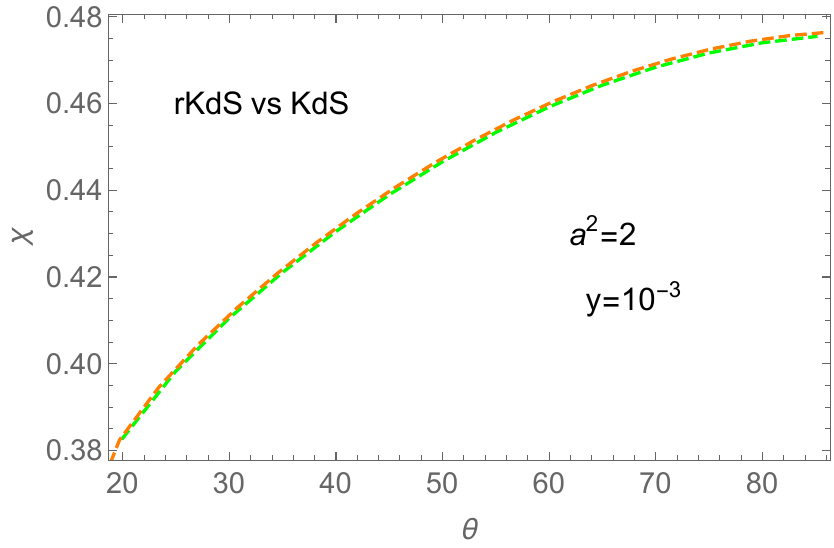}&\includegraphics[width=0.24\textwidth]{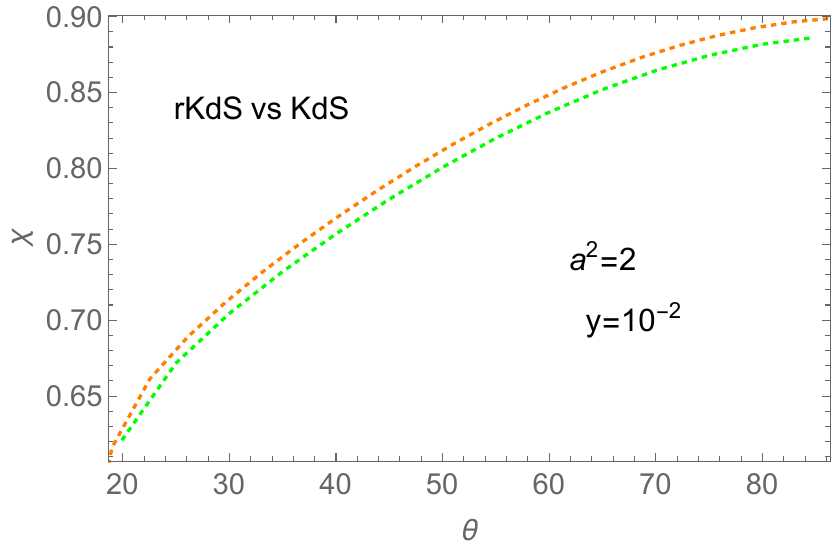}	
	\end{tabular}	
	\caption{Comparison of the dependence of the magnitude of the angular radius $\chi$, characterizing the angular size of the shadow of the superspinar in the observer's sky, on the latitudinal coordinate $\theta_{0}$ of the observer for rKdS and KdS geometries for the case of spacetimes without polar SPOs. A slight magnifying effect is seen with increasing observer latitude, and with increasing cosmological parameter $y$, being slightly more pronounced for KdS spacetimes.}     
	\label{fig_chi_theta_2}	
\end{figure*}

The angular radius $\chi$, which characterizes the observed angular dimensions of the superspinar shadow, is given in Fig. \ref{fig_chi_theta_2} for a selected representative value of the spin parameter $a^2$ and some values of the cosmological parameter $y$.

\begin{figure*}[h!]
	\centering
	\begin{tabular}{ccc}
		\includegraphics[width=0.33\textwidth]{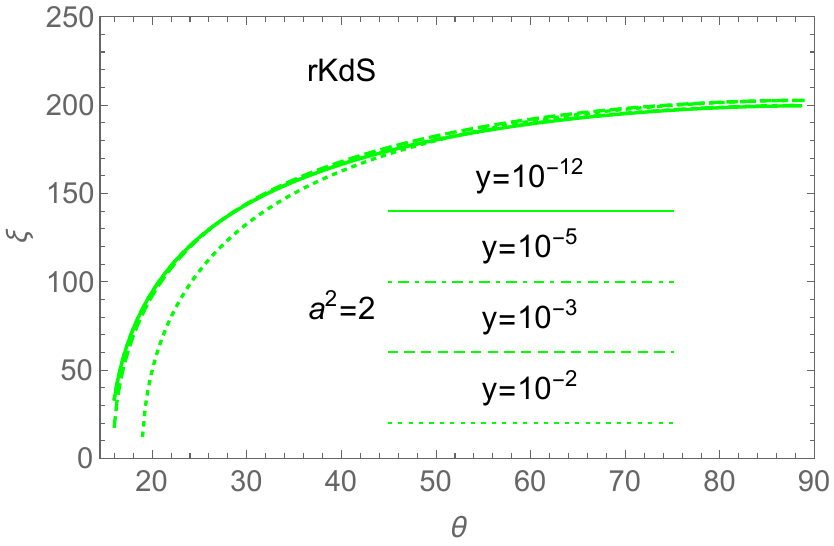}&\includegraphics[width=0.33\textwidth]{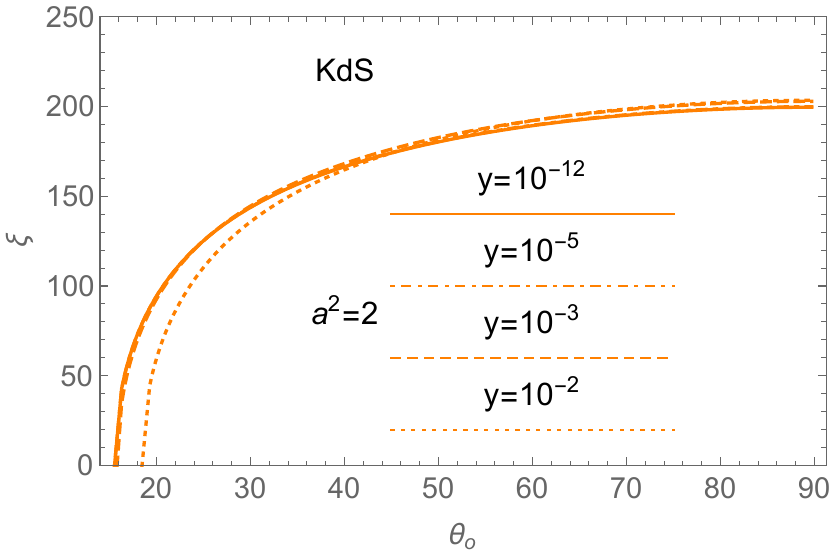}&\includegraphics[width=0.33\textwidth]{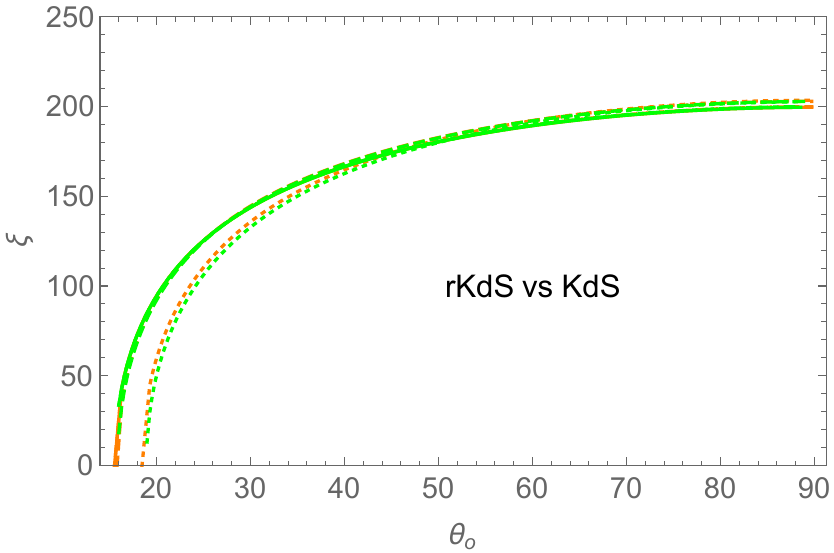}		
	\end{tabular}	
	\caption{Comparison of the dependence of the magnitude of the central angle $\xi$ on the latitudinal coordinate $\theta_{0}$ of the observer for rKdS (green) and KdS (orange) geometries for the case of spacetimes without polar SPOs. The magnitudes of the angles on both axes are expressed in degrees.}     
	\label{fig_ksi_theta_2}	
\end{figure*}

Figure \ref{fig_ksi_theta_2} shows the central angle $\xi$ versus the latitudinal coordinate $\theta_{o}$ of the observer for some selected values of the spacetime parameters corresponding to spacetimes with no polar SPOs. Here again, in agreement with the KdS spacetime case, there is a minimum observer latitude $\theta_{min(arc)}$, such that for $0\leq \theta_{o}\leq \theta_{min(arc)}$ $\xi=0$--i.e., the light arc appears only for $\theta_{min(arc)}>0$--whose central angle $\xi$ increases with increasing observer latitude, reaching a maximum $\xi_{eq}$ in the equatorial plane.

\begin{figure*}[h!]
	\centering
	\begin{tabular}{ccc}
		\includegraphics[width=0.33\textwidth]{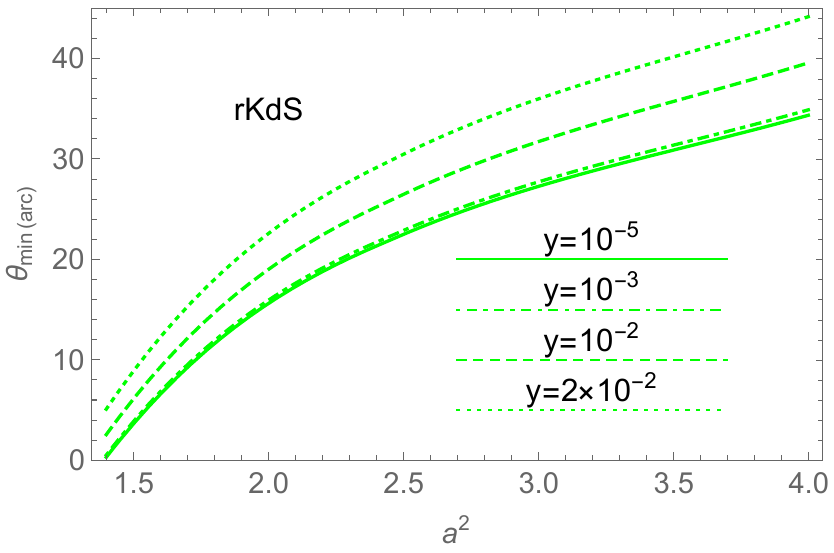}&\includegraphics[width=0.33\textwidth]{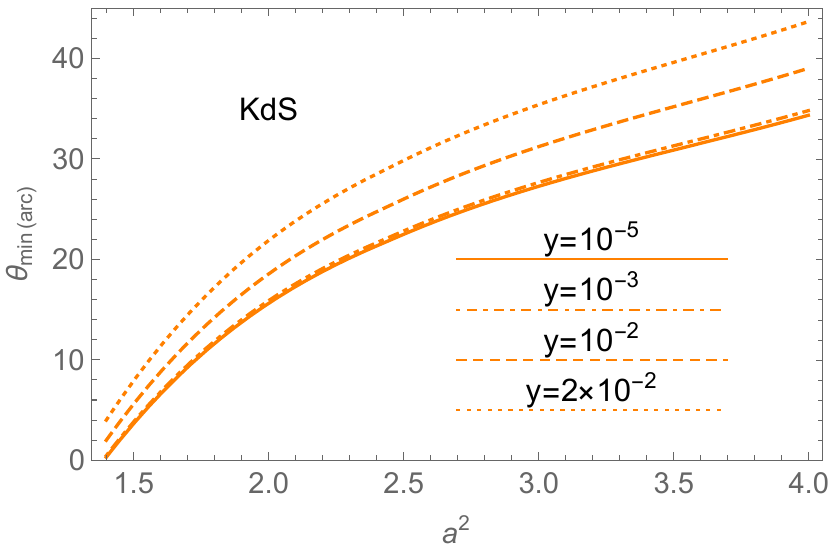}&\includegraphics[width=0.33\textwidth]{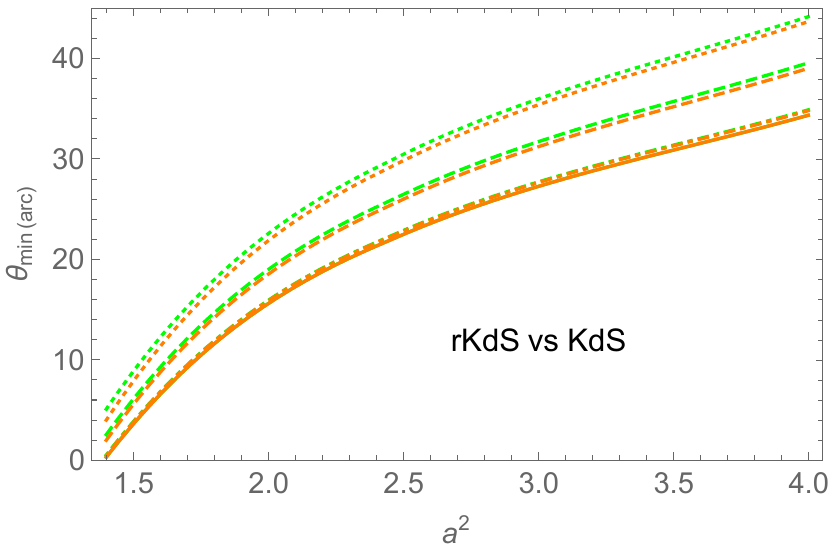}		
	\end{tabular}	
	\caption{Comparison of the dependence of the magnitude of the critical angle $\theta_{min(arc)}$ on the spin parameter $a^2$ for rKdS (green) and KdS (orange) geometries for the case of spacetimes without polar SPOs. The magnitudes of the angles on the vertical axes are expressed in degrees.}     
	\label{fig_thetaminarc_a2}	
\end{figure*}

\begin{figure*}[h!]
	\centering
	\begin{tabular}{ccc}
		\includegraphics[width=0.33\textwidth]{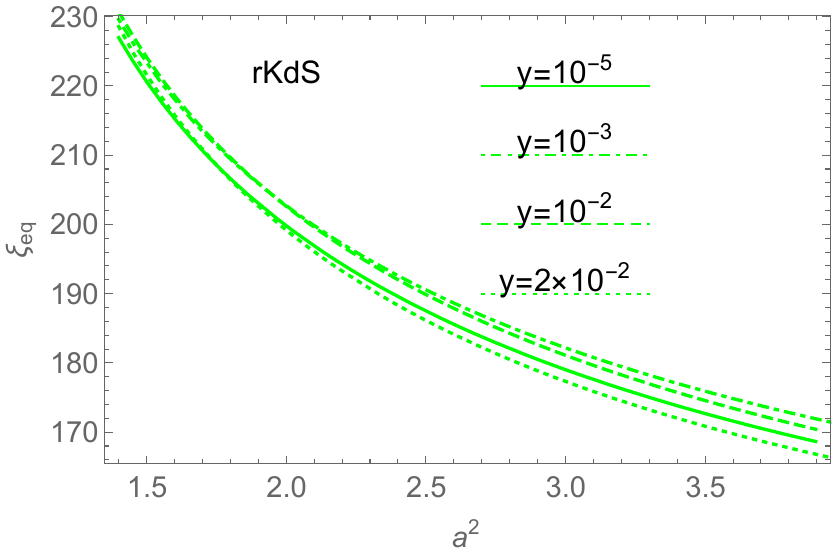}&\includegraphics[width=0.33\textwidth]{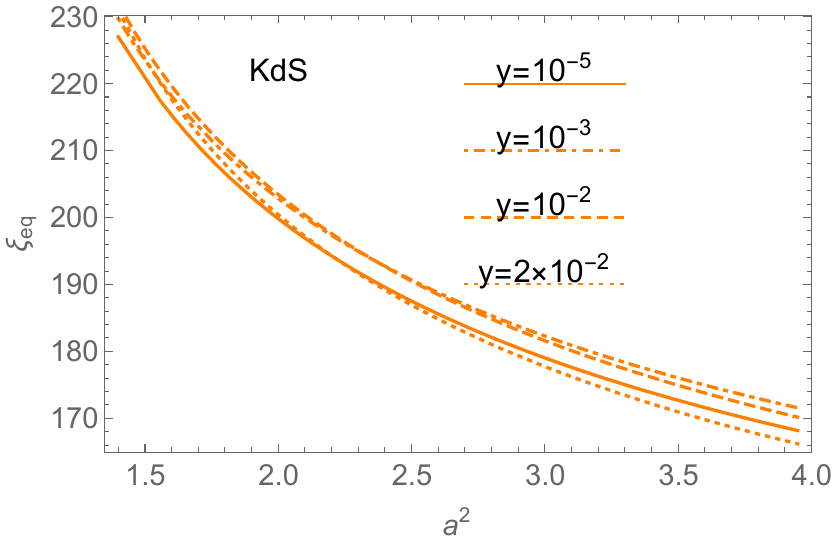}&\includegraphics[width=0.33\textwidth]{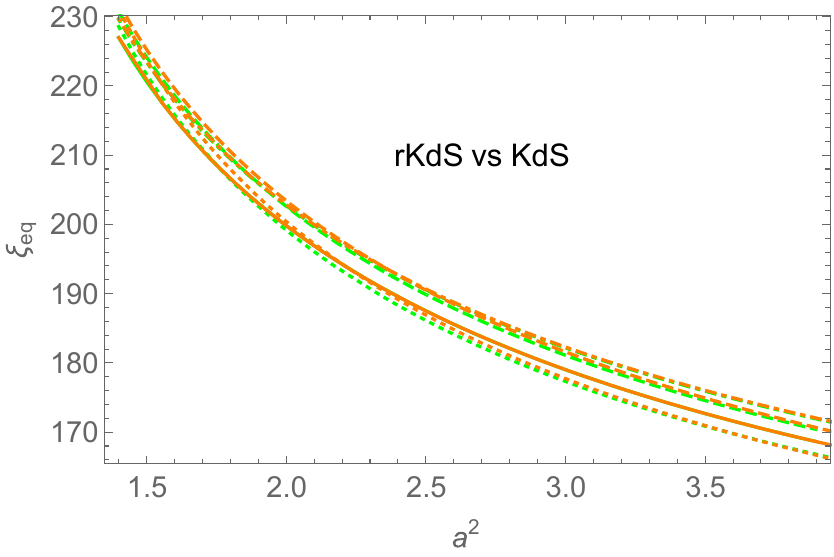}		
	\end{tabular}	
	\caption{Comparison of the maximum central angle $\xi_{eq}$ corresponding to observation in the equatorial plane $\theta_{0}=90\dgr$ for rKdS and KdS geometries as it occurs in spacetimes without polar orbits (see Fig. \ref{fig_ksi_theta_2}). The magnitudes of the angles on both axes are expressed in degrees.}     
	\label{fig_ksieq_a2}	
\end{figure*}

The dependence of the characteristic angles $\theta_{min(arc)}$ and $\xi_{eq}$ on the spin parameter $a^2$ for selected values of the cosmological parameter $y$ are compared for both rKdS and KdS geometries in Figs. \ref{fig_thetaminarc_a2} and \ref{fig_ksieq_a2}.

\clearpage
\subsubsection{Appearance of observables corresponding to both types of spacetimes with and without polar orbits}

\begin{figure*}[b!]
	\centering
	\begin{tabular}{ccc}
		\includegraphics[width=0.33\textwidth]{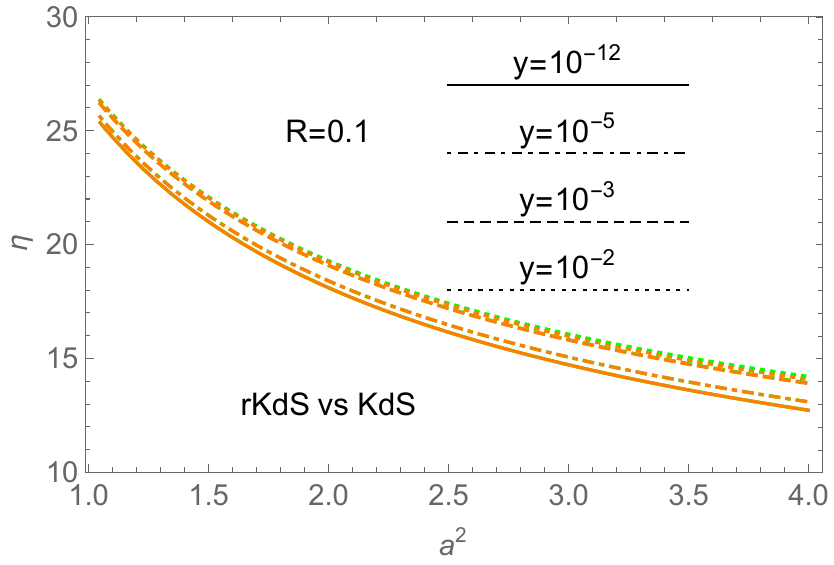}&\includegraphics[width=0.33\textwidth]{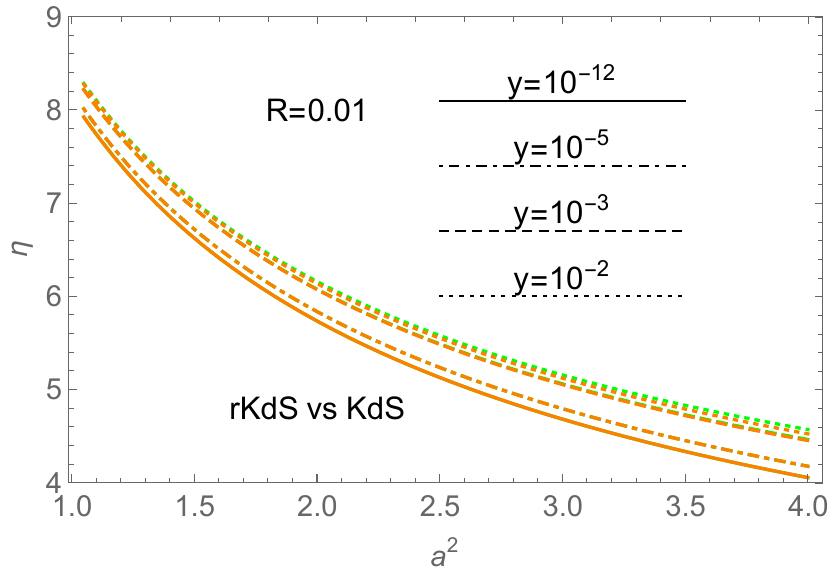}&\includegraphics[width=0.33\textwidth]{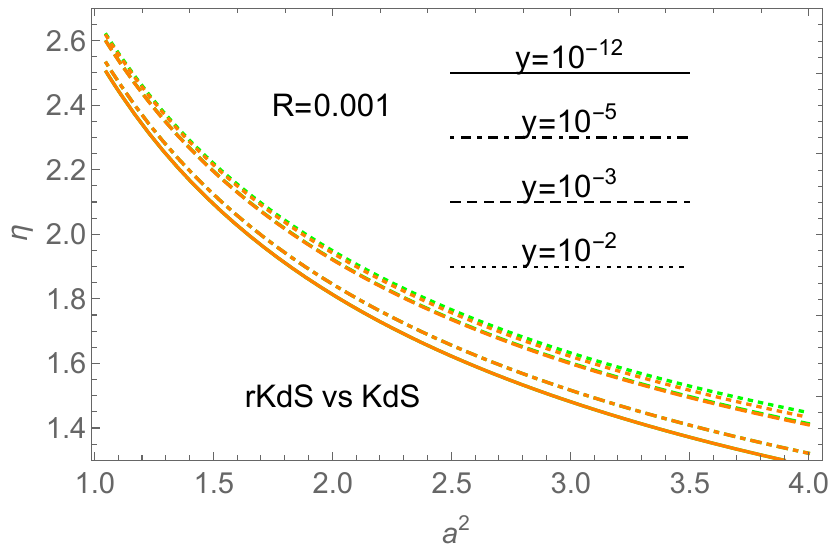}		
	\end{tabular}	
	\caption{Comparison of the peak angle $\eta$ corresponding to the shadow of the superspinar with the surface at the marked radius $R$ for both rKdS (green curves) and KdS (orange curves) geometries. The curves are almost indistinguishable for a given $R$, so we only present them directly in the comparison mode.}     
	\label{fig_eta_a2}	
\end{figure*}

\begin{figure*}[h!]
	\centering
	\begin{tabular}{cccc}
		\includegraphics[width=0.24\textwidth]{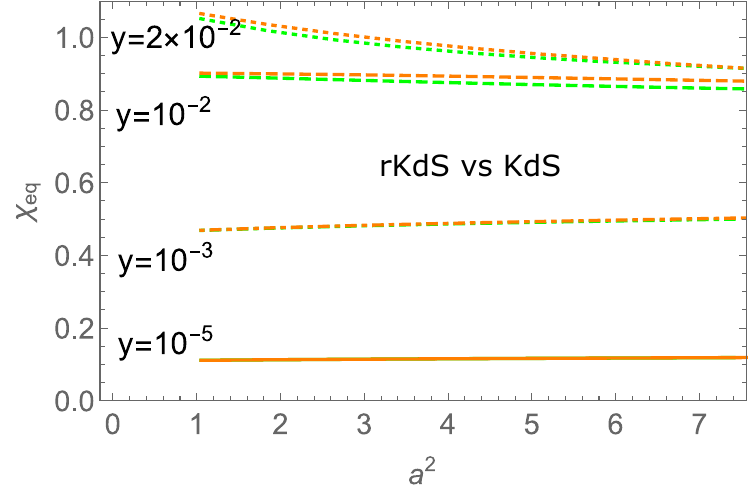}&\includegraphics[width=0.24\textwidth]{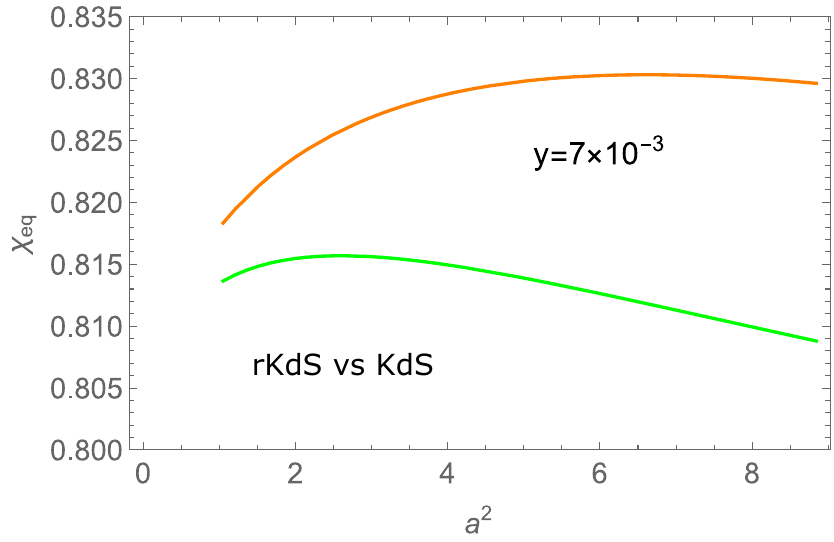}&\includegraphics[width=0.24\textwidth]{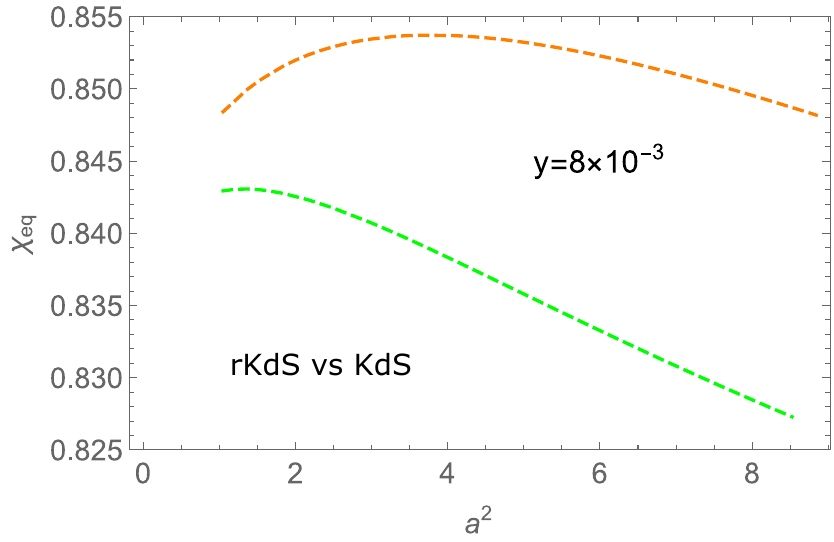}&\includegraphics[width=0.24\textwidth]{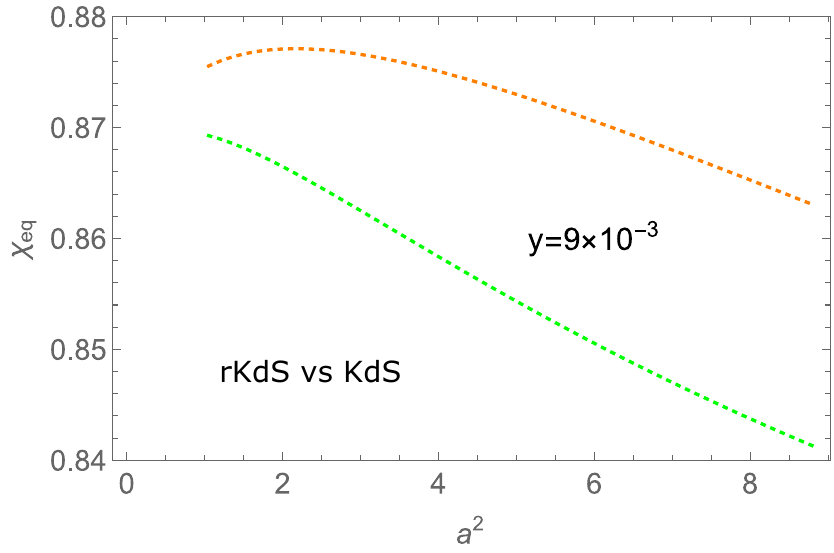}		
	\end{tabular}	
	\caption{Comparison of the angular radius $\chi_{eq}$ of the osculating circle as seen in the equatorial plane $\theta_{0}=90\dgr$ for rKdS (green) and KdS (orange) geometries. The figure on the left shows the magnification effect of the cosmological constant, while the remaining figures compare in detail the dependence on the spin parameter.}     
	\label{fig_chieq_a2}	
\end{figure*}

We compare the behavior of the peak angle $\eta$ of the silhouette of the superspinar and the central angle $\xi_{eq}$ of the light arc observed from the equatorial plane as a function of the spin parameter $a^2$ in Figs. \ref{fig_eta_a2} and \ref{fig_chieq_a2} over a wider range of its values corresponding to both types of spacetimes with and without polar SPOs, since their presence in these cases does not cause qualitative changes.

The plots in Fig. \ref{fig_eta_a2} are constructed for different values of the radius $R$ of the surface of the superspinar, which we consider as the absorbing surface. It can be seen that the observed angular dimensions of the shadows are almost identical for the same values of $a^2, y, R$ for both geometries.

From Fig. \ref{fig_chieq_a2}, for the same parameter $a^2$, the magnifying effect of the cosmological constant $y$ can be seen, which is slightly more pronounced in the case of the standard KdS spacetime. While the effect grows monotonically with increasing $y$ for a given $a^2$, the dependence on the spin parameter $a^2$ for a given $y$ is more complicated. For very small values of $y$, it initially increases with increasing $a^2$ , while for larger values of $y$ it decreases with increasing $a^2$. A qualitative change occurs for $y$ around $7\times 10^{-3}$ for both geometries, as the three plots on the right show in detail.

\clearpage

\subsection{Shadows in radial geodesic frames}
In this subsection, we compare the effect of the cosmological constant on the observed shadow of a superspinar that is radially drifted by the cosmic repulsion. Here, again in agreement with our work \cite{Stu-Char:2024:PHYSR4:}, we consider the radial motion of the observer having only the radial nonzero component of the observer's four-velocity measured locally with respect to the LNRF passing through the momentarily considered point of the spacetime.

\subsubsection{Radial geodesic frames}
Building on previous work \cite{Stu-Char-Sche:2018:EPJC:}, we define a radial geodetic reference frame (RGF) as a frame that is associated with a radially falling or escaping observer, for which the only nonzero velocity component, measured with respect to an LNRF currently orbiting the observer's radial coordinate, is the radial component $v^{(r)}$.
 
The locally measured three-velocity components of the observer $v^{(i)}$, $i=r,\theta, \phi$, are related via the locally measured four-velocity components $u^{(a)}$ to the coordinate four-velocity components $u^{\mu}=\din x^{\mu}/\dbe \tau$ by a general relation
\be
v^{(i)}=\frac{u^{(i)}}{u^{(t)}}=\frac{\omega^{(i)}_{\mu}u^{\mu}}{\omega^{(t)}_{\nu}u^{\nu}}. \label{v(i)}
\ee
Since the proper time $\tau$ is related to the affine parameter $\lambda$ by $\tau=m \lambda$, therefore $p^{\mu}=m u^{\mu}$, and we can rewrite Eq. (\ref{v(i)}) as
\be
 v^{(i)}=\frac{\omega^{(i)}_{\mu}p^{\mu}}{\omega^{(t)}_{\nu}p^{\nu}}, \label{vi(p)}
 \ee
 where $p^{\mu}$'s values are determined by Eqs. (\ref{CarterR})--(\ref{CarterT}).
 However, it is convenient to express the azimuthal component $v^{(\phi)}$ using Eqs. (\ref{p(a)contra})--(\ref{p(a)cov vs p(a)contra}) in terms of the covariant components of the form
 \be
 v^{(\phi)}=\frac{p^{(\phi)}}{p^{(t)}}=-\frac{p_{(\phi)}}{p_{(t)}}=-\frac{e^{\mu}_{(\phi)}p_{\mu}}{e^{\nu}_{(t)}p_{\nu}}=\frac{e^{\phi}_{(\phi)}\Phi}{e^{t}_{(t)}\cale-e^{\phi}_{(t)}\Phi}, \label{v(phi)}
 \ee
 from which it is immediately apparent that the condition $v^{(\phi)}=0$ implies
 \be
 \Phi=0.\label{Phi=0}
 \ee
 
 Using relation (\ref{v(i)}), we can derive that
 \be
 v^{(\theta)}=\sqrt{\frac{A}{\Delta}}\frac{p^{\theta}}{p^{t}}, \label{v(theta)}
 \ee
 from which, along with the associated relation (\ref{CarterW}), it can be seen that the condition $v^{(\theta)}=0$ implies $W=0$. This, following Ref. \cite{Stu-Char-Sche:2018:EPJC:}, can be is expressed by using the rescaled motion constant $Q$ and Eq.~(\ref{Phi=0}) in the form
 \be
 Q=Q_{\theta}(\theta;E^2,a^2)\equiv a^2 \cos^2\theta(1-E^2).           \label{Q=Q(theta)}
 \ee
 In general, the latitudinal motion of test particles is given by the condition $W\geq0$, which can be written in the form $Q\geq Q_{\theta}(\theta;E,a^2)$ for a particle with $\Phi=0$ using the motion constants introduced above. Perturbation analysis can then show that stable/unstable motion along a hyperplane of constant latitude $\theta_{0}$ is admissible for motion constants $E^2, Q$ corresponding to local minima/maxima of the function $Q_{\theta}(\theta;E^2,a^2)$. Stable motion therefore exists in the equatorial plane, $\theta_{0}=\pi/2$, for $E^2<1$ and $Q=0$, or along the axis of rotation, $\theta_{0}=0$, for $E^2>1$ and $Q=a^2(1-E^2)<0$, while the unstable motion exists along the spin axis for $E^2<1$ and $Q=a^2(1-E^2)>0$, or at any latitudinal coordinate $0\leq \theta_{0}\leq \pi/2$ for $E^2=1$ and $Q=0$, or in the equatorial plane for $E^2>1$ and $Q=0$. Here and in the following, we restrict our discussion to the interval $\theta_{0}\in [0;\pi/2]$, due to the symmetry with respect to the equatorial plane.
 
 There is therefore a difference from the KdS spacetimes, where there is a small interval of energies $I^{-1}\leq E^2\leq I$ and corresponding values of the motion constant $Q(E^2,\theta_{0})\leq 0$~\cite{Stu-Char-Sche:2018:EPJC:}, for which stable motion is possible for any latitude $0\leq \theta_{0}\leq \pi/2$. 
 
 Since our intention is to study motion along an arbitrary constant coordinate $\theta_{0}$, we restrict ourselves in the following to the case $E^2=1, Q=0$. With these motion constants, the radial function $R$ in Eq. (\ref{CarterR}) simplifies significantly into the form
 \be
 R(r) = (r^2+a^2)(r^2+a^2-\Delta). \label{R(E=1,Q=0)} 
 \ee
 The locally measured radial velocity $v^{(r)}$ according to the general relation (\ref{v(i)}) is given by 
 \be
 v^{(r)}=\frac{\sqrt{A}}{\Delta}\frac{p^{r}}{p^{t}},
 \ee
 which can be simplified using Eq. (\ref{R(E=1,Q=0)}) to a form that can again be obtained directly by removing the boxed expressions in its KdS version derived in Ref. \cite{Stu-Char:2024:PHYSR4:}:
 \be
 v^{(r)}=\sqrt{1-\frac{\rho^2 \Delta}{A}}. \label{v(r)}
 \ee
 The rKdS result is formally identical to that obtained in the case of pure Kerr and also braneworld Kerr spacetimes \cite{Sche-Stu:2009:IJMPD:,Stu:1980:BULAI:}, however, there is a difference in the definition of the $\Delta$ term.
 
\begin{figure}
 	\includegraphics[width=\linewidth]{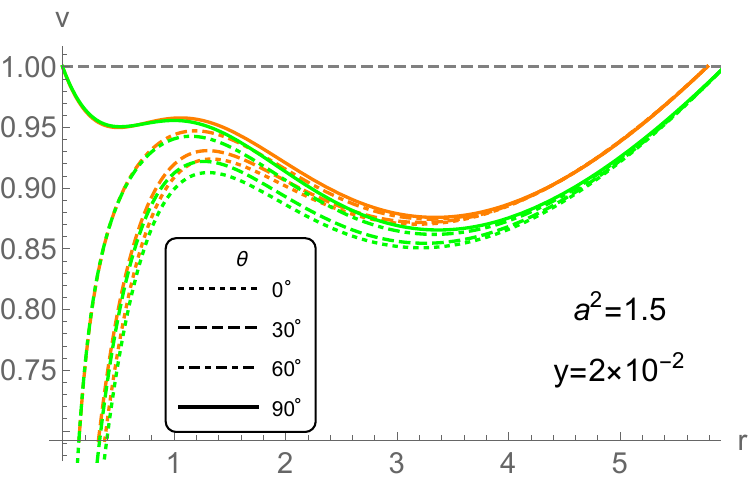}
 	\caption{Comparison of radial velocity versus radial coordinate for the indicated latitudes of motion for rKdS (green curves) and KdS (orange curves) geometries.    
 	} \label{fig_rad_vel_0p02}
 \end{figure}

 A comparison of the radial velocity versus the radial coordinate for different latitudes of motion for both rKdS and KdS geometries is shown in Fig.~\ref{fig_rad_vel_0p02}.

\clearpage

\subsubsection{Components of the four-momentum of a photon measured in RGFs}
The four-momentum photon components $k^{(\hat{a})}$ measured locally in the RGF are related to the LNRF components $k^{(b)}$ by a standard Lorentz transformation
\be
k^{(\hat{a})}=\Lambda^{\!(\hat{a})}_{\:(b)}k^{(b)}, \label{khata}
\ee
where $\Lambda^{\!(\hat{a})}_{\:(b)}$ is the local Lorentz transformation matrix 
\be
\Lambda^{(\!\hat{a})}_{\:(b)}=\left( \begin{array}{cccc}
	\gamma & -\gamma v & 0 & 0\\
	-\gamma v & \gamma & 0 & 0\\
	0 & 0 & 1 & 0\\
	0 & 0 & 0 & 1
\end{array}	\right), \label{Ldirect}
\ee
and 
\be
\gamma=\frac{1}{\sqrt{1-v^2}}
\ee
is the Lorentz factor (see Ref. \cite{Stu-Char:2024:PHYSR4:}), where for brevity we have used the symbol $v$ to denote the relative velocity of the LNRF and RGF systems instead of $v^{(r)}$. 

\subsubsection{Directional angles of photons and celestial coordinates in the RGFs}
The ratios of the $k^{(\hat{a})}$ components analogous to relations (\ref{cosalpha}-\ref{sinalphasinbeta}) define the directional angles $(\hat{\alpha}, \hat{\beta})$ of the photon measured locally in the RGF. This transformation leads to the well-known relation for aberration
\be
\cos \hat{\alpha} =\frac{\cos \alpha-v}{1-v \cos \alpha}, \label{cosbaralpha}
\ee
and
\be
\hat{\beta}=\beta.\label{barbeta=tildebeta}
\ee 
The celestial coordinates $\bar{\alpha}, \bar{\beta}$ on a small section of the sky of a RGF observer are then related to the directional angles $\hat{\alpha}, \hat{\beta}$ by the relations
\bea
\bar{\alpha}&=&- k^{(\hat{\phi})}/k^{(\hat{t})}=-\sin \hat{\alpha} \sin \hat{\beta}  \nonumber \\
\bar{\beta}&=& k^{(\hat{\theta})}/k^{(\hat{t})}=\sin \hat{\alpha} \cos \hat{\beta} . \label{stereograph2}
\eea

\subsubsection{Astronomical observables}
In this subsection, we compare the realistic shapes of the superspinar shadows within the KdS and rKdS metrics assuming cosmological distances at which the effect of the radial velocity of the observer with respect to the source due to the expansion of the Universe starts to take effect. We use the currently estimated value of the cosmological constant $\Lambda=1.1\times 10^{-52}\text{m}^{-2}$. To model the shadows generated by the (r)KdS metric, which describes an isolated rotating BH/NS (superspinar) in a de Sitter background with no surrounding matter, we choose as a representative case the observation of an object at a distance of $300 \text{Mpc}$. This distance corresponds approximately to the opposite edge of the local cosmological void known as the Local Void~\cite{KeenanBargerCowie2013,supercluster}, a large subdense region adjacent to a local group of galaxies estimated to be several hundred Mpc in size. In the following calculations we will consider a compact object of mass of the order of $10^{10}\,M_{\odot}$--i.e., comparable to the most massive black hole currently known in the core of the quasar TON 618, whose estimated mass is approximately $6.6 \times 10^{10}\,M_{\odot}$ and distance $3.31\,\text{Gpc}$. Such an extreme mass not only allows the effects of gravitational curvature near the superspinar surface to be well resolved, but also ensures that the shadow will be observable from cosmological distances within the chosen model.

\begin{figure*}[h!]
 	\centering
 	\begin{tabular}{cc}
 		\includegraphics[width=0.45\textwidth]{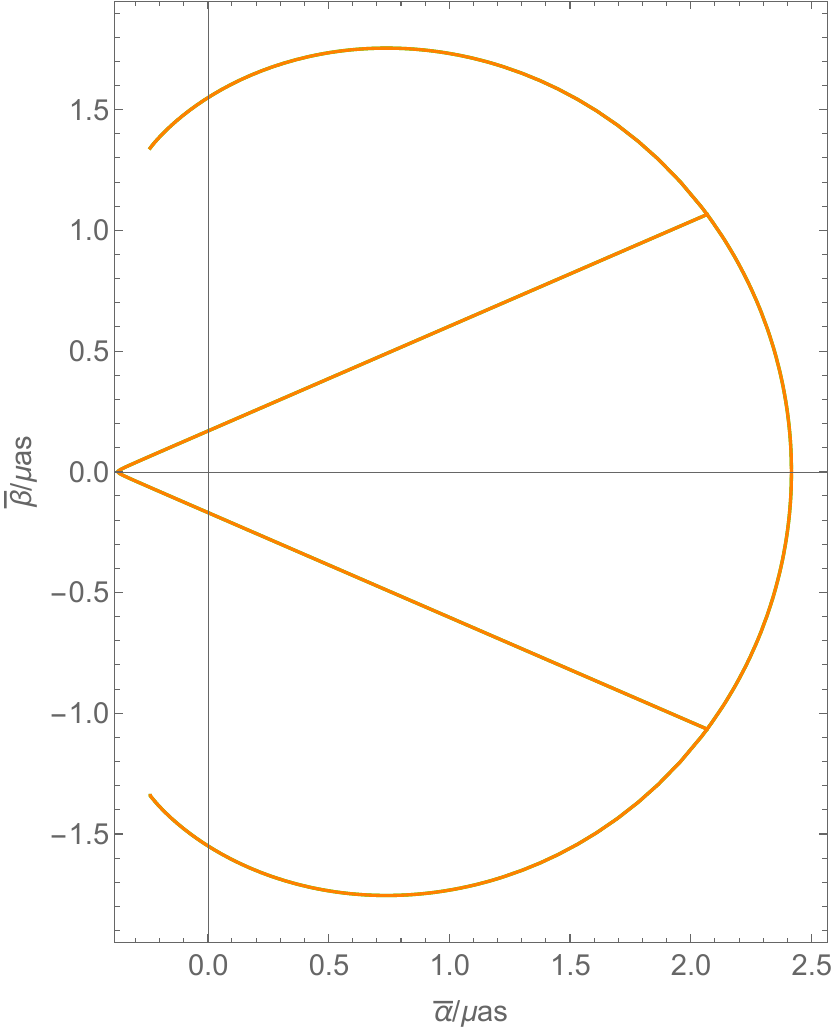}&\includegraphics[width=0.45\textwidth]{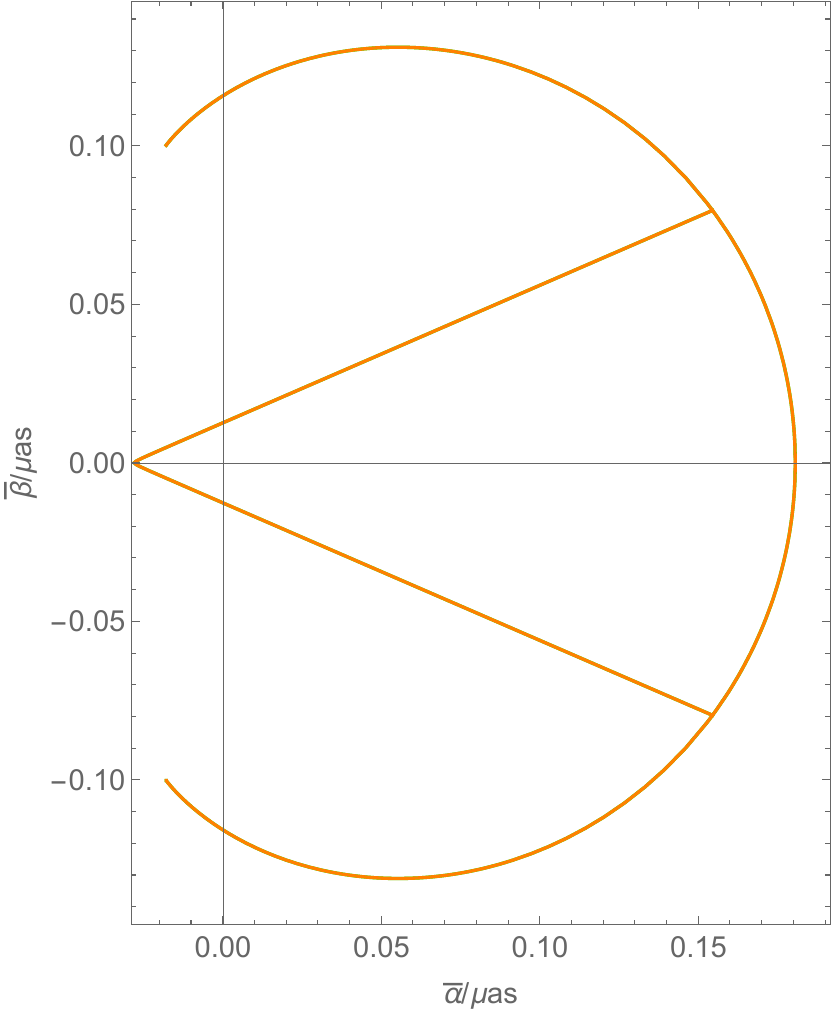}\\
 		(a)&(b)		
 	\end{tabular}	
 	\caption{(a) Shadow of the superspinar with mass $M=10^{10}\,M_{\odot}$, corresponding to the mass of the most massive black hole currently known in the core of the quasar TON 618, observed from the equatorial plane from the radial distance $r_{o(dim)}=300\,\text{Mpc}$ in standard dimensional units--i.e., $r_{o}=6.1\times 10^{11}$. The distance chosen corresponds to the dimensions of the local cosmological void known as the Local Void. (b) The same, done for the actual quasar distance $r_{o(dim)}=3.31 \text{Gpc}$--i.e., $r_{o}=6.6\times 10^{12}$. Panels (a) and (b) show the constructions for both rKdS and KdS geometries overlapping each other. }     
 	\label{fig_shads}	
 \end{figure*}

In Ref. \cite{Stu-Char:2024:PHYSR4:}, for the considered value of the cosmological constant $\Lambda$, we expressed the cosmological parameter $y$ by a multiple $n_{\odot}$ of the solar mass of the form
\be
 y=8\times10^{-47}n^{2}_{\odot}. \label{ynsun}
\ee
In the present intended case, $n_{\odot}=10^{10}$, hence
\be
y=8 \times 10^{-27}. \label{y1}
\ee
It can easily be shown that for such small values of the parameter $y$, the radial velocity [Eq.(\ref{v(r)})] can be expressed in a very good approximation of the form
\be
v^{(r)}\approx r \sqrt{y}, \label{vr_approx}
\ee 
 which is valid for any latitude and $0\lesssim r \leq r_{c}$. Moreover, the same approximate expression as Eq. (\ref{vr_approx}) holds under these conditions for the radial velocity $v^{(r)}$ in the case of the KdS metric. Furthermore, in both rKdS and KdS metrics, under the conditions considered, the radius of the cosmological horizon is
 \be
 r_{c}\approx \frac{1}{\sqrt{y}}. \label{rc_approx}
 \ee
 The cosmological horizon and radial velocity at a given radius $r_{o}$ of the observer are therefore the same for both geometries, and from Eqs. (\ref{vr_approx}) and (\ref{rc_approx}), we apparently can write
 \be
 v^{(r)}=\frac{r_{o}}{r_{c}},\quad 0\lesssim r_{o} \leq r_{c}. \label{vr=ro/rc}
 \ee 
 For the value given by Eq. (\ref{y1}), we have 
 \be
 r_{c}=1.1\times 10^{13},
 \ee
 and the dimensional radial distance $r_{o(dim)}=300\,\text{Mpc}$ and the considered mass $M=10^{10}\,M_{\odot}$ give in our dimensionless units the value $r_{o}=6.1\times 10^{11}$. The radial velocity is then $v^{(r)}=0.05$. The resulting images obtained for the two geometries are practically indistinguishable, as shown in Fig. \ref{fig_shads}(a). For comparison, in Fig.~\ref{fig_shads}(b) we present the resulting images for $r_{o}=6.6\times 10^{12}$, corresponding to the estimated real distance of the TON 618 quasar in our dimensionless units.  We construct the images for observations from the equatorial plane because, as can be seen from Figs. \ref{fig_chi_theta_1.2} and \ref{fig_chi_theta_2}, they correspond to the largest angular dimensions.
  
\begin{figure*}[h!]
	\centering
	\begin{tabular}{ccc}
		\includegraphics[width=0.33\textwidth]{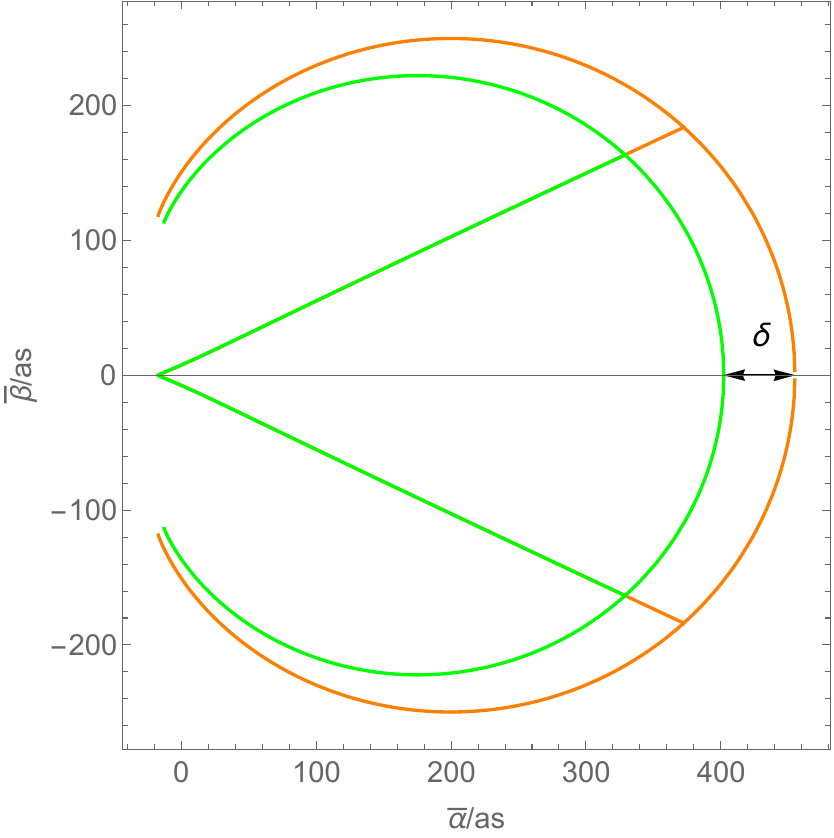}&\includegraphics[width=0.33\textwidth]{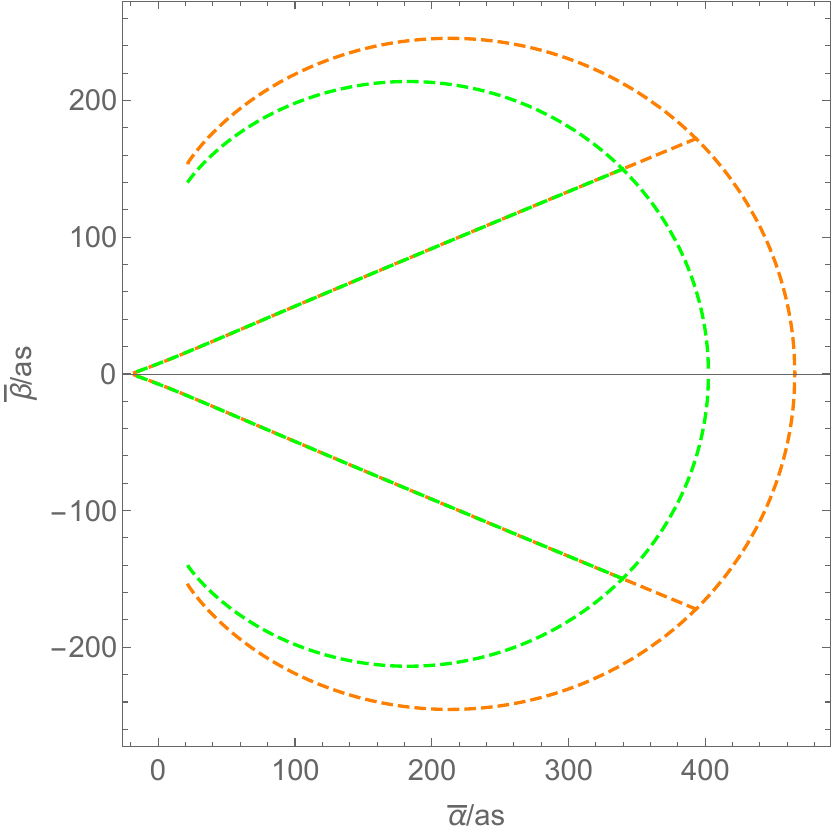}&\includegraphics[width=0.33\textwidth]{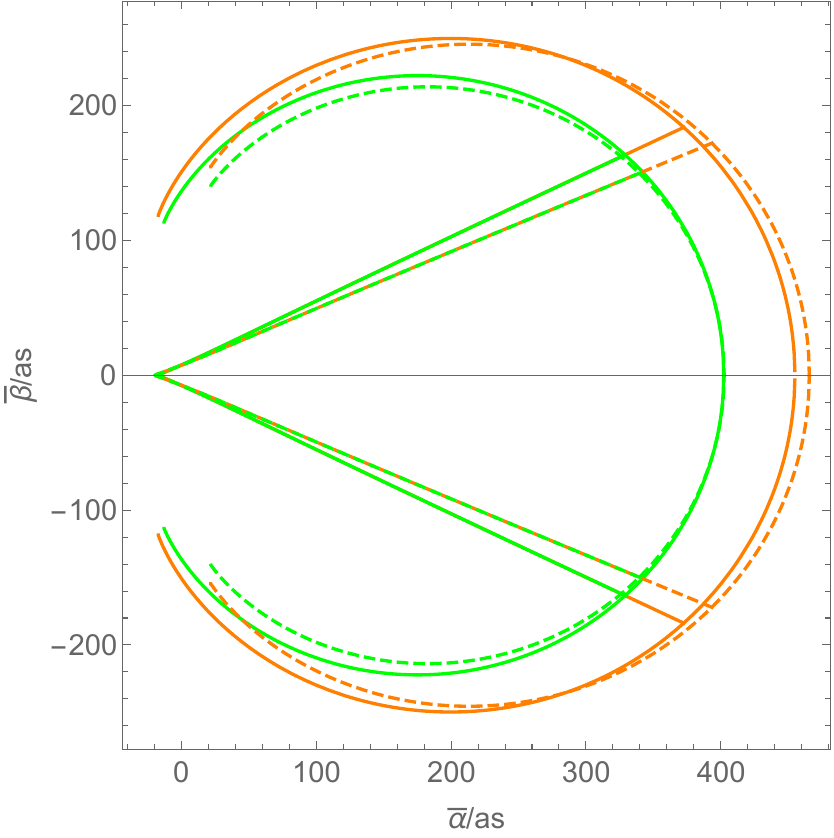}\\
		(a)&(b)&(c)
		\end{tabular}	
	\caption{Comparison of superspinar shadows as seen by a radially escaping observer at $r_{o}=0.85 r_{c}$ in rKdS (green curves) and KdS (orange curves) spacetimes for the parameter sets (a) $y = 0.02, a^{2}=1.2$ and (b) $y = 0.02, a^{2}=1.5$, and (c) with both cases overlaid to highlight the effect of $a^{2}$. In the rKdS/KdS case, $r_{c}=5.90/5.77$. These illustrations are intended solely to clarify the qualitative differences introduced by the new rKdS geometry, since the chosen value $y=0.02$ corresponds to a cosmological constant many orders of magnitude larger than current observational estimates and therefore does not represent a physically realistic scenario. In the left panel (a), the bidirectional arrow marks the angular separation $\delta$ between the left edges of the rKdS and KdS shadows; this quantity is introduced to evaluate, in the following analysis, for which values of $y$ (and thus of the cosmological constant) the difference becomes comparable to the angular resolution of current observational facilities. }   
	\label{fig_shads_comp}	
\end{figure*}

 To further illustrate the qualitative differences between the rKdS and KdS geometries, we present in Fig. \ref{fig_shads_comp} a set of shadow constructions computed for deliberately exaggerated values of the cosmological parameter $y$. Although these configurations do not correspond to observationally plausible scenarios, they provide a clearer visual contrast between the two spacetimes. In particular, we introduce the angular separation $\delta$ between the left edges of the two shadows, as indicated in the left panel. This quantity will be examined below in order to determine for which values of $y$ (and thus of the cosmological constant) its magnitude becomes comparable to the angular resolution achievable by current observational facilities.

\clearpage

 \begin{figure}
	\includegraphics[width=\linewidth]{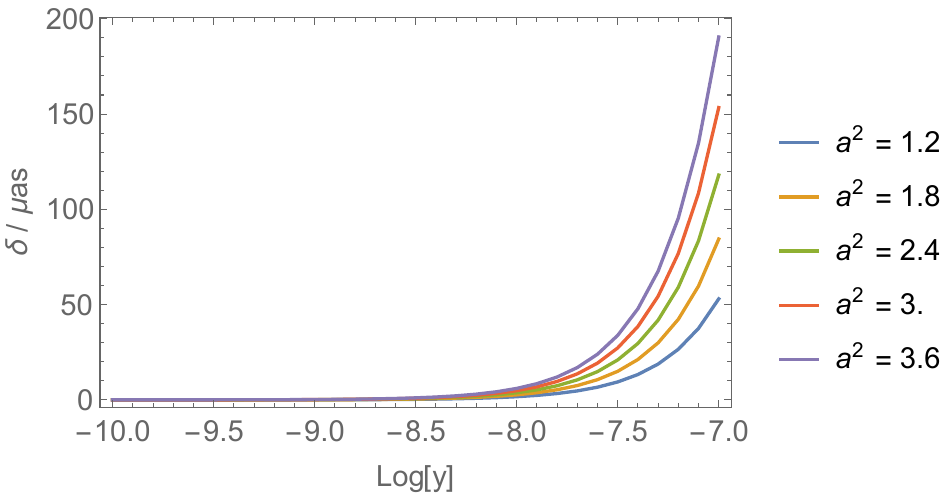}
	\caption{Comparison of the angular displacement~$\delta$ of the left edges of superspinar shadows for rKdS and KdS geometries, computed for radially escaping observers near the cosmological horizon, as a function of the cosmological parameter~$y$ for several values of the spin parameter~$a^{2}$.
	} \label{fig_deltas}
\end{figure}

From Fig.~\ref{fig_deltas}, we observe that if the angular resolution is limited to several tens of microarcseconds, as in the case of the Event Horizon Telescope, then noticeable differences between the rKdS and KdS shadows would become distinguishable only for cosmological parameters of the order of $y \gtrsim 10^{-8}$. If a more detailed substructure of the shadow -- potentially differing between the two geometries -- were accessible, such a value of $y$ would therefore be required in order to determine which geometry the observed image corresponds to. For the assumed superspinar mass $M = 10^{10}\,M_{\odot}$, this threshold corresponds to the cosmological constant 
\be \Lambda = \frac{3y}{M^2}= \frac{3\times 10^{-8}}{(10^{10}\times 1.5\times 10^3 \text{m})^2}\approx 10^{-34} \text{m}^{-2}. \label{lambda_1044}
\ee
We see that to obtain resolvable differences between the two geometries at the required level, the central object would need to be more massive by about 9 orders of magnitude--i.e., $M \sim 10^{19}\,M_{\odot}$. Such an extreme mass would vastly exceed that of any known astrophysical structure, surpassing even the total mass of the most massive galaxy clusters and approaching the mass scale of a substantial fraction of the observable Universe contained within a few hundred megaparsecs. This conclusion is not alleviated by the increasing angular separation between the two geometries observed at higher values of the spin parameter $a^{2}$, as shown in Fig.~\ref{fig_deltas}; even in these cases, the required mass scale remains far beyond any astrophysically conceivable configuration.

  \subsection{Light-escape cones in CGFs}
  In this subsection, we compare the LECs created for both rKdS and KdS geometries for a light source connected with the circular geodesic frame (CGF) orbiting the innermost and outermost stable equatorial circular orbits of test particles. The research of LECs in such reference frames is important for studying optical phenomena related to processes in Keplerian discs, such as self-illumination, self-eclipse, or self-heating \cite{Stu-Sche:2010:CQG:}, \cite{Stu-Char-Sche:2018:EPJC:}. For this reason, it is advisable to focus on the area of the disk that overlaps with the area of the SPOs.

\subsubsection{Significant particle and photon orbits}
It is generally assumed that the boundaries of the Keplerian disk are given by marginally stable circular orbits of test particles in the equatorial plane. When we simultaneously solve the conditions
\be
R(r)=0 \quad \oder{R}{r}=0 \label{conditions_co}
\ee
for the existence of circular orbits with respect to specific energy 
\be
E=E_{\pm}(r;a,y)=\frac{1-2/r-yr^2\pm a\sqrt{1/r^3-y}}{1-3/r\pm2a\sqrt{1/r^3-y}} \label{Epm}
\ee
and specific angular momentum
\be \Phi=\Phi_{\pm}(r;a,y)=\frac{-2a/r-yar^2\pm(r^2+a^2)\sqrt{1/r^3-y}}{[1-3/r\pm2a\sqrt{1/r^3-y}]^{1/2}}, \label{Phipm}
\ee
then by substituting the expressions in Eqs. (\ref{Epm}) and (\ref{Phipm}) into the condition
\be
\frac{\mathrm{d^2}R}{\mathrm{d}r^2}=0 \label{conditions_ms}
\ee
for marginally stable orbits and restricting them to the equatorial plane where $Q=0$, we can express the radii $r_{msc}$ of marginally stable circular orbits of test particles as solutions of the equations (see Ref. \cite{Slany:2023:PHYSR4:}) \footnote{Building on the work of \cite{Stu-Sla:2004:PHYSR4:} and \cite{Slany:2023:PHYSR4:}, in this subsection we will express all the relevant radii using the characteristic functions of the form $a^2(r;y)$.}  
\be
a^2=a^2_{ms(1,2)}(r;y)\equiv \frac{r}{9} [4(1-yr^3)^{3/2}\pm \sqrt{\cald}]^2, \label{a2ms(1,2)}
\ee
where
\be
\cald \equiv (1-4yr^3)[3r-2-yr^3(11-4yr^3)]. \label{cald}
\ee
Here the functions $a^2_{ms(1,2)}(r;y)$ correspond in some manner to marginally stable orbits of the so-called plus-family and minus-family orbits (for details, see Ref. \cite{Slany:2023:PHYSR4:}). 

The definition of the functions $a^{2(KdS)}_{ms(1,2)}(r;y)$, which are counterparts to functions defined in Eq. (\ref{a2ms(1,2)}) in standard KdS geometry is rather complex and can be found in Ref. \cite{Stu-Sla:2004:PHYSR4:}.

The area of possible occurrence of self-illuminating effects of the Keplerian disk is determined by its overlap with the area of stable spherical photon orbits. 

In the equatorial plane, the SPO region is defined by equatorial circular photon orbits (ECPOs) determined by the condition (\ref{qspo=0}), which can be equivalently expressed by the equation
\be
a^2=a^2_{ECPO}(r;y)\equiv \frac{r(r-3)^2}{4(1-yr^3)}. \label{a2ECPO}
\ee
The function analogous to that in Eq. (\ref{a2ECPO}) in the KdS case reads \footnote{There is another function describing ECPO, but it is not relevant to our considerations because it corresponds to parameters under which the Keplerian disk does not exist.}
\bea
&&a^{2(KdS)}_{ECPO}(r;y)\equiv \\
&&\frac{2-3yr^2-yr^3-2\sqrt{(1-yr^3)(1-3yr^3)}}{y^2r^3}.\nonumber \label{a2ECPOKdS}
\eea
The boundaries of the Keplerian disk region with the possible occurrence of the above-mentioned optical effects are determined by marginally stable SPOs, whose radii $r_{ms}$ we can express, in analogy with Eq. (\ref{a2ms(1,2)}), as the solution to Eq. (\ref{rms_conditions}) in the form
\be
a^2=a^{2}_{ms(SPO)}(r;y)\equiv \frac{3r(1-yr^3)+r^2(r-3)}{1-4yr^3}. \label{a2ms(SPO)}
\ee
The corresponding formula for marginally stable SPOs in standard KdS spacetime is
\bea
&&a^{2(KdS)}_{ms(SPO)}(r;y)\equiv \\
&&\frac{1-3yr^2-2yr^3+\sqrt{(1-4yr^3)(1-6yr^2-3y^2r^4)}}{2y^2r^3}.\nonumber \label{a2ms(SPO)KdS}
\eea

The SPO region with negative covariant energy $E<0$ is also of particular importance, where rotational energy can be expected to be extracted from the central gravitational object via the so-called Penrose radiation process \cite{Kol-Tur-Stu:2021:PHYSR4:}. The boundary of this region is given by the radii of photons with zero energy $E=0$, which corresponds to divergence of the function $X_{\spo}(r)$, which is given by $r^{\pm}_{d(SPO)}$, defined in Eq. (\ref{rdexpm}) and discussed in Sec. \ref{ssec_class_criteria}.

In the context of the Keplerian disk, only the radius $r^{+}_{d(SPO)}$ is relevant, which we refer to as $r_{E=0}$ in the following text.

The radii $r_{E=0}$ in the KdS case can be determined by the function
\be
a^2=a^{2(KdS)}_{E=0}(r;y)\equiv \frac{r-2yr^3-1}{yr}, \label{a2(E=0)(KdS)}
\ee
which we introduce in order to compare the radii of orbits with negative energy with other outstanding orbits determined by the functions introduced above, and between the two rKdS and KdS geometries.
 
\begin{figure*}[b!]
	\centering
	\begin{tabular}{ccc}
		\includegraphics[width=0.33\textwidth]{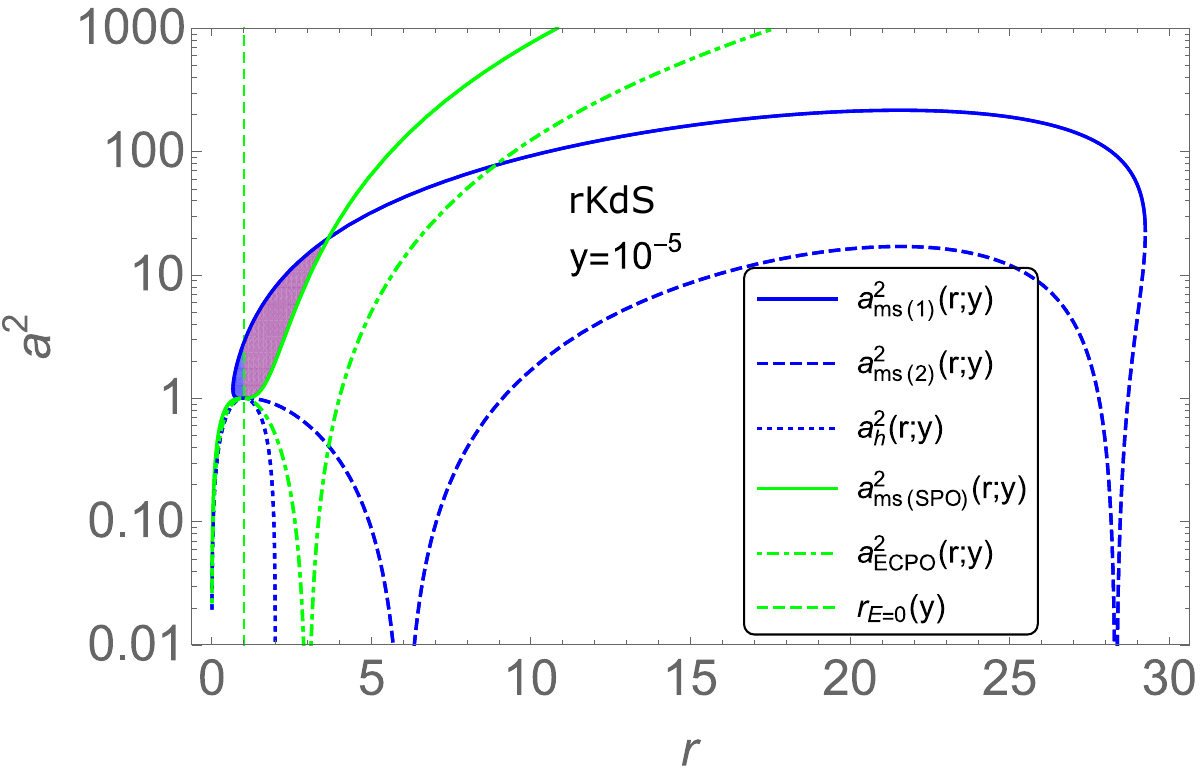}&\includegraphics[width=0.33\textwidth]{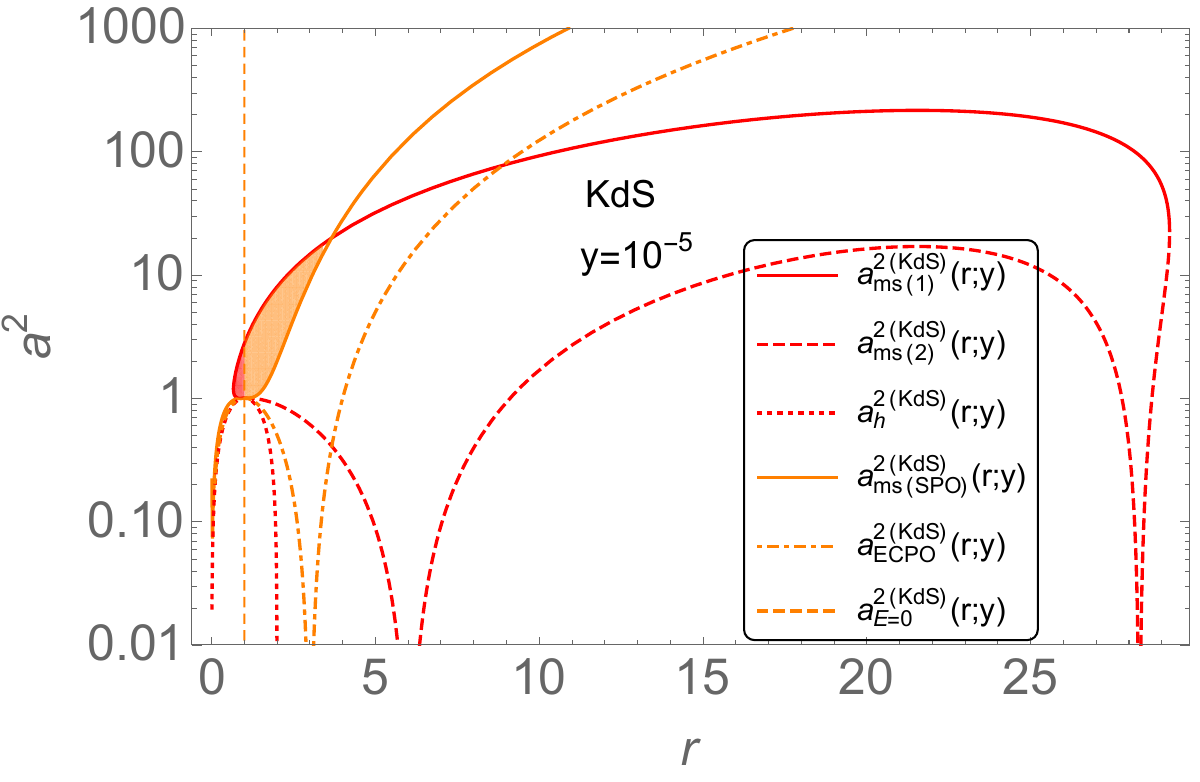}&\includegraphics[width=0.33\textwidth]{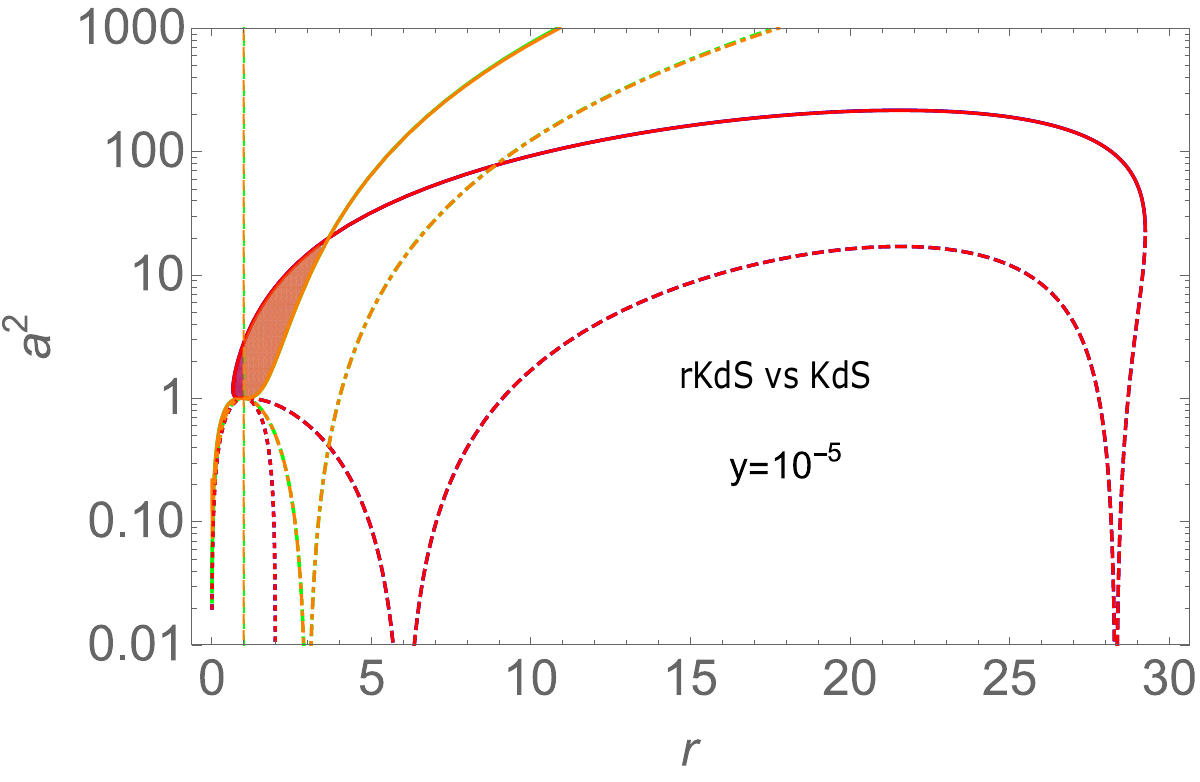}\\
		\includegraphics[width=0.33\textwidth]{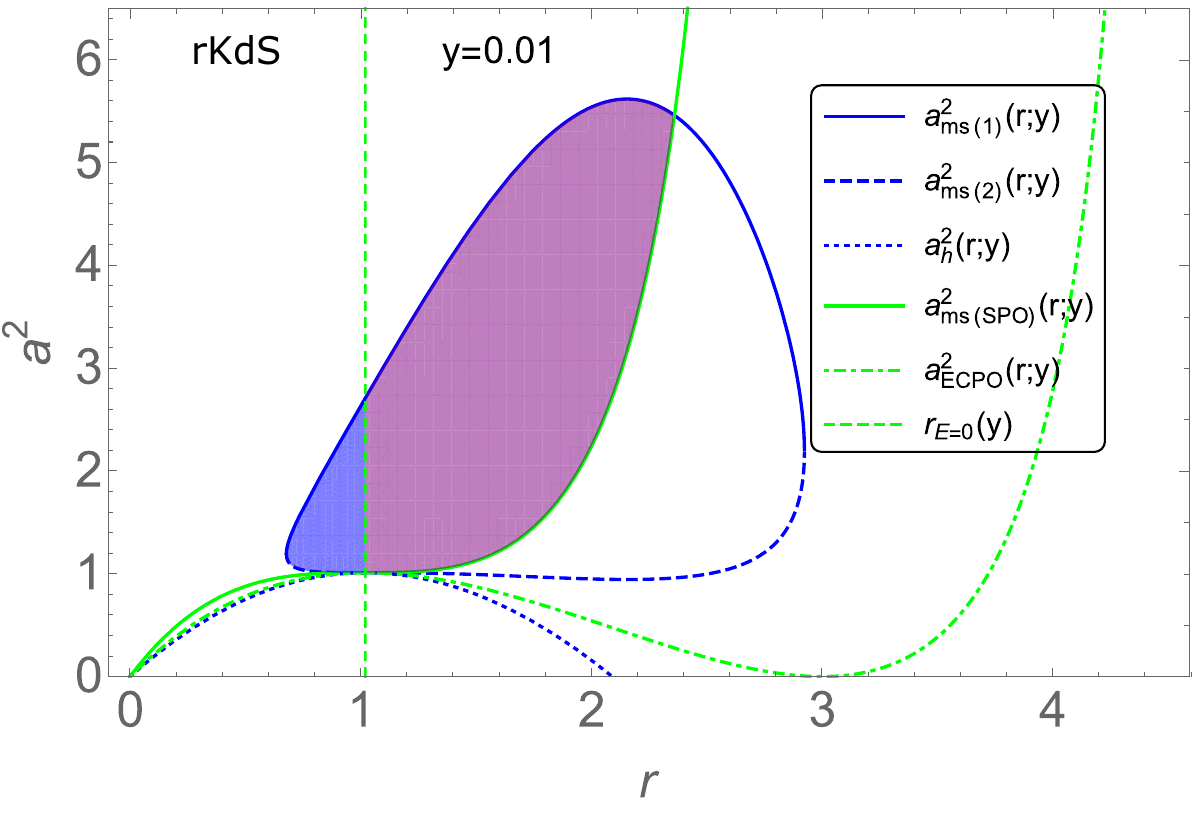}&\includegraphics[width=0.33\textwidth]{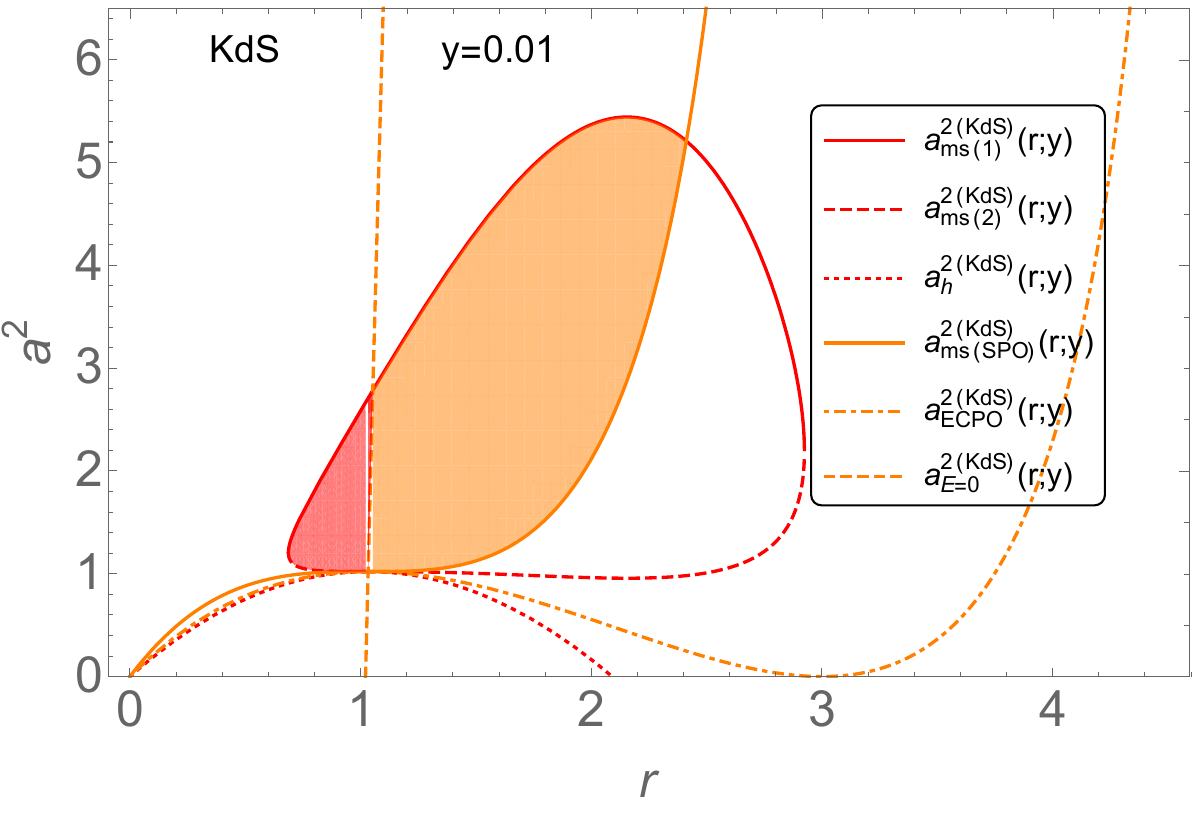}&\includegraphics[width=0.33\textwidth]{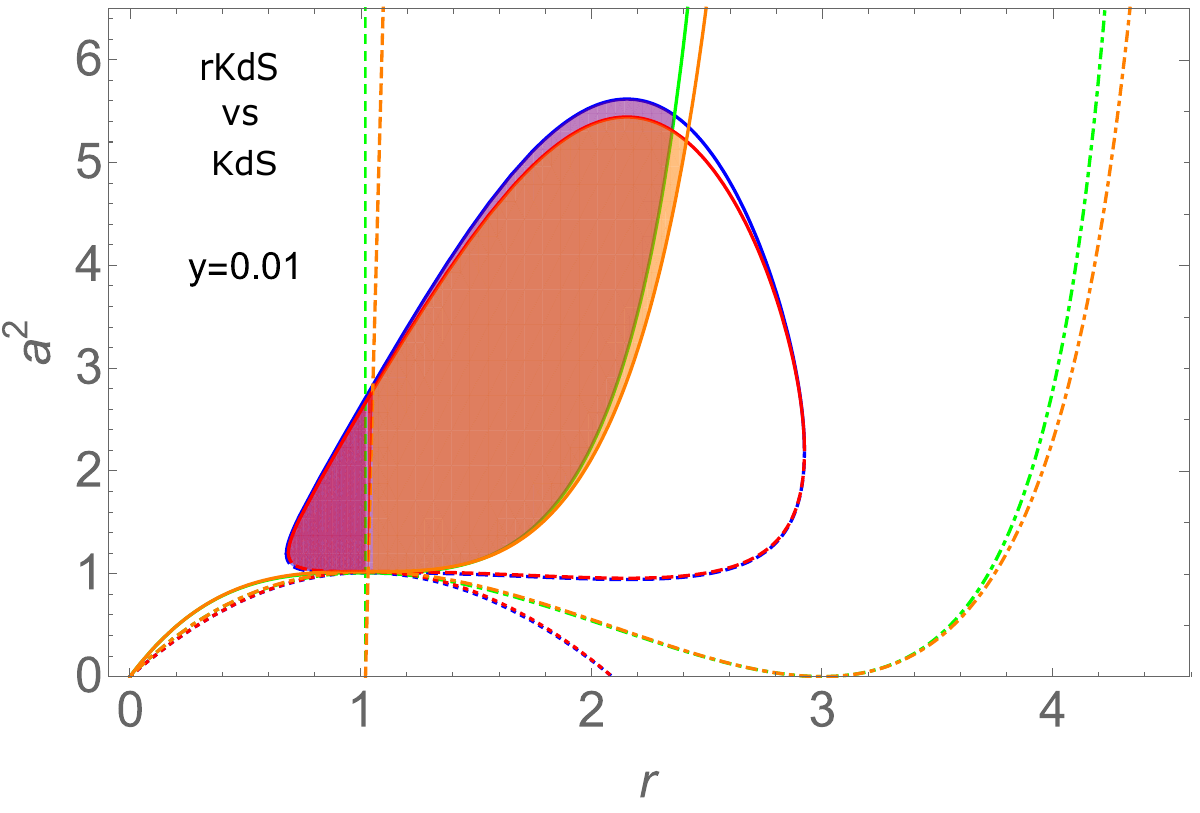}		
	\end{tabular}	
	\caption{Comparison of the relative positions of the marginally stable orbits of test particles and spherical photon orbits for both rKdS and KdS geometries. The radii of the marginally stable orbits of test particles, which determine the possible extent of the Keplerian disk, are given by the functions $a^2_{ms(1,2)}(r;y)/a^{2(KdS)}_{ms(1,2)}(r;y)$ (blue/red solid and dashed curves) for rKdS/KdS spacetime. The region of SPOs is limited by the radii of equatorial circular photon orbits,  which are determined by the function $a^2_{ECPO}(r;y)/a^{2(KdS)}_{ECPO}(r;y)$ (green/orange dot-dashed curve). The radii of marginally stable photon orbits are determined by the function $a^2_{ms(SPO)}(r;y)/a^{2(KdS)}_{ms(SPO)}(r;y)$ (solid green/orange curve). The region of possible overlap between the Keplerian disk and stable SPOs is highlighted in purple/orange. The area of overlap between the Keplerian disk and photons with negative energy, whose radii are determined by the function $a^2_{E=0}(r;y)/a^{2(KdS)}_{E=0}(r;y)$ (green/orange dashed curve), is marked in blue/red. The position of BH horizons is determined by the function $a^{2}_{h}(r;y)/a^{2(KdS)}_{h}(r;y)$ (blue/red dotted curve). The upper row corresponds to the case $y<12/15^4=2.37\times 10^{-4}$, which is a limit for the existence of stable minus-family orbits for both rKdS and KdS geometries, that are always counterrotating with respect to the LNRFs (area under the convex part of the dashed blue/orange curve): for more details, see Refs. \cite{Stu-Sla:2004:PHYSR4:,Slany:2023:PHYSR4:}.  }     
	\label{fig_a2ms}

\end{figure*}

The behavior of the characteristic functions $a^2_{ms(1,2)}(r;y)$, $a^2_{ECPO}(r;y)$, $a^2_{ms(SPO)}(r;y)$ and the radii $r_{E=0}$ of the rKdS spacetime and their KdS counterparts $a^{2(KdS)}_{ms(1,2)}(r;y)$, $a^{2(KdS)}_{ECPO}(r;y)$, $a^{2(KdS)}_{ms(SPO)}(r;y)$ and $a^{2(KdS)}_{E=0}(r;y)$  is compared in Fig. \ref{fig_a2ms}. To clearly distinguish the naked singularity spacetimes from the black-hole (BH) spacetimes, we also include the event horizon functions, $a^2_h(r;y)$ and $a^{2(KdS)}_{h}(r;y)$. They are defined by the relations
\be
a^2_h(r;y)\equiv yr^4+2r-r^2 \label{a2h_rKdS}
\ee
for the rKdS case, and
\be
a^{2(KdS)}_{h}(r;y)\equiv \frac{yr^4+2r-r^2}{1-yr^2} \label{a2h_KdS}
\ee
for the KdS case.

Figure \ref{fig_a2ms} shows that although the corresponding functions for rKdS and KdS spacetimes differ considerably in their definition, they almost overlap for small values of the parameter $y$.

\clearpage

\subsubsection{Circular geodesic frame}
\begin{figure}
	\includegraphics[width=\linewidth]{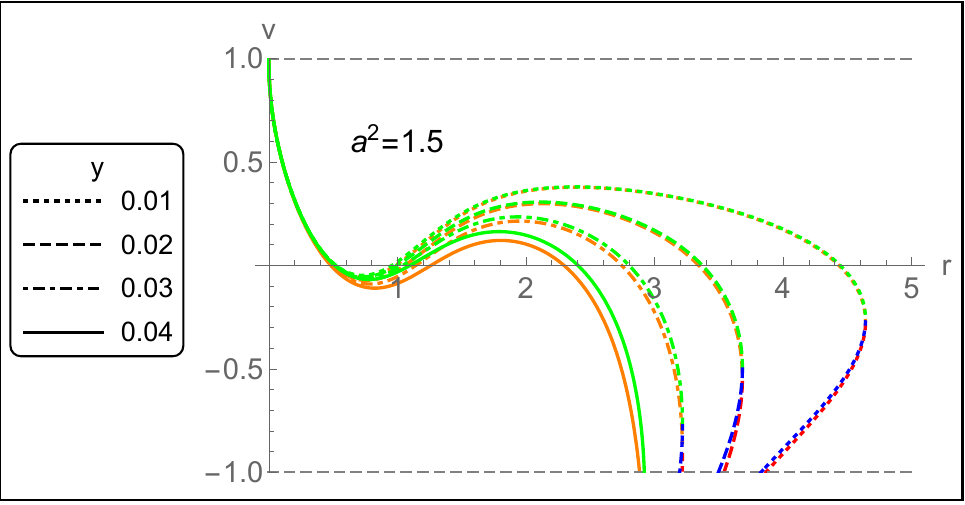}
	\caption{Comparison of orbital velocity versus radial coordinate for the indicated values of the cosmological parameter $y$ for rKdS and KdS spacetime. Green/blue curves correspond to the plus/minus-family orbits in the rKdS spacetime, and orange/red ones correspond to the plus/minus-family orbits in the KdS spacetime.     
	} \label{fig_circ_vel_pm}
\end{figure}

The CGF is defined such that it moves in the azimuthal direction with respect to the LNRF with locally measured velocity $v^{(\phi)}$, which can be expressed. By combining the relationships  (\ref{v(phi)}), (\ref{Epm}), and (\ref{Phipm}), we obtain an expression that can again be rewritten in a hybrid form, allowing the comparison of rKdS and KdS cases:
\be
v^{(\phi)}_{\pm}=\frac{A}{r^2\sqrt{\Delta}}(\Omega_{\pm}-\Omega_{LNRF}), \label{vphi}
\ee 
which is formally the same relation as in the case of KdS spacetime \cite{Stu-Char-Sche:2018:EPJC:,Slany:2023:PHYSR4:}. The term $\Omega_{\pm}$ stands for the Keplerian frequency defined by
\be
\Omega_{\pm}=\din \phi/\din t = p^{\phi}/p^{t}=\frac{1}{a\pm r^{3/2}/\sqrt{1-yr^3}}, \label{Omegapm}
\ee
where $p^{\phi}$ and $p^{t}$ are given by Eqs. (\ref{CarterPhi}), (\ref{CarterT}) and related relations, in which we have substituted the specific energy and angular momentum, given by the relations (\ref{Epm}) and (\ref{Phipm}). The plus/minus signs in Eq. (\ref{vphi}) do not directly reflect membership of the plus/minus-family orbits, but rather  the orientation relative to the LNRF (for details, see Ref. \cite{Slany:2023:PHYSR4:}). The Keplerian frequency is identical for both rKdS and KdS spacetimes, so the only difference in orbital velocities $v^{(\phi)}$ for both spacetimes is in the definition for expressions $A$ and $\Delta$. A comparison of orbital velocities for both types of spacetime at a selected spin value corresponding to naked singularities/superspins is shown in Fig. \ref{fig_circ_vel_pm} for various cosmological parameters $y$. It can be seen that the differences between the two spacetimes are only noticeable at values of $y$ many orders of magnitude greater than astrophysically relevant values.

\subsubsection{Directional angles of a photon measured in CGFs}
The components of photon four-momentum measured locally in CGF, which we denote by $k^{(\tilde{a})}$, are linked to the LNRF components by transformation equations analogous to Eq. (\ref{khata}), where the local Lorentz transformation matrix is given by the relation
\be
\Lambda^{(\tilde{a})}_{(b)}=\left( \begin{array}{cccc}
	\gamma & 0 & 0 & -\gamma v_{\pm}\\
	0 & 1 & 0 & 0\\
	0 & 0 & 1 & 0\\
	-\gamma v_{\pm} & 0 & 0 & \gamma
\end{array}	\right) 
\ee
and the Lorentz factor is 
\be
\gamma=(1-v_{\pm}^2)^{-1/2}
\ee
For brevity, we have again omitted the superscript $(\phi)$.

The directional angles $\tilde{\alpha}, \tilde{\beta}$ measured locally in CGFs are then given by the relations

\begin{align}
	\cos \tilde{\alpha}
	&= \frac{k^{(\tilde{r})}}{k^{(\tilde{t})}}
	= \frac{\cos \alpha}{\gamma(1 - v_{\pm} \sin \alpha \sin \beta)}, \label{costildealpha} \\
	\sin \tilde{\alpha} \cos \tilde{\beta}
	&= \frac{k^{(\tilde{\theta})}}{k^{(\tilde{t})}}
	= \frac{\sin \alpha \sin \beta - v_{\pm}}{1 - v_{\pm} \sin \alpha \sin \beta}, \label{sintildealphacostildebeta} \\
	\sin \tilde{\alpha} \sin \tilde{\beta}
	&= \frac{k^{(\tilde{\phi})}}{k^{(\tilde{t})}}
	= \frac{\sin \alpha \sin \beta}{\gamma(1 - v_{\pm} \sin \alpha \sin \beta)}. \label{sintildealphasintildebeta}
\end{align}
All the directional angles defined in this paper are linked through equalities (\ref{CarterPhiphot}), (\ref{CarterTphot}), (\ref{k(a)contra}), and (\ref{k(a)cov}) to the motion constants of the photon $X$, or alternatively $\ell$, and $q$. In the equatorial plane, where $q=0$, it may therefore be inspiring, such as in Refs. \cite{Stu-Char-Sche:2018:EPJC:,Stu-Hle:2000:CLAQG:}, to study the dependence of the impact parameter $\ell$ on the directional angles. Using the relations 
\bea
\Phi &=& k_{\phi}=k_{(\tilde{a})} \omega^{(\tilde{a})}_{\phi}, \label{Phi}\\
-E &=& k_{t} = k_{(\tilde{b})} \omega^{(\tilde{b})}_{t}, \label{kt}
\eea
where 
\be
\omega^{(\tilde{a})}_{\mu}=\Lambda^{(\tilde{a})}_{(b)} \omega^{(b)}_{\mu}, \label{omegatilde}
\ee
one can derive
\bea
&&\ell(\tilde{\alpha})=\\
&&\hspace{-1.2em}\frac{A(\sin \tilde{\alpha} \sin \tilde{\beta}+v_{\pm})}{r^2\sqrt{\Delta}+v_{\pm}A\Omega+(v_{\pm}r^2\sqrt{\Delta}+A\Omega_{LNRF})\sin \tilde{\alpha} \sin \tilde{\beta}}.\nonumber \label{l(tildealpha)}
\eea

\begin{figure*}[b!]
	\centering
	\begin{tabular}{ccc}
		\includegraphics[width=0.33\textwidth]{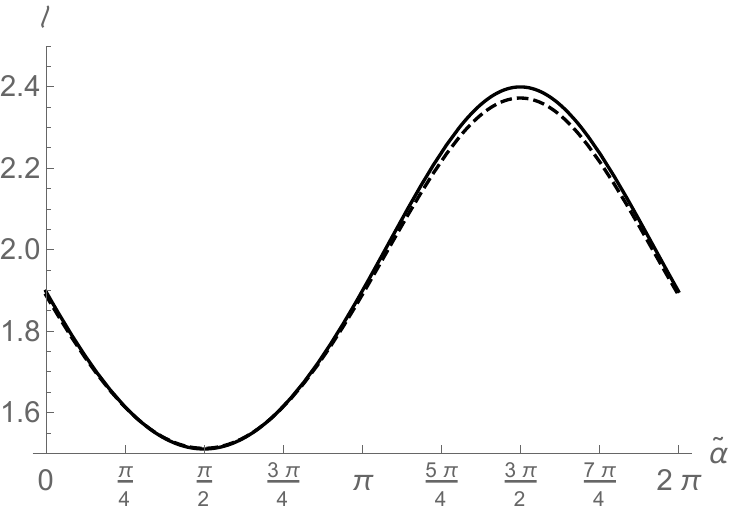}&\includegraphics[width=0.33\textwidth]{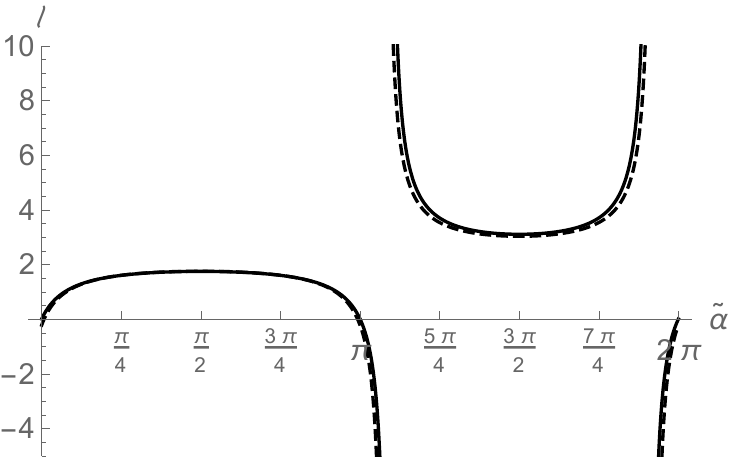}&\includegraphics[width=0.33\textwidth]{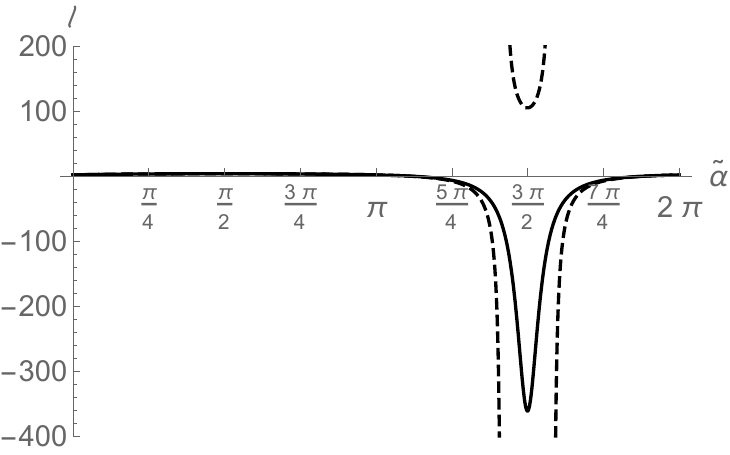}\\
		(a) $r=0.75$& (b) $r=1$& (c) $r=2.24$\\
		\includegraphics[width=0.33\textwidth]{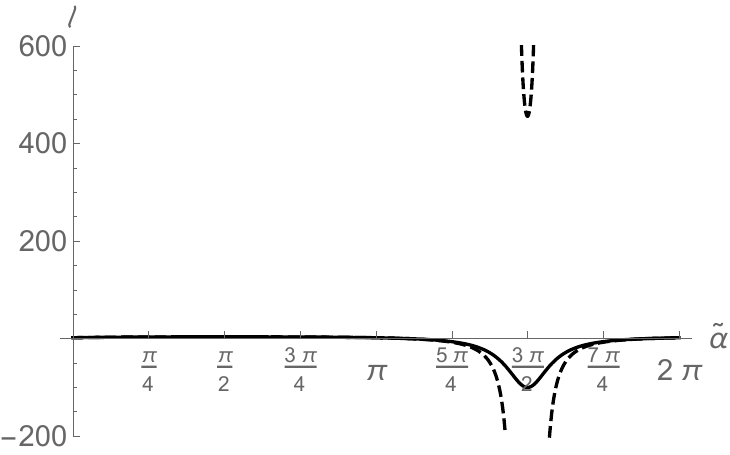}&\includegraphics[width=0.33\textwidth]{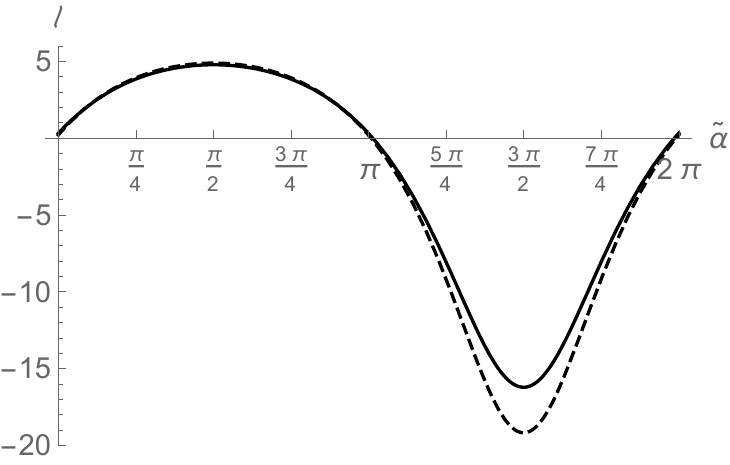}&\includegraphics[width=0.33\textwidth]{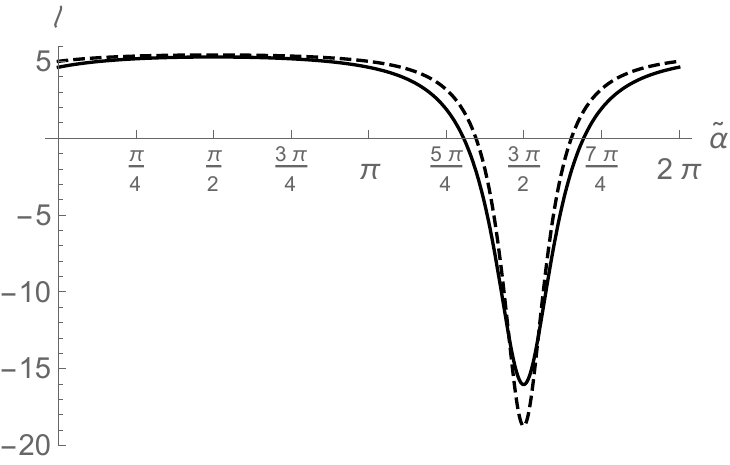}\\
		(d) $r=2.3$ & (e) $r=3.3$ & (f) $r=3.6$\\		
	\end{tabular}	
	\caption{Comparison of angular profiles of the $\ell(\tilde{\alpha})$ function for rKdS (solid curves) and KdS (dashed curves) geometries of the naked singularity/superspinar type at significant values of the radial coordinate $r$ (see Table \ref{tab_IVa_orbs}). In case (a), $v=v_{+}<0$. In cases (c), (d), and (e), $v=v_{+}>0$. In case (b), $v\approx 0$. In case (f), $v=v_{-}<0$.}     
	\label{fig_l(tildealpha)}
	
\end{figure*}

In Fig. \ref{fig_l(tildealpha)}, we compare the dependence of the function $\ell(\tilde{\alpha})$ on the direction angle $\alpha$ for both rKdS and KdS spacetimes, where $\alpha$ ranges from $0\leq \alpha \leq 2\pi$. For a fixed value of $\beta = \pi/2$, this corresponds to the rotation of the wave 3-vector $\vec{k}$ of the photon by a full angle in the equatorial plane. For the corresponding formula for $\ell(\tilde{\alpha})$ in the KdS case, see Ref. \cite{Stu-Char-Sche:2018:EPJC:}.

We compare the angular profile of the function $\ell(\tilde{\alpha})$ on radial coordinates, which are relevant in terms of the motion of photons and test particles within the Keplerian disk. These outstanding values are listed in Table~\ref{tab_IVa_orbs}. The values of the spacetime parameters $a^2=1.5$ and $y=0.02$ were chosen to correspond to the geometries of (r)KdS naked singularities (NSs) and superspinars. The selected, astrophysically extreme value of the cosmological parameter $y$ serves to illustrate a situation where the quantitative differences in the behavior of the characteristic functions $a^2(r;y)$ shown in Fig. \ref{fig_a2ms} become more pronounced. At the same time, as can be seen in Fig. \ref{fig_circ_vel_pm}, with increasing $y$, the values of orbital velocities in the marginally stable orbit region of the Keplerian disk decrease more significantly. The selected value of $y$ is therefore a compromise giving a value of $y$ that allows the influence of aberration to be demonstrated. In the case under consideration, all particle orbits belong to the plus-family orbits.

\renewcommand{\arraystretch}{2.0}
\begin{table*}[t]
	\caption{Significant radii related to the test particle and photon motion in the rKdS/KdS spacetime of class \textbf{IV}.}\label{tab_IVa_orbs}
	\resizebox{\textwidth}{!}{
	\begin{tabular}{|c|c|c|c|c|c|c|c|c|c|}
		\hline
		$y=0.02$\quad$a^2=1.5$&$r^{+}_{(ms)i}$&$r_{E=0}$&$r^{SPO}_{ms}$&$r^{+}_{erg}/r^{+}_{(ms)o}$&$r^{+}_{(ms)o}/r^{+}_{erg}$&$r^{-}_{ph}$&$r_{s}$&$r^{-}_{erg}$&$r_{c}$\\
		\hline
		rKdS&$0.73$&$1.05$&$1.75$&$2.22$&$2.31$&$3.50$&$3.68$&$5.70$&$5.94$\\
		\hline
		KdS&$0.75$&$1.08$&$1.78$&$2.31$&$2.32$&$3.54$&$3.68$&$5.51$&$5.78$\\
		\hline		
	\end{tabular}
}							
\end{table*}

\begin{figure*}[t]
	\centering

	\vspace{0.5em}
	
	\setlength{\tabcolsep}{4pt}
	\renewcommand{\arraystretch}{1.15}
	
	\begin{tabular}{|c|c|c|}
		\hline
		\multicolumn{1}{|c|}{\bet{c} $a^2 = 1.5$\\ $y = 0.02$ \ent } &
		\textbf{LNRF} & \textbf{CGF} \\
		\hline
		\raisebox{3cm}{\textbf{rKdS}} &
		\includegraphics[width=0.45\textwidth]{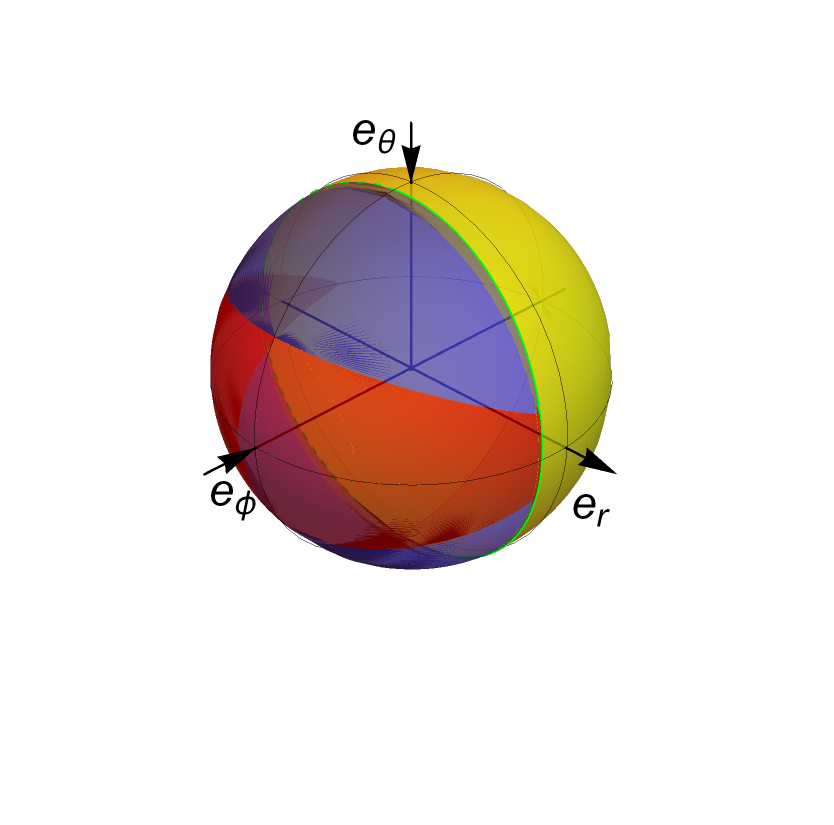} &
		\includegraphics[width=0.45\textwidth]{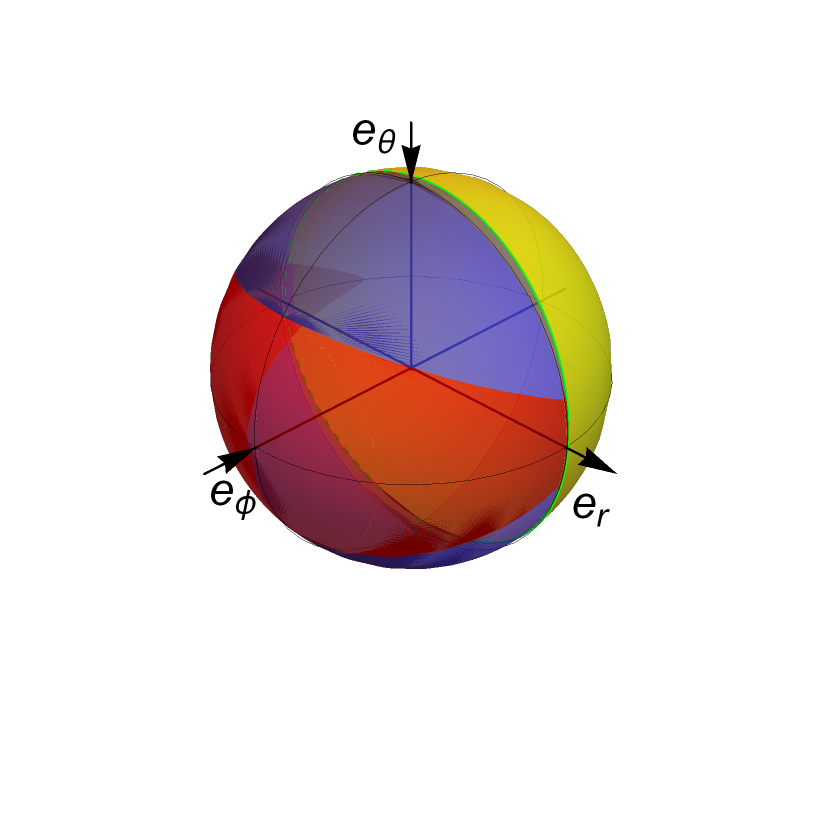} \\
		\hline
		\raisebox{3cm}{\textbf{KdS}} &
		\includegraphics[width=0.45\textwidth]{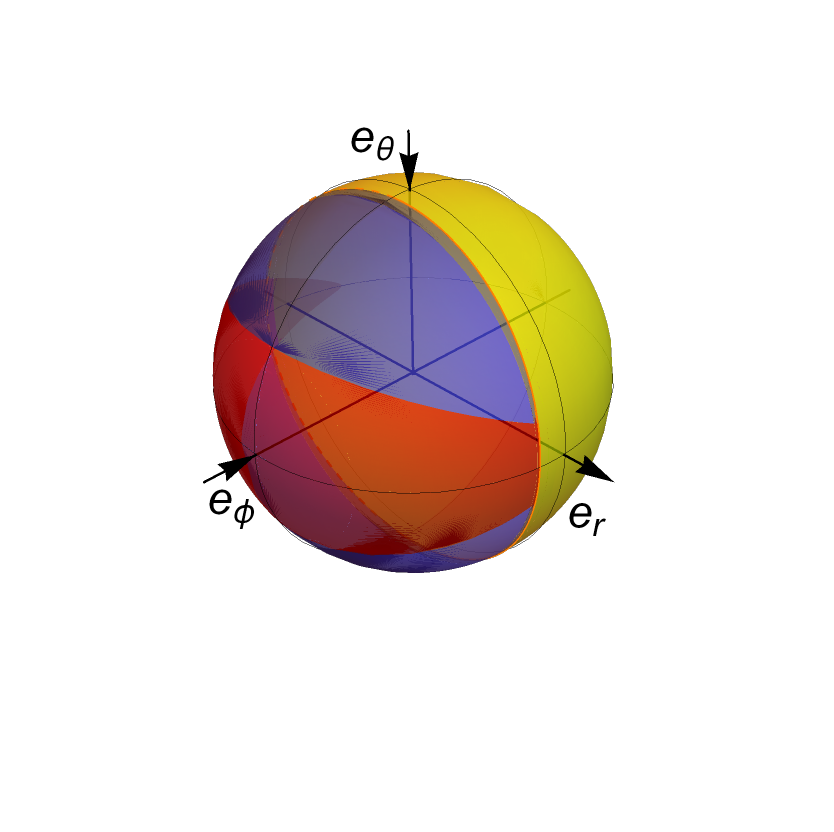} &
		\includegraphics[width=0.45\textwidth]{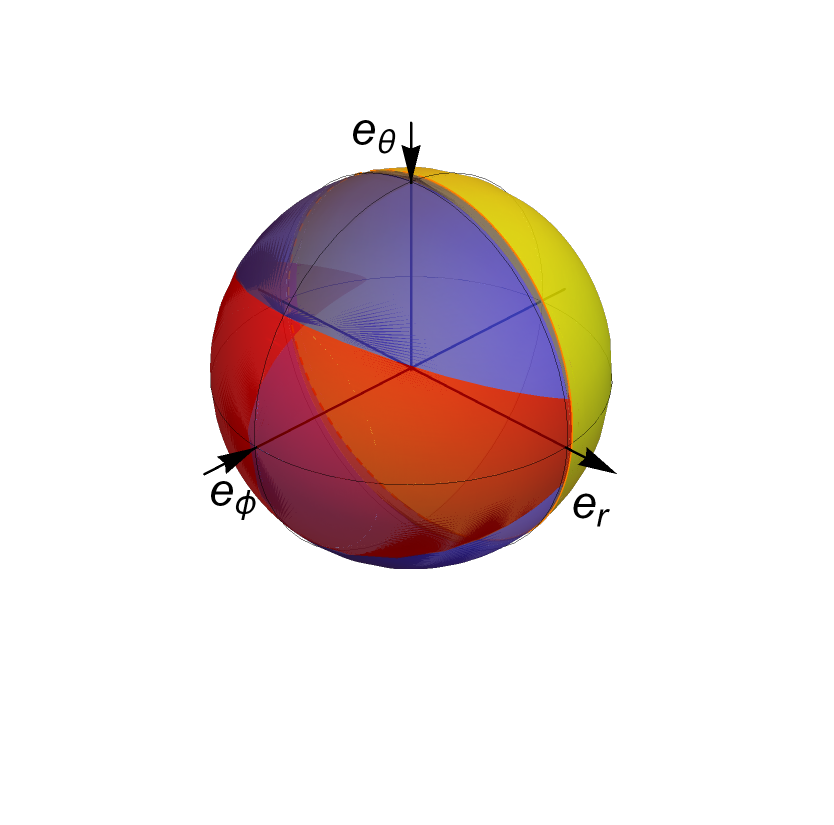} \\
		\hline
	\end{tabular}
	\caption{%
	Comparison of light-escape cones for rKdS and KdS spacetime in LNRF and CGF systems from the view in the positive azimuthal direction. The radial coordinate of the emitter is $r_{e}=0.75$, which corresponds to the inner marginally stable orbit $r^{+}_{(ms)i}$ of the KdS spacetime, which is slightly above $r^{+}_{(ms)i}$ of the rKdS spacetime with the given spacetime parameters (see Table \ref{tab_IVa_orbs} ). The yellow areas correspond to escaping photons, the gray areas to bound photons, the blue areas to photons with negative covariant energy, and the red areas to photons that strike the surface of the superspinar, whose radius is $r=0.1$. The green/orange lines correspond to photons with motion constants $X_{\spo}, q_{\spo}$ that reach SPO either directly (solid part of the curve) or after reaching a certain radial turning point (dashed part).  The orbital velocity of the source is $v=v^{(\phi)}_{+}=-0.05/-0.07$ for the rKdS/KdS spacetime (see Fig. \ref{fig_circ_vel_pm}),; i.e., the source is counterrotating.
	}\label{fig_LECs_ims_f}	
\end{figure*}

\begin{figure*}[t]
	\centering
	
	\vspace{0.5em}
	
	\setlength{\tabcolsep}{4pt}
	\renewcommand{\arraystretch}{1.15}
	
	\begin{tabular}{|c|c|c|}
		\hline
		\multicolumn{1}{|c|}{\bet{c} $a^2 = 1.5$\\ $y = 0.02$ \ent } &
		\textbf{LNRF} & \textbf{CGF} \\
		\hline
		\raisebox{3cm}{\textbf{rKdS}} &
		\includegraphics[width=0.45\textwidth]{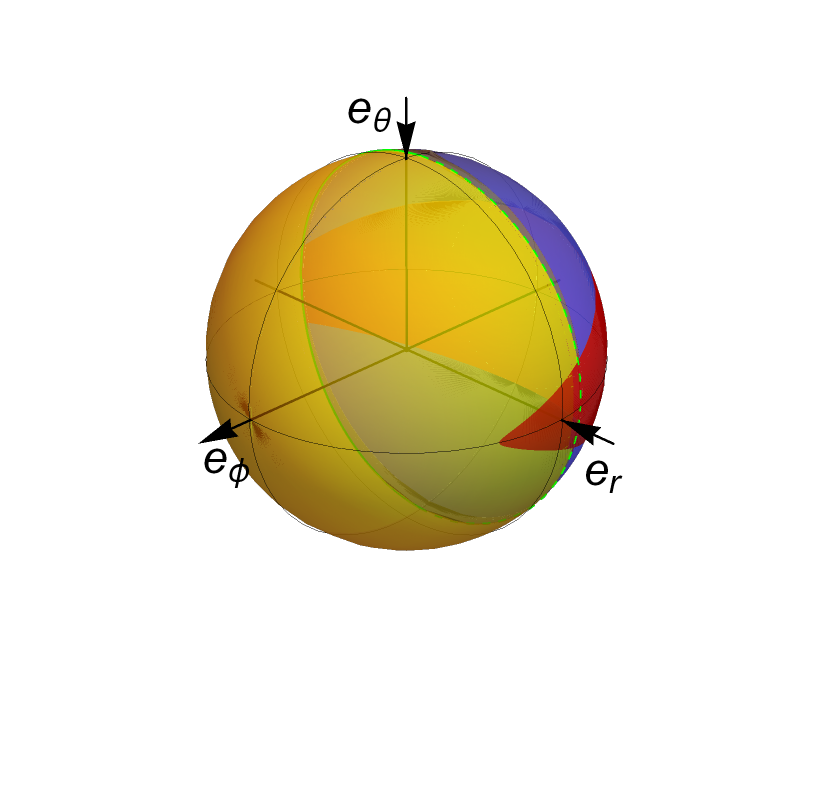} &
		\includegraphics[width=0.45\textwidth]{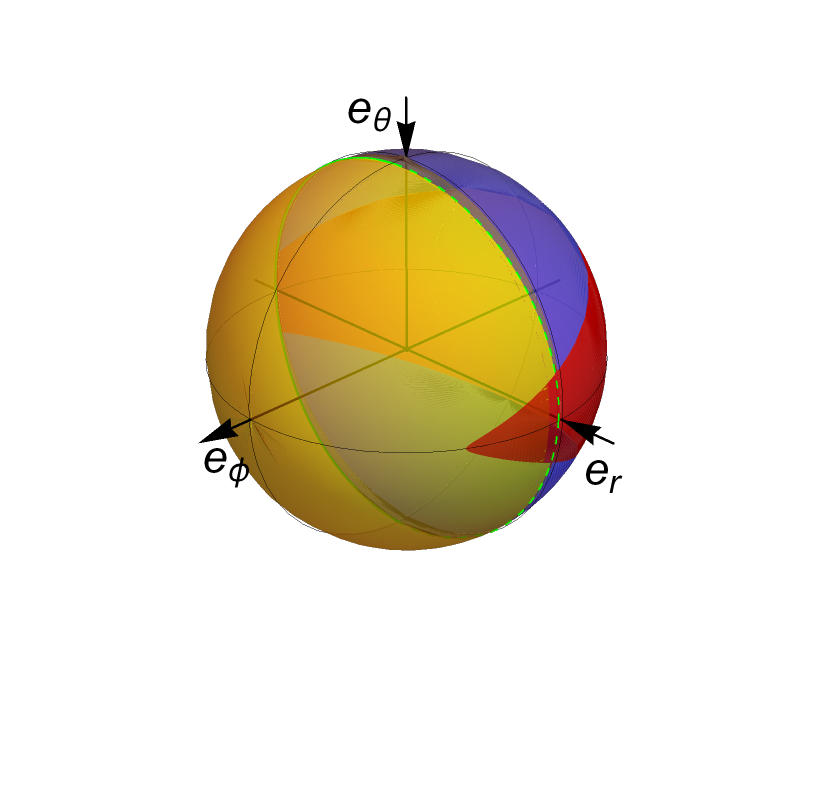} \\
		\hline
		\raisebox{3cm}{\textbf{KdS}} &
		\includegraphics[width=0.45\textwidth]{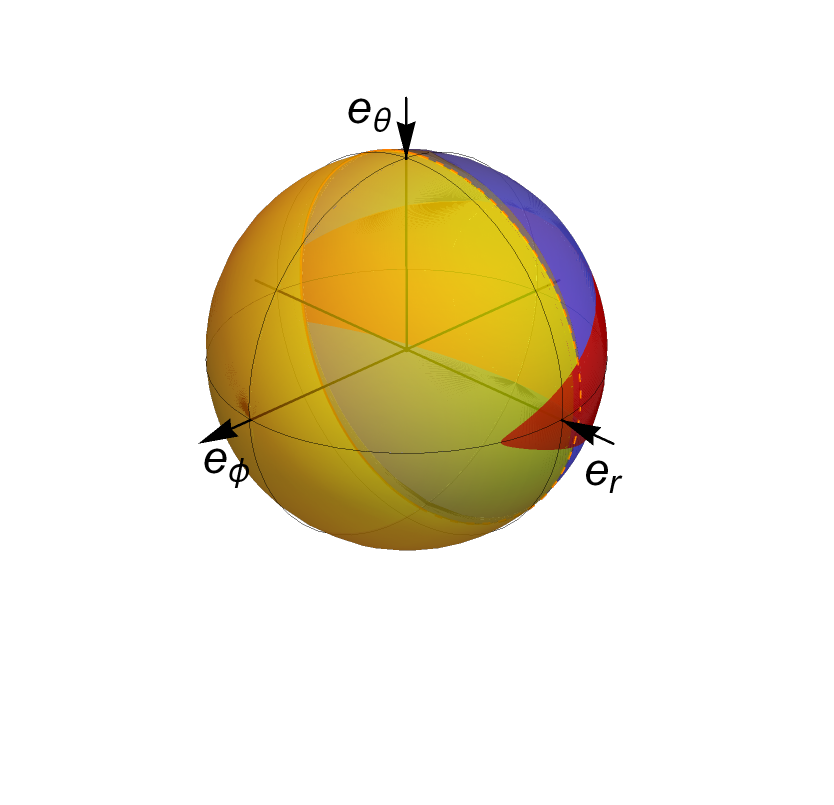} &
		\includegraphics[width=0.45\textwidth]{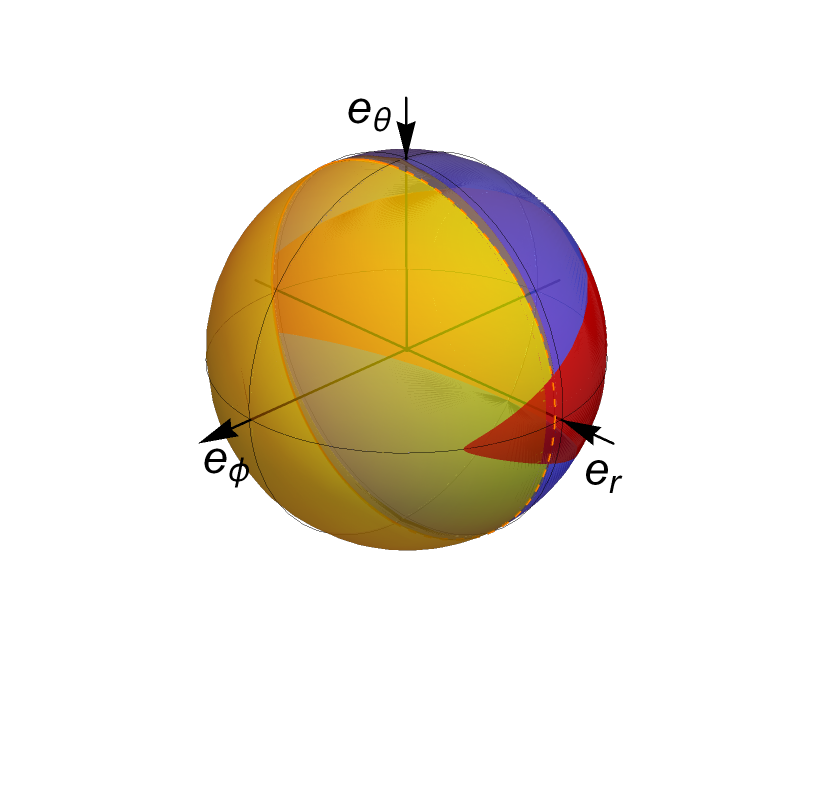} \\
		\hline
	\end{tabular}
	\caption{%
		Continuation of the previous figure for the view in the negative azimuthal direction.
	}\label{fig_LECs_ims_b}	
\end{figure*}

\begin{figure*}[t]
	\centering
	
	\vspace{0.5em}
	
	\setlength{\tabcolsep}{4pt}
	\renewcommand{\arraystretch}{1.15}
	\begin{tabular}{|c|c|c|}
	\hline
	\multicolumn{1}{|c|}{\bet{c} $a^2 = 1.5$\\ $y = 0.02$ \ent } &
	\textbf{LNRF} & \textbf{CGF} \\
	\hline
	\raisebox{3cm}{\textbf{rKdS}} &
	\includegraphics[width=0.45\textwidth]{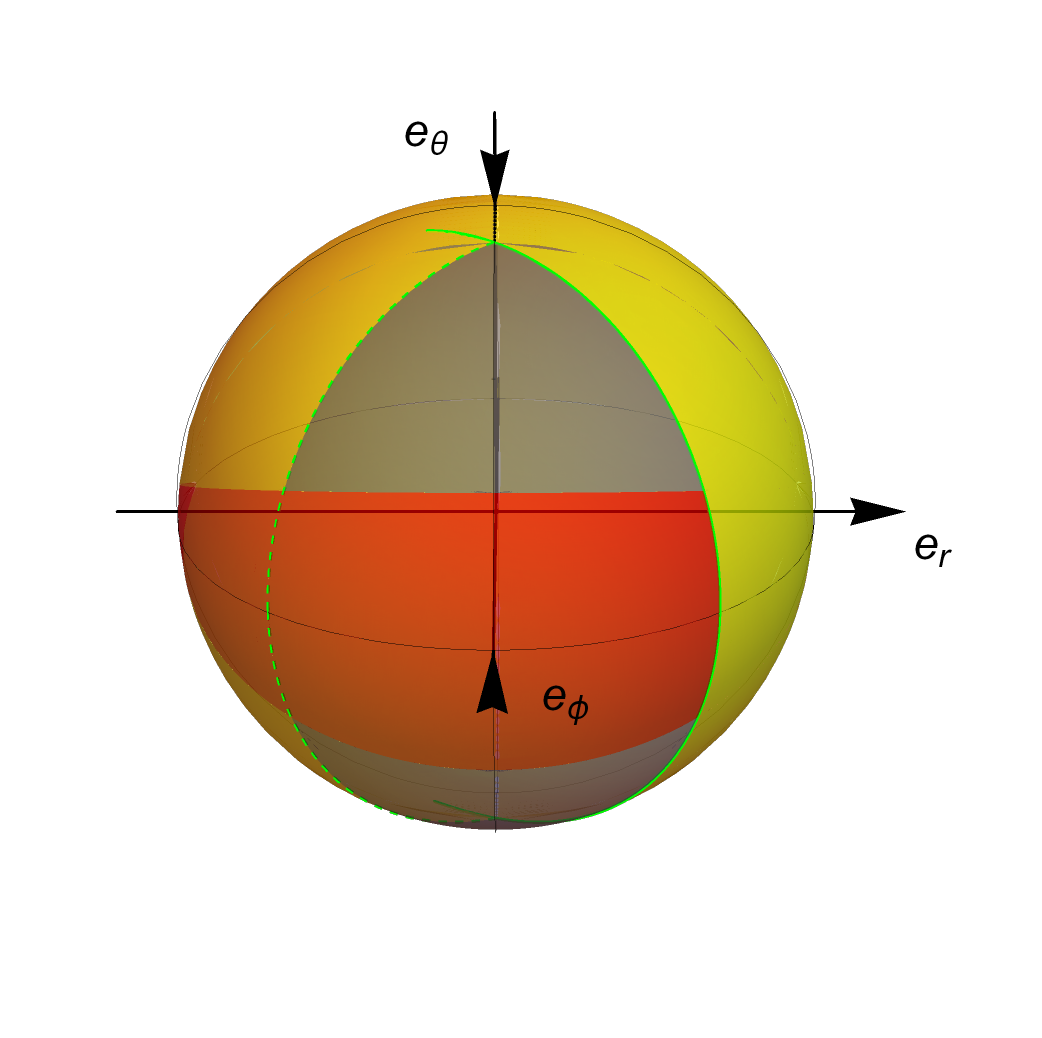} &
	\includegraphics[width=0.45\textwidth]{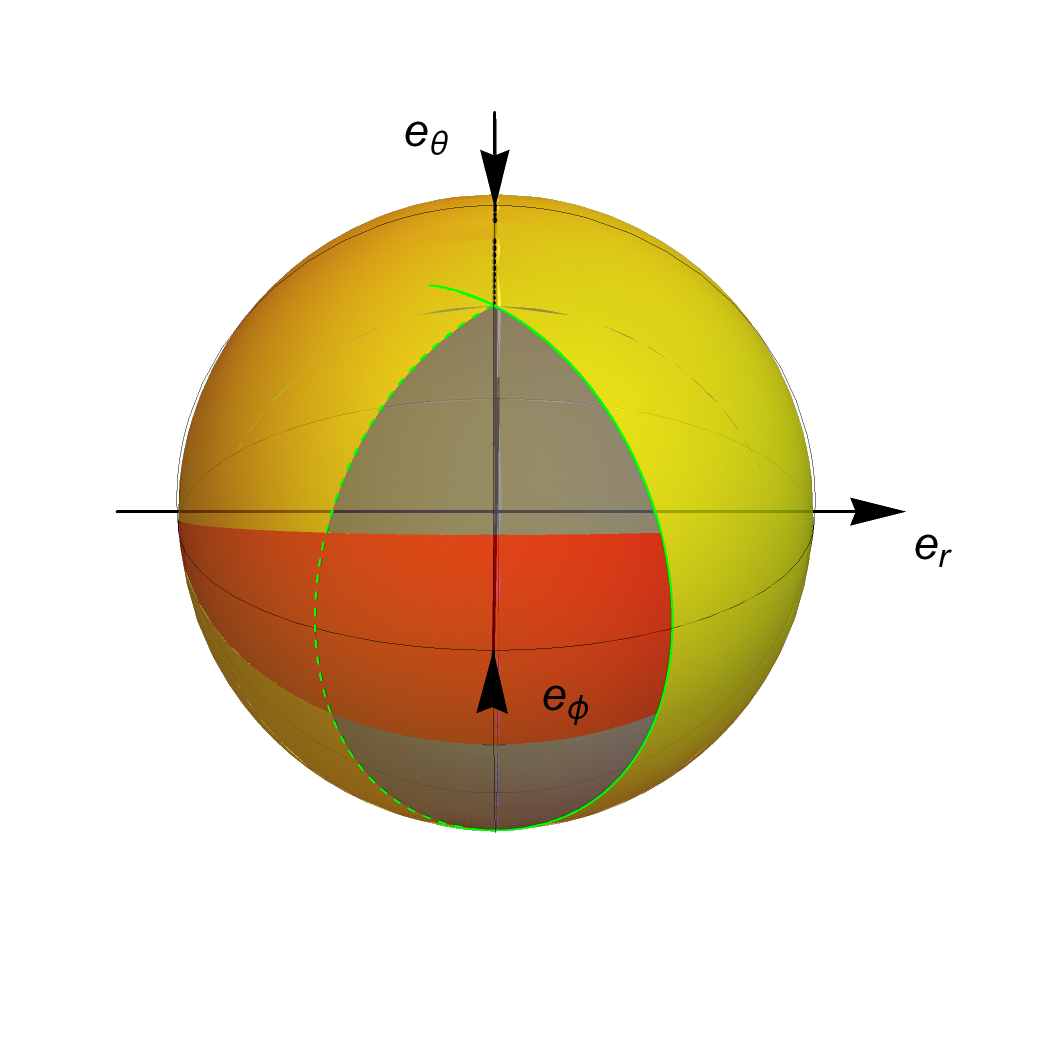} \\
	\hline
	\raisebox{3cm}{\textbf{KdS}} &
	\includegraphics[width=0.45\textwidth]{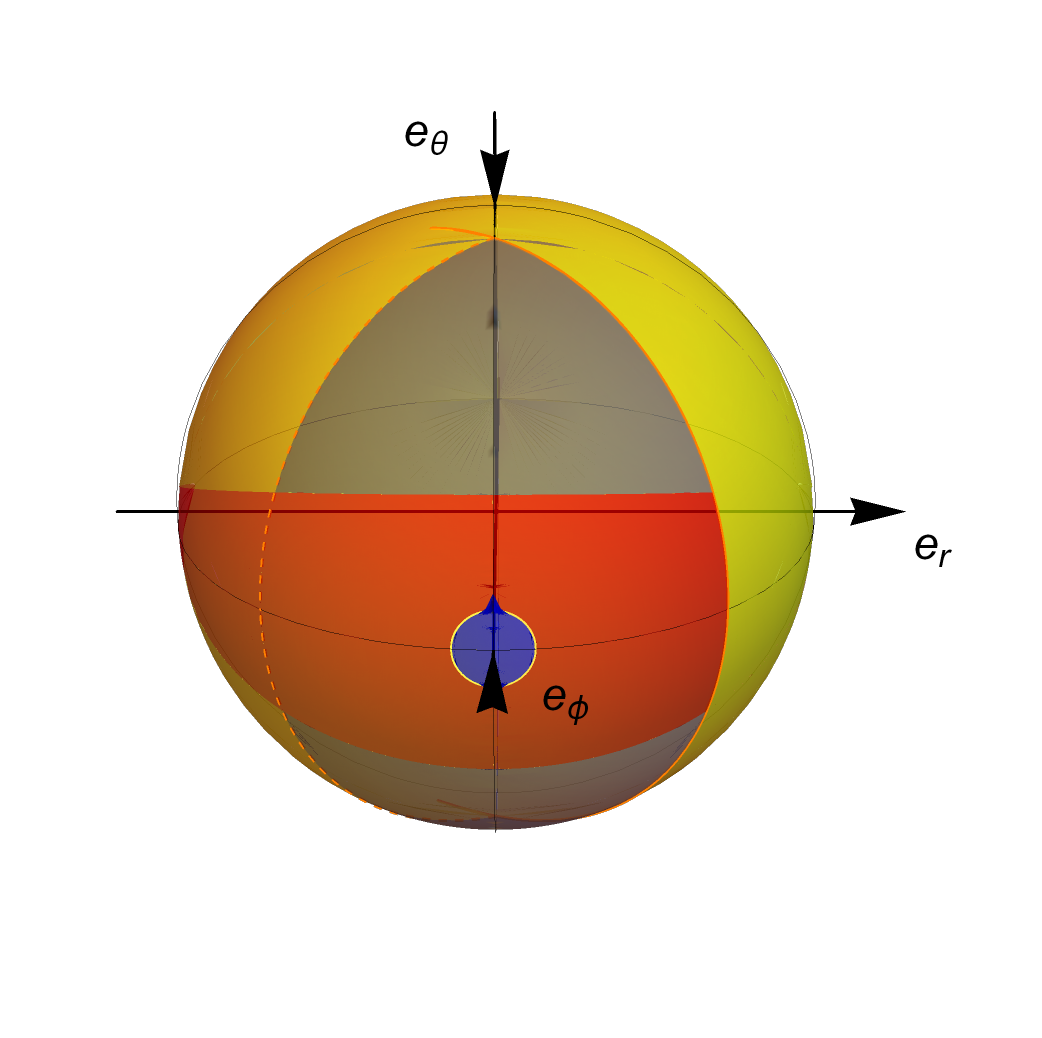} &
	\includegraphics[width=0.45\textwidth]{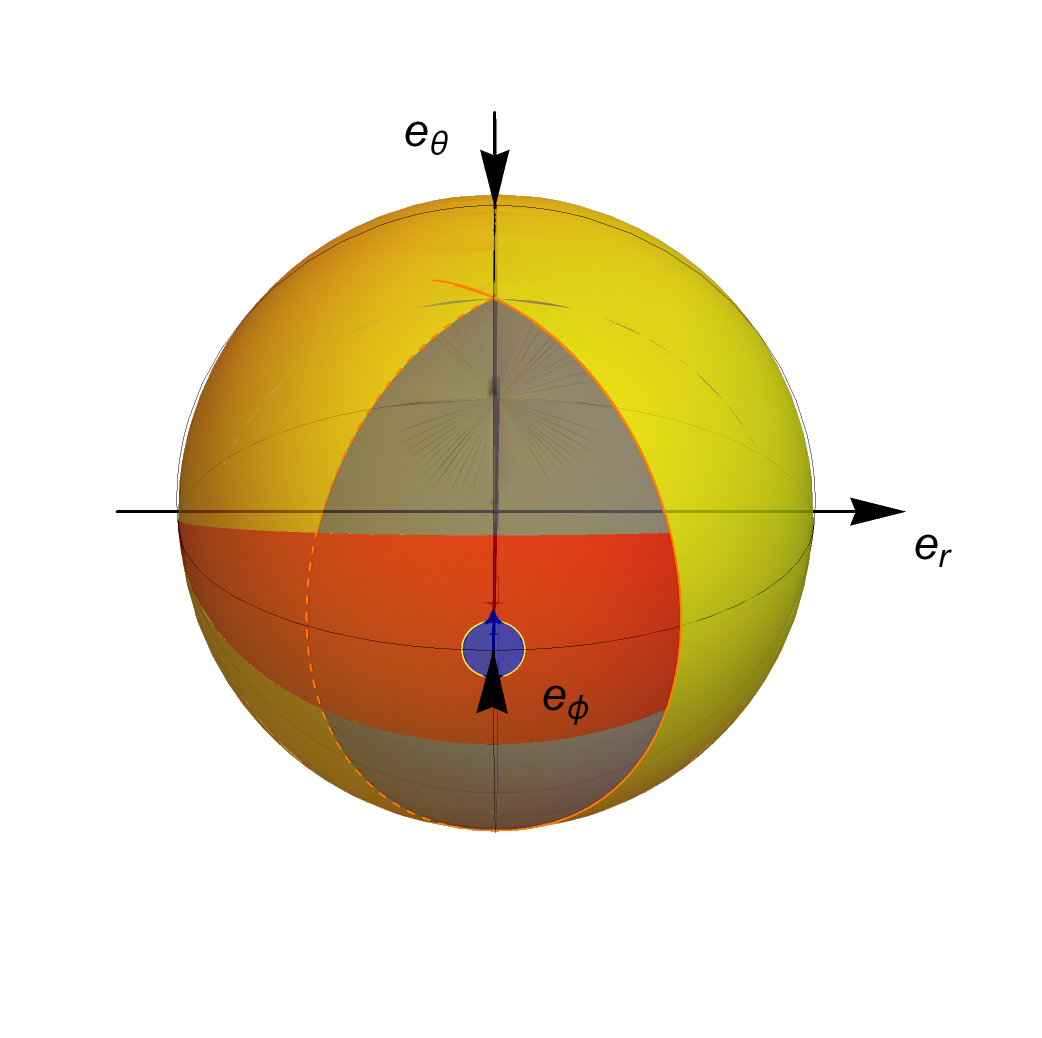} \\
	\hline
\end{tabular}
\caption{
	Comparison of light-escape cones for rKdS and KdS spacetime in LNRF and CGF systems from the view in the positive azimuthal direction. The radial coordinate of the emitter is $r_{e}=2.31$, which corresponds to the outer marginally stable orbit $r^{+}_{(ms)o}$ of both rKdS and KdS spacetimes, which is slightly above/below the boundary of the inner ergosphere $r^{+}_{erg}$ of the rKdS/KdS spacetime with the given spacetime parameters (see Table~\ref{tab_IVa_orbs} ). The colours have the same meaning as in Fig. \ref{fig_LECs_ims_f}. The orbital velocity of the source is $v=v^{(\phi)}_{+}=0.30/0.29$ for the rKdS/KdS spacetime (see Fig. \ref{fig_circ_vel_pm}); i.e., the source is corotating.}
\label{fig_LECs_oms_f}	
\end{figure*}

\begin{figure*}[t]
	\centering
	
	\vspace{0.5em}
	
	\setlength{\tabcolsep}{4pt}
	\renewcommand{\arraystretch}{1.15}
	\begin{tabular}{|c|c|c|}
		\hline
		\multicolumn{1}{|c|}{\bet{c} $a^2 = 1.5$\\ $y = 0.02$ \ent } &
		\textbf{LNRF} & \textbf{CGF} \\
		\hline
		\raisebox{3cm}{\textbf{rKdS}} &
		\includegraphics[width=0.45\textwidth]{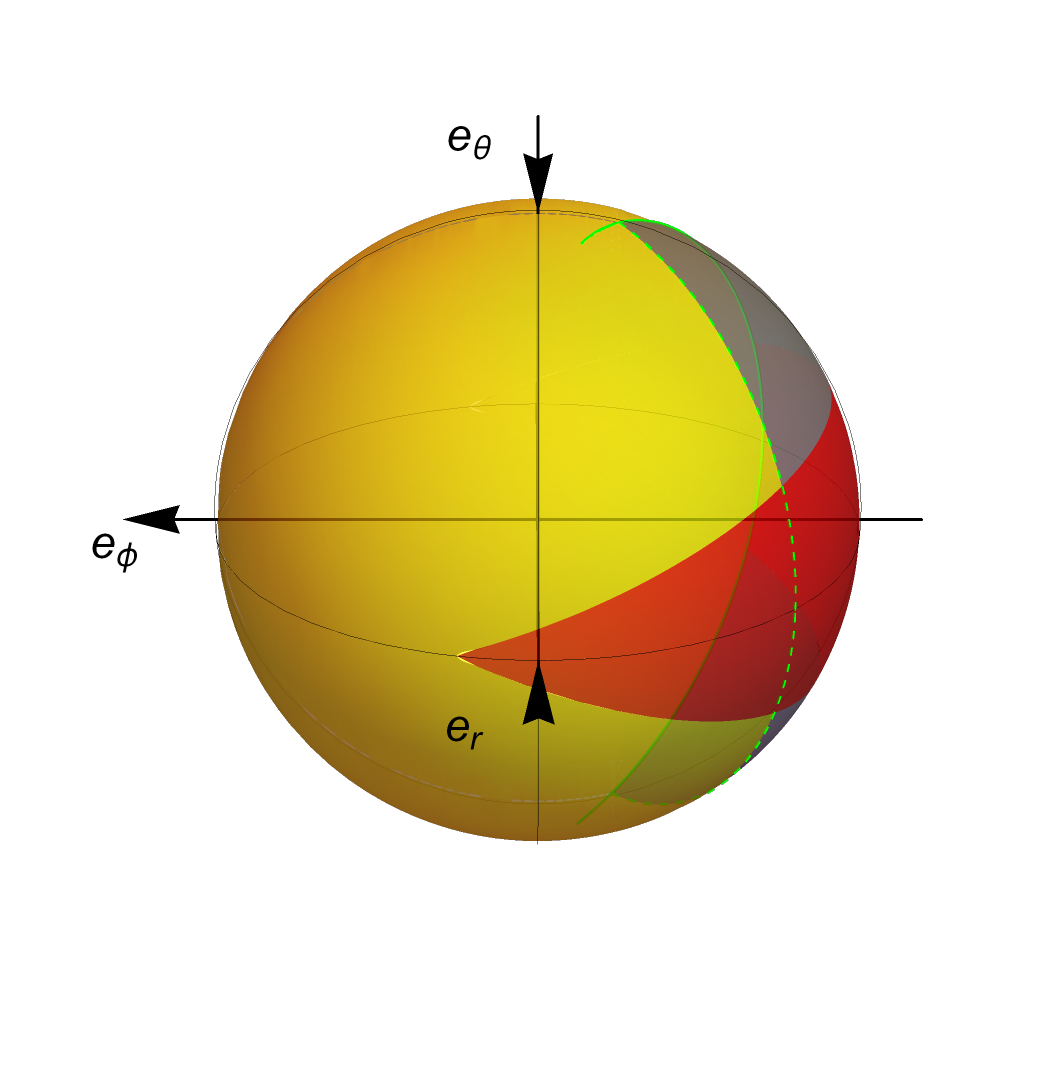} &
		\includegraphics[width=0.45\textwidth]{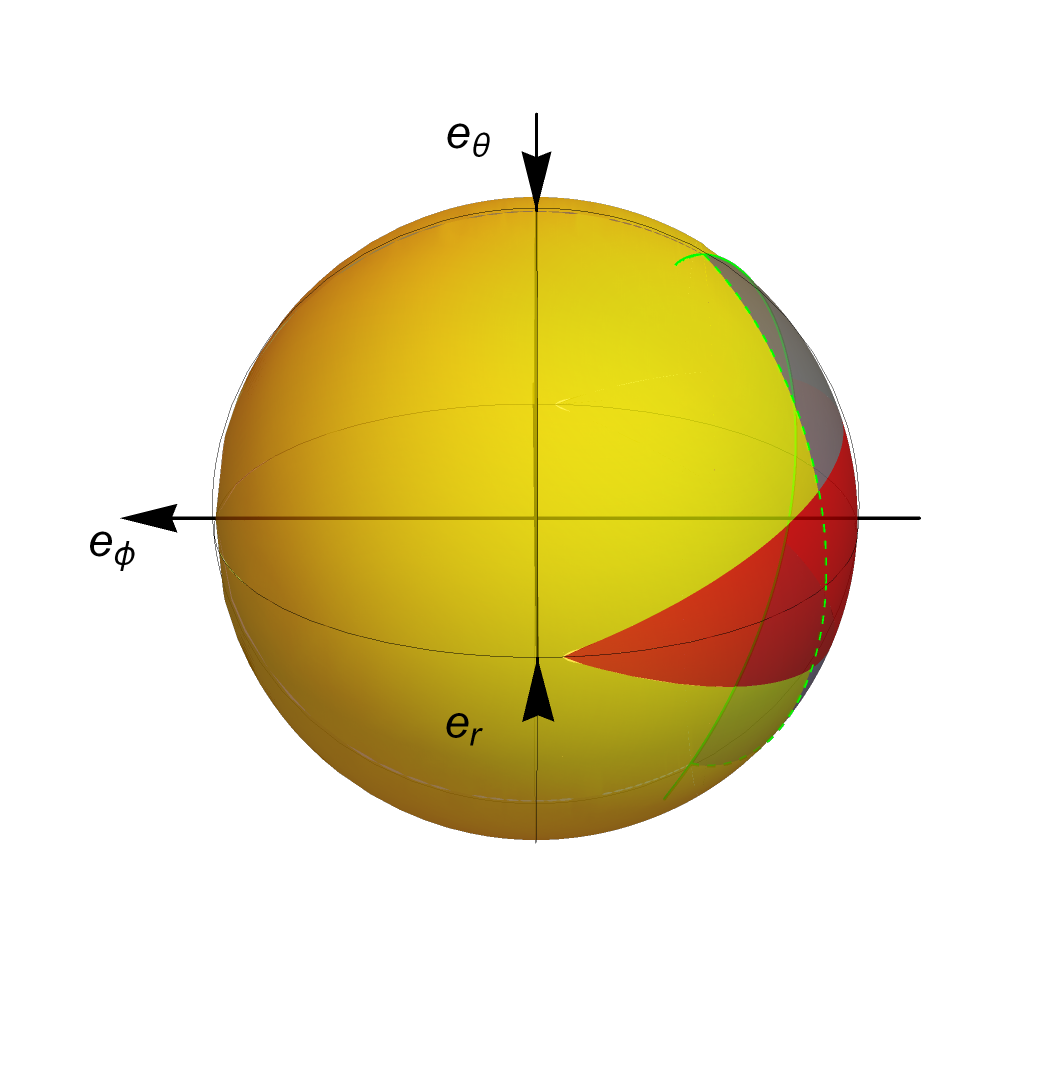} \\
		\hline
		\raisebox{3cm}{\textbf{KdS}} &
		\includegraphics[width=0.45\textwidth]{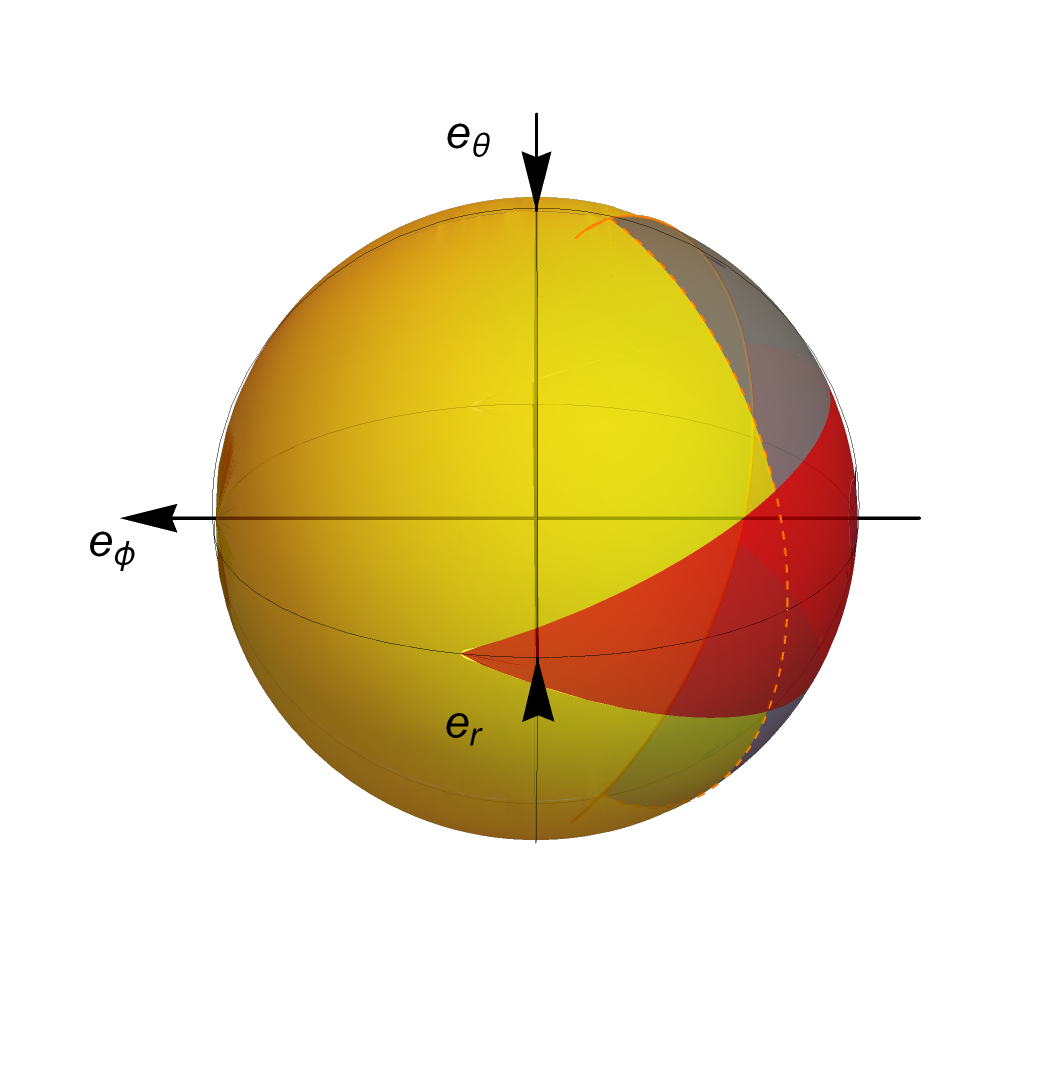} &
		\includegraphics[width=0.45\textwidth]{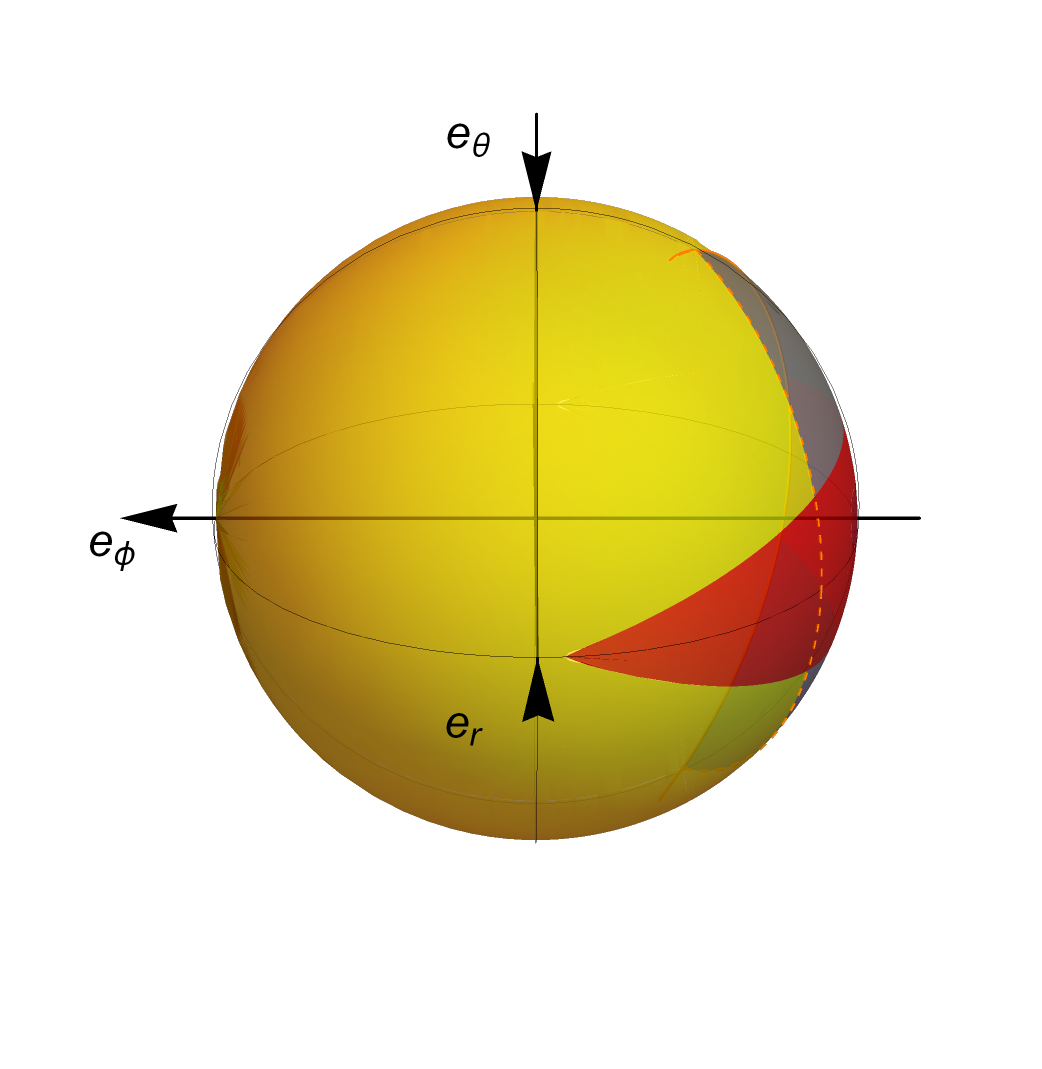} \\
		\hline
	\end{tabular}
	\caption{%
	Continuation of the previous figure for the view in the positive radial direction.}
	\label{fig_LECs_oms_b}	
\end{figure*}
\clearpage   

\section{Conclusions} \label{sec5}
We analyzed the behavior of photon motion in the rKdS metric and then classified the rKdS spacetimes into seven different classes. We found differences with respect to KdS spacetimes, mainly of a quantitative nature. An example of a qualitative difference is the absence of spacetime of class VIII, which exists in the KdS case where it corresponds to a different behavior of the photon motion in the latitudinal direction, while in the rKdS
 metric this behavior coincides with the Kerr case.

To assess the differences between the two geometries, we examined the shadow structures of the superspinar observed in the LNRF at a static radius that is identical for both geometries. We chose prominent astronomically observable quantities and compared these within the two geometries as a function of various parameters. However, significant differences occur only for values of the cosmological parameter $y$ that are hopelessly far from the value corresponding to the current estimates of the cosmological constant $\Lambda=1.1\times 10^{-52}\,\text{m}^{-2}$.

Further, we performed a shadow construction approximating the real observation, considering the effect of cosmological expansion. From the results obtained, it can be concluded that the observations do not seem to be able to distinguish whether the light propagation corresponds to one or the other geometry. Sufficiently large values for observation could be obtained for solutions of the f(R) gravity \cite{nojiri2024blackholesshadowsfr,Dastan2022} giving in the spacetime geometry a term corresponding to an effective cosmological constant that could exceed the observed one in an appropriate way. 

Finally, we constructed and compared light-escape cones for both geometries for a source located at the inner and outer marginally stable equatorial circular orbits of test particles, which define the boundaries of the Keplerian disk. As shown in Figs. \ref{fig_LECs_ims_f}--\ref{fig_LECs_oms_b}, the differences between LECs are only quantitative for both geometries, but are only noticeable for extreme values of the cosmological parameter, which exceed astrophysically relevant values by many orders of magnitude. 
 
 \section*{ACKNOWLEDGEMENTS}
 The authors acknowledge support of the Research Centre for Theoretical Physics and Astrophysics, Institute of Physics, Silesian University in Opava, and the Internal Grant System of Silesian University in Opava (IGS/26/2026). 
 
 \section*{DATA AVAILABILITY}
  The data are not publicly available. The data are available from the authors upon reasonable request.       

\bibliographystyle{unsrt}
\bibliography{bibliography_rKdS}

\begin{thebibliography}{10}

\bibitem{Bar:1973:BlaHol:}
J.~M. Bardeen.
\newblock {Timelike and null geodesics in the Kerr metric}.
\newblock In C.~DeWitt and B.~S. DeWitt, editors, {\em Black Holes (Les Astres
  Occlus)}, pages 215--239, New York, 1973. Gordon and Breach Science
  Publishers.

\bibitem{Cun-Bar:1973:ApJ:}
C.~T. {Cunningham} and J.~M. {Bardeen}.
\newblock {The Optical Appearance of a Star Orbiting an Extreme Kerr Black
  Hole}.
\newblock {\em Astrophys. J.}, 173:L137, May 1972.

\bibitem{Lum:1979:ASTRA:}
J.-P. {Luminet}.
\newblock {Image of a spherical black hole with thin accretion disk}.
\newblock {\em Astronomy and Astrophysics}, 75:228--235, May 1979.

\bibitem{Vie:1993:ASTRA:}
S.~U. {Viergutz}.
\newblock {Image generation in Kerr geometry. I. Analytical investigations on
  the stationary emitter-observer problem}.
\newblock {\em Astronomy and Astrophysics}, 272:355, May 1993.

\bibitem{FalckeMeliaAgol2000}
Heino Falcke, Fulvio Melia, and Eric Agol.
\newblock {Viewing the shadow of the black hole at the Galactic center}.
\newblock {\em Astrophysical Journal Letters}, 528:L13--L16, 2000.

\bibitem{BeckwithDone2005}
Kris Beckwith and Chris Done.
\newblock {Extreme gravitational lensing near rotating black holes}.
\newblock {\em Monthly Notices of the Royal Astronomical Society},
  359:1217--1228, 2005.

\bibitem{SotiriouFaraoni2010}
T.~P. Sotiriou and V.~Faraoni.
\newblock {f(R) Theories of Gravity}.
\newblock {\em Reviews of Modern Physics}, 82:451--497, 2010.

\bibitem{DeFeliceTsujikawa2010}
A.~De~Felice and S.~Tsujikawa.
\newblock {f(R) theories}.
\newblock {\em Living Reviews in Relativity}, 13:3, 2010.

\bibitem{NojiriOdintsov2011}
S.~Nojiri and S.~D. Odintsov.
\newblock {Unified cosmic history in modified gravity: from F(R) theory to
  Lorentz non-invariant models}.
\newblock {\em Physics Reports}, 505:59--144, 2011.

\bibitem{RandallSundrum1999}
Lisa Randall and Raman Sundrum.
\newblock {Large mass hierarchy from a small extra dimension}.
\newblock {\em Physical Review Letters}, 83:3370--3373, 1999.

\bibitem{Dadhich2000}
Naresh Dadhich, Roy Maartens, Petros Papadopoulos, and Vahid Rezania.
\newblock {Black holes on the brane}.
\newblock {\em Physics Letters B}, 487:1--6, 2000.

\bibitem{AlievGumrukcuoglu2005}
Anzhong~A. Aliev and A.~Emrah G{\"u}mr{\"u}k{\c c}{\"u}o{\u g}lu.
\newblock {Charged rotating black holes on a 3-brane}.
\newblock {\em Physical Review D}, 71:104027, 2005.

\bibitem{Reuter1998}
Martin Reuter.
\newblock {Nonperturbative evolution equation for quantum gravity}.
\newblock {\em Physical Review D}, 57:971--985, 1998.

\bibitem{BonannoReuter2000}
Alfio Bonanno and Martin Reuter.
\newblock {Renormalization group improved black hole spacetimes}.
\newblock {\em Physical Review D}, 62:043008, 2000.

\bibitem{ReuterSaueressig2012}
Martin Reuter and Frank Saueressig.
\newblock {Quantum Einstein gravity}.
\newblock {\em New Journal of Physics}, 14:055022, 2012.

\bibitem{SaueressigEtAl2015}
Frank Saueressig, Natalia Alkofer, Giulio D'Odorico, and Francesca Vidotto.
\newblock {Black holes in asymptotically safe gravity}.
\newblock {\em arXiv:1503.06472}, 2015.

\bibitem{Eichhorn2023}
Astrid Eichhorn and Aaron Held.
\newblock {\em {Black Holes in Asymptotically Safe Gravity and Beyond}}, pages
  131--183.
\newblock Springer Nature Singapore, Singapore, 2023.

\bibitem{Lovelock1971}
D.~Lovelock.
\newblock {The Einstein Tensor and Its Generalizations}.
\newblock {\em Journal of Mathematical Physics}, 12:498--501, 1971.

\bibitem{BoulwareDeser1985}
David~G. Boulware and Stanley Deser.
\newblock {String-generated gravity models}.
\newblock {\em Physical Review Letters}, 55:2656--2660, 1985.

\bibitem{Abd-etal:2015:EPJC:}
Ahmadjon Abdujabbarov, Farruh Atamurotov, Naresh Dadhich, Bobomurat Ahmedov,
  and Zden{\v{e}}k Stuchl{\'\i}k.
\newblock {Energetics and optical properties of 6-dimensional rotating black
  hole in pure Gauss--Bonnet gravity}.
\newblock {\em The European Physical Journal C}, 75(8):1--11, 2015.

\bibitem{BroderickLoeb2009}
Avery Broderick and Abraham Loeb.
\newblock {Imaging the Black Hole Silhouette of M87: Implications for Jet
  Formation and Black Hole Spin}.
\newblock {\em The Astrophysical Journal}, 697:1164--1179, 05 2009.

\bibitem{Broderick2014}
Avery~E. Broderick, Tim Johannsen, Abraham Loeb, and Dimitrios Psaltis.
\newblock {Testing the No-Hair Theorem with Event Horizon Telescope
  Observations of Sagittarius A*}.
\newblock {\em The Astrophysical Journal}, 784(1):7, feb 2014.

\bibitem{Nguyen2023}
Bao Nguyen, Pierre Christian, and Chi-kwan Chan.
\newblock {Shadow Geometry of Kerr Naked Singularities}.
\newblock {\em The Astrophysical Journal}, 954(1):78, aug 2023.

\bibitem{ShaikhJoshi2018}
Rajibul Shaikh, Prashant Kocherlakota, Ramesh Narayan, and Pankaj~S Joshi.
\newblock {Shadows of spherically symmetric black holes and naked
  singularities}.
\newblock {\em Monthly Notices of the Royal Astronomical Society},
  482(1):52--64, 10 2018.

\bibitem{Cunha2015}
Pedro V.~P. Cunha, Carlos A.~R. Herdeiro, Eugen Radu, and Helgi~F. R\'unarsson.
\newblock {Shadows of Kerr Black Holes with Scalar Hair}.
\newblock {\em Phys. Rev. Lett.}, 115:211102, Nov 2015.

\bibitem{Vincent2016}
F~H Vincent, Z~Meliani, P~Grandclément, E~Gourgoulhon, and O~Straub.
\newblock {Imaging a boson star at the Galactic center}.
\newblock {\em Classical and Quantum Gravity}, 33(10):105015, apr 2016.

\bibitem{Ohgami2015}
Takayuki Ohgami and Nobuyuki Sakai.
\newblock {Wormhole shadows}.
\newblock {\em Phys. Rev. D}, 91:124020, Jun 2015.

\bibitem{Sokoliuk2022}
O.~{Sokoliuk}, S.~{Praharaj}, A.~{Baransky}, and P.~K. {Sahoo}.
\newblock {Accretion flows around exotic tidal wormholes. I. Ray-tracing}.
\newblock {\em Astronomy and Astrophysics}, 665:A139, September 2022.

\bibitem{Tangphati2023}
Takol Tangphati, Phongpichit Channuie, Kazuharu Bamba, and Davood Momeni.
\newblock {Shadows and photon spheres in static and rotating traversable
  wormholes}.
\newblock {\em Nucl. Phys. B}, 1014:116876, 2025.

\bibitem{Younsi2016}
Ziri Younsi, Alexander Zhidenko, Luciano Rezzolla, Roman Konoplya, and Yosuke
  Mizuno.
\newblock {New method for shadow calculations: Application to parametrized
  axisymmetric black holes}.
\newblock {\em Phys. Rev. D}, 94:084025, Oct 2016.

\bibitem{Uniyal2025}
Akhil Uniyal, Indu Dihingia, Yosuke Mizuno, and Luciano Rezzolla.
\newblock {The future ability to test theories of gravity with black-hole
  shadows}.
\newblock {\em Nature Astronomy}, 10:165--172, 11 2025.

\bibitem{NojiriOdintsov2025}
Shin'ichi Nojiri and S.~D. Odintsov.
\newblock {Black holes and their shadows in $F(R)$ gravity}.
\newblock {\em Physics of the Dark Universe}, 47:101785, 2025.

\bibitem{RodriguezChagoyaOrtiz2024}
B.~Rodr{\'\i}guez, J.~Chagoya, and C.~Ortiz.
\newblock {Shadows of black holes in dynamical Chern-Simons modified gravity}.
\newblock {\em arXiv preprint arXiv:2403.13062 [gr-qc]}, 2024.

\bibitem{Jafarzade:2023dak}
Khadije Jafarzade, Seyed~Hossein Hendi, Mubasher Jamil, and Sebastian
  Bahamonde.
\newblock {Kerr{\textendash}Newman black holes in Weyl{\textendash}Cartan
  theory: Shadows and EHT constraints}.
\newblock {\em Phys. Dark Univ.}, 45:101497, 2024.

\bibitem{Liu_2024}
Wentao Liu, Di~Wu, Xiongjun Fang, Jiliang Jing, and Jieci Wang.
\newblock {Kerr-MOG-(A)dS black hole and its shadow in scalar-tensor-vector
  gravity theory}.
\newblock {\em Journal of Cosmology and Astroparticle Physics}, 2024(08):035,
  aug 2024.

\bibitem{Luo2024}
Shu Luo and Chenghao Li.
\newblock {Black hole shadow of quantum Oppenheimer-Snyder--de Sitter
  spacetime}.
\newblock {\em Phys. Rev. D}, 110:124042, Dec 2024.

\bibitem{Chur-Kon-Stu-Zhi:2021:JCAP:}
M.S. Churilova, R.A. Konoplya, Z.~Stuchl{\'{\i}}k, and A.~Zhidenko.
\newblock Wormholes without exotic matter: quasinormal modes, echoes and
  shadows.
\newblock {\em Journal of Cosmology and Astroparticle Physics}, 2021(10):010,
  Oct 2021.

\bibitem{Aki-etal:2019:ApJLa}
Kazunori Akiyama et~al.
\newblock {First M87 Event Horizon Telescope Results. I. The Shadow of the
  Supermassive Black Hole}.
\newblock {\em The Astrophysical Journal Letters}, 875(1):L1, Apr 2019.

\bibitem{Aki-etal:2019:ApJLb}
Kazunori Akiyama et~al.
\newblock {First M87 Event Horizon Telescope Results. II. Array and
  Instrumentation}.
\newblock {\em The Astrophysical Journal Letters}, 875(1):L2, Apr 2019.

\bibitem{Akiyama_2019:ApJLc}
Kazunori Akiyama et~al.
\newblock {First M87 Event Horizon Telescope Results. III. Data Processing and
  Calibration}.
\newblock {\em The Astrophysical Journal Letters}, 875(1):L3, Apr 2019.

\bibitem{Akiyama_2019:ApJLd}
Kazunori Akiyama et~al.
\newblock {First M87 Event Horizon Telescope Results. IV. Imaging the Central
  Supermassive Black Hole}.
\newblock {\em The Astrophysical Journal Letters}, 875(1):L4, Apr 2019.

\bibitem{Akiyama_2019:ApJLe}
Kazunori Akiyama et~al.
\newblock {First M87 Event Horizon Telescope Results. V. Physical Origin of the
  Asymmetric Ring}.
\newblock {\em The Astrophysical Journal Letters}, 875(1):L5, Apr 2019.

\bibitem{Akiyama:2019eap}
Kazunori Akiyama et~al.
\newblock {First M87 Event Horizon Telescope Results. VI. The Shadow and Mass
  of the Central Black Hole}.
\newblock {\em Astrophys. J. Lett.}, 875(1):L6, 2019.

\bibitem{Akiyama_2021:ApJLa}
Kazunori Akiyama et~al.
\newblock {First M87 Event Horizon Telescope Results. VII. Polarization of the
  Ring}.
\newblock {\em The Astrophysical Journal Letters}, 910(1):L12, Mar 2021.

\bibitem{Akiyama_2021:ApJLb}
Kazunori Akiyama et~al.
\newblock {First M87 Event Horizon Telescope Results. VIII. Magnetic Field
  Structure near The Event Horizon}.
\newblock {\em The Astrophysical Journal Letters}, 910(1):L13, Mar 2021.

\bibitem{Horava2009Quantum}
Petr Ho\v{r}ava.
\newblock {Quantum Gravity at a Lifshitz Point}.
\newblock {\em Phys. Rev. D}, 79:084008, 2009.

\bibitem{KehagiasSfetsos2009}
Alexios Kehagias and Konstantinos Sfetsos.
\newblock {The Black Hole and FRW Geometries of Non-Relativistic Gravity}.
\newblock {\em Phys. Lett. B}, 678:123--126, 2009.

\bibitem{GimonHorava2009}
E.~G. Gimon and P.~Horava.
\newblock {Astrophysical Violations of the Kerr Bound as a Possible Signature
  of String Theory}.
\newblock {\em Physics Letters B}, 672:299--302, 2009.

\bibitem{Aliev2005}
A.~N. Aliev and A.~E. G{\"u}mr{\"u}k{\c c}{\"u}o{\u g}lu.
\newblock {Gravitational Field Equations on and off a 3-Brane World}.
\newblock {\em Physical Review D}, 71:104027, 2005.

\bibitem{Stu-Hle-Tru:2011:CLAQG:}
Zden{\v{e}}k Stuchl{\'\i}k, S.~Hled{\'i}k, and K.~Truparov\'{a}.
\newblock {Evolution of Kerr superspinars due to accretion counterrotating thin
  discs}.
\newblock {\em Classical Quantum Gravity}, 28(15):155017, August 2011.

\bibitem{Stu-Sche:2010:CQG:}
Zden{\v{e}}k Stuchl{\'\i}k and J.~Schee.
\newblock {Appearance of Keplerian discs orbiting Kerr superspinars}.
\newblock {\em Classical Quantum Gravity}, 27(21):215017 (39~pages), November
  2010.

\bibitem{Stu-Sche:2012:CQG:}
Zdeněk Stuchl{\'{\i}}k and Jan Schee.
\newblock {Observational phenomena related to primordial Kerr superspinars}.
\newblock {\em Classical and Quantum Gravity}, 29(6):065002, 2012.

\bibitem{Stu-Sche:2013:CLAQG:}
Zden{\v{e}}k Stuchl{\'\i}k and J.~Schee.
\newblock {Ultra-high-energy collisions in the superspinning Kerr geometry}.
\newblock {\em Classical Quantum Gravity}, 30(7):075012, April 2013.

\bibitem{Nedkova2013}
Petya~G. Nedkova, Vassil~K. Tinchev, and Stoytcho~S. Yazadjiev.
\newblock {Shadow of a rotating traversable wormhole}.
\newblock {\em Phys. Rev. D}, 88:124019, Dec 2013.

\bibitem{Abdujabbarov:2016efm:}
Ahmadjon Abdujabbarov, Bakhtinur Juraev, Bobomurat Ahmedov, and Zden\v{e}k
  Stuchl\'\i{}k.
\newblock {Shadow of rotating wormhole in plasma environment}.
\newblock {\em Astrophys. Space Sci.}, 361(7):226, 2016.

\bibitem{Abd-Ahm-Dad-Ata:2017:PHYSR4:}
Ahmadjon Abdujabbarov, Bobomurat Ahmedov, Naresh Dadhich, and Farruh
  Atamurotov.
\newblock {Optical properties of a braneworld black hole: Gravitational lensing
  and retrolensing}.
\newblock {\em Physical Review D}, 96:084017, Oct 2017.

\bibitem{Wielgus_2020}
Maciek Wielgus, Jiří Horák, Frederic Vincent, and Marek Abramowicz.
\newblock {Reflection-asymmetric wormholes and their double shadows}.
\newblock {\em Physical Review D}, 102(8), Oct 2020.

\bibitem{Schee_2022}
Jan Schee and Zdeněk Stuchlík.
\newblock {Appearance of Keplerian discs orbiting on both sides of
  reflection-symmetric wormholes}.
\newblock {\em Journal of Cosmology and Astroparticle Physics}, 2022(01):054,
  jan 2022.

\bibitem{Riess1998}
Adam~G. Riess, Alexei~V. Filippenko, Peter Challis, Alejandro Clocchiatti, Alan
  Diercks, Peter~M. Garnavich, Ronald~L. Gilliland, Craig~J. Hogan, Saurabh
  Jha, Robert~P. Kirshner, Bruno Leibundgut, Mark~M. Phillips, David Reiss,
  Brian~P. Schmidt, Robert~A. Schommer, R.~Chris Smith, J.~Spyromilio,
  Christopher Stubbs, Nicholas~B. Suntzeff, and John Tonry.
\newblock {Observational Evidence from Supernovae for an Accelerating Universe
  and a Cosmological Constant}.
\newblock {\em The Astronomical Journal}, 116(3):1009--1038, 1998.

\bibitem{Perlmutter1999}
Saul Perlmutter, Greg Aldering, Gerson Goldhaber, Robert~A. Knop, Peter Nugent,
  Peter~G. Castro, Susana Deustua, Sebastien Fabbro, Ariel Goobar, Donald~E.
  Groom, Isobel Hook, Andrew~G. Kim, Myungshin Kim, Jae Lee, Nicolas~J. Nunes,
  Reynald Pain, Carl~R. Pennypacker, Robert Quimby, Chris Lidman, Richard~S.
  Ellis, Mike Irwin, Richard~G. McMahon, Pilar Ruiz-Lapuente, Nicholas~A.
  Walton, Bradley Schaefer, Benjamin~J. Boyle, Alexei~V. Filippenko, Thomas
  Matheson, Andrew~S. Fruchter, Nino Panagia, Heidi~Jo Newberg, and Warrick~J.
  Couch.
\newblock {Measurements of $\Omega$ and $\Lambda$ from 42 High-Redshift
  Supernovae}.
\newblock {\em The Astrophysical Journal}, 517(2):565--586, 1999.

\bibitem{Kra-Tur:1995:GENRG2:}
L.~M. Krauss and M.~S. Turner.
\newblock {The cosmological constant is back}.
\newblock {\em Gen. Relativity Gravitation}, 27(11):1137--1144, November 1995.

\bibitem{Kra:1998:ASTRJ2:}
L.~M. Krauss.
\newblock The end of the age problem, and the case for a cosmological constant
  revisited.
\newblock {\em Astrophys. J.}, 501(2):461--466, 1998.

\bibitem{Stu-etal:2020:Universe:}
Zdeněk Stuchlík, Martin Kološ, Jiří Kovář, Petr Slaný, and Arman
  Tursunov.
\newblock {Influence of cosmic repulsion and magnetic fields on accretion disks
  rotating around Kerr black holes}.
\newblock {\em Universe}, 6(2):26, 2020.

\bibitem{Stu-Sla-Hle:2000:ASTRA:}
Zden{\v{e}}k Stuchl{\'\i}k, P.~Slan{\'y}, and S.~Hled{\'i}k.
\newblock {Equilibrium configurations of perfect fluid orbiting
  Schwarz\-schild--de~Sitter black holes}.
\newblock {\em Astronomy and Astrophysics}, 363(2):425--439, November 2000.

\bibitem{Stu:2005:MODPLA:}
Zden{\v{e}}k Stuchl{\'\i}k.
\newblock {Influence of the Relict Cosmological Constant on Accretion Discs}.
\newblock {\em Modern Phys. Lett. A}, 20(8):561--575, March 2005.

\bibitem{Stu-Sche:2011:JCAP:}
Zden{\v{e}}k Stuchl{\'\i}k and J.~Schee.
\newblock {Influence of the cosmological constant on the motion of Magellanic
  Clouds in the gravitational field of Milky Way}.
\newblock {\em Journal of Cosmology and Astroparticle Physics}, 9:018,
  September 2011.

\bibitem{Stu-Nov-Hla:2025:}
Zdeněk Stuchlík, Jan Novotný, and Jan Hladík.
\newblock {Dark matter halos modeled by polytropic spheres influenced by the
  relict cosmological constant and trapping polytropes forming supermassive
  black holes}.
\newblock {\em Astronomy and Astrophysics}, 702:A2, September 2025.

\bibitem{Stu-Char-Sche:2018:EPJC:}
Zden{\v{e}}k Stuchl{\'i}k, Daniel Charbul{\'a}k, and Jan Schee.
\newblock {Light escape cones in local reference frames of Kerr--de Sitter
  black hole spacetimes and related black hole shadows}.
\newblock {\em The European Physical Journal C}, 78(3):180, Mar 2018.

\bibitem{Stu-Char:2024:PHYSR4:}
Zden\v{e}k Stuchl\'\i{}k and Daniel Charbul\'ak.
\newblock {Shadows of Kerr\textendash{}de Sitter naked singularities}.
\newblock {\em Phys. Rev. D}, 109(6):064008, 2024.

\bibitem{Char-Stu:2024:PHYSR4:}
Daniel Charbul\'ak and Zden{\v{e}}k Stuchl\'{\i}k.
\newblock {Light escape cones in the locally nonrotating reference frames of
  the Kerr--de Sitter superspinars and related superspinar shadows}.
\newblock {\em Phys. Rev. D}, 110:124050, Dec 2024.

\bibitem{Ovalle_2021}
J.~Ovalle, E.~Contreras, and Z.~Stuchl{\'{i}}k.
\newblock {Kerr–de Sitter black hole revisited}.
\newblock {\em Physical Review D}, 103(8), Apr 2021.

\bibitem{Omw-etal:2022:EPJC:}
Eunice Omwoyo, Humberto Belich, J{\'{u}}lio~C. Fabris, and Hermano Velten.
\newblock {Remarks on the black hole shadows in Kerr-de Sitter space times}.
\newblock {\em The European Physical Journal C}, 82(5), May 2022.

\bibitem{Omwoyo2023}
Eunice Omwoyo, Humberto Belich, J.~C. Fabris, and Hermano Velten.
\newblock {Black hole lensing in Kerr-de Sitter spacetimes}.
\newblock {\em The European Physical Journal Plus}, 138(11):1043, 2023.

\bibitem{Slany:2023:PHYSR4:}
P.~Slan{\'y}.
\newblock {Test particle motion along equatorial circular orbits in the
  revisited Kerr-de Sitter spacetime}.
\newblock {\em Phys. Rev. D}, 108(8):084026, 2023.

\bibitem{Char-Stu:2017:EPJC:}
Daniel Charbul{\'a}k and Zden{\v{e}}k Stuchl{\'i}k.
\newblock {Photon motion in Kerr--de Sitter spacetimes}.
\newblock {\em The European Physical Journal C}, 77(12):897, Dec 2017.

\bibitem{Planck2018}
Planck Collaboration.
\newblock {Planck 2018 results. VI. Cosmological parameters}.
\newblock {\em Astronomy \& Astrophysics}, 641:A6, 2020.

\bibitem{ShemmerNetzerMaiolino2004}
O.~Shemmer, H.~Netzer, R.~Maiolino, E.~Oliva, and S.~Croom.
\newblock {Near-infrared spectroscopy of high-redshift active galactic nuclei.
  I. A metallicity--accretion rate relationship}.
\newblock {\em The Astrophysical Journal}, 614:547--557, 2004.

\bibitem{Ziolkowski:2008aq}
Janusz Ziolkowski.
\newblock {Masses of Black Holes in the Universe}.
\newblock {\em Chin. J. Astron. Astrophys. Suppl.}, 8:273--280, 2008.

\bibitem{Mis-Tho-Whe:1973:Gravitation:}
C.~W. {Misner}, K.~S. {Thorne}, and J.~A. {Wheeler}.
\newblock {\em {Gravitation}}.
\newblock 1973.

\bibitem{Bic-Stu:1976:BULAI:}
J.~Bi{\v{c}}{\'a}k and Zden{\v{e}}k Stuchl{\'\i}k.
\newblock On the latitudinal and radial motion in the field of a rotating black
  hole.
\newblock {\em Bull. Astronom. Inst. Czechoslovakia}, 27(3):129--133, 1976.

\bibitem{Sche-Stu:2009:IJMPD:}
J.~{Schee} and Z.~{Stuchl{\'{\i}}k}.
\newblock {Optical phenomena in the field of braneworld Kerr black holes}.
\newblock {\em International Journal of Modern Physics D}, 18(06):983--1024,
  2009.

\bibitem{Stu:1980:BULAI:}
Zden{\v{e}}k Stuchl{\'\i}k.
\newblock {Equatorial circular orbits and the motion of the shell of dust in
  the field of a rotating naked singularity}.
\newblock {\em Bull. Astronom. Inst. Czechoslovakia}, 31:129--144, 1980.

\bibitem{KeenanBargerCowie2013}
Ryan~C. Keenan, Amy~J. Barger, and Lennox~L. Cowie.
\newblock {Evidence for a ~300 Megaparsec Scale Under-density in the Local
  Galaxy Distribution}.
\newblock {\em The Astrophysical Journal}, 775:62, 2013.

\bibitem{supercluster}
R.~Tully, Helene Courtois, Yehuda Hoffman, and Daniel Pomarède.
\newblock {The Laniakea supercluster of galaxies}.
\newblock {\em Nature}, 513, 09 2014.

\bibitem{Stu-Sla:2004:PHYSR4:}
Zden{\v{e}}k Stuchl{\'\i}k and P.~Slan{\'y}.
\newblock {Equatorial circular orbits in the Kerr--de~Sitter spacetimes}.
\newblock {\em Physical Review D}, 69:064001, 2004.

\bibitem{Kol-Tur-Stu:2021:PHYSR4:}
Martin Kološ, Arman Tursunov, and Zdeněk Stuchlík.
\newblock {Radiative Penrose process: Energy gain by a single radiating charged
  particle in the ergosphere of rotating black hole}.
\newblock {\em Physical Review D}, 103(2), Jan 2021.

\bibitem{Stu-Hle:2000:CLAQG:}
Zden{\v{e}}k Stuchl{\'\i}k and S.~Hled{\'i}k.
\newblock {Equatorial photon motion in the Kerr--Newman spacetimes with a
  non-zero cosmological constant}.
\newblock {\em Classical Quantum Gravity}, 17(21):4541--4576, November 2000.

\bibitem{nojiri2024blackholesshadowsfr}
Shin'ichi Nojiri and S.~D. Odintsov.
\newblock {Black holes and their shadows in $F(R)$ gravity}, 2024.

\bibitem{Dastan2022}
Sara Dastan, Reza Saffari, and Saheb Soroushfar.
\newblock {Shadow of a charged rotating black hole in f(R) gravity}.
\newblock {\em The European Physical Journal Plus}, 137, 09 2022.

\bibitem{RatraPeebles1988}
Bharat Ratra and P.~J.~E. Peebles.
\newblock {Cosmological consequences of a rolling homogeneous scalar field}.
\newblock {\em Physical Review D}, 37:3406--3427, 1988.

\bibitem{Caldwell1998}
Robert~R. Caldwell, Rahul Dave, and Paul~J. Steinhardt.
\newblock {Cosmological imprint of an energy component with general equation of
  state}.
\newblock {\em Physical Review Letters}, 80:1582--1585, 1998.

\bibitem{Far:2000:PHYSR4:}
V.~{Faraoni}.
\newblock {Inflation and quintessence with nonminimal coupling}.
\newblock {\em Phys. Rev. D}, 62(2):023504, July 2000.

\bibitem{Far-Jens:2006:CQG:}
Valerio Faraoni and Michael~N Jensen.
\newblock Extended quintessence, inflation and stable de sitter spaces.
\newblock {\em Classical and Quantum Gravity}, 23(9):3005, 2006.

\bibitem{Caldwell2002}
Robert~R. Caldwell.
\newblock {A phantom menace? Cosmological consequences of a dark energy
  component with super-negative equation of state}.
\newblock {\em Physics Letters B}, 545:23--29, 2002.

\bibitem{Carroll2003}
Sean~M. Carroll, Mark Hoffman, and Mark Trodden.
\newblock {Can the dark energy equation-of-state parameter w be less than
  $-1$?}
\newblock {\em Physical Review D}, 68:023509, 2003.

\bibitem{Riess2016}
Adam~G. Riess et~al.
\newblock {A 2.4\% determination of the local value of the Hubble constant}.
\newblock {\em Astrophysical Journal}, 826:56, 2016.

\bibitem{Verde2019}
Licia Verde, Tommaso Treu, and Adam~G. Riess.
\newblock {Tensions between the early and late Universe}.
\newblock {\em Nature Astronomy}, 3:891--895, 2019.

\bibitem{PhysRevD.103.L121302}
Rong-Gen Cai, Zong-Kuan Guo, Li~Li, Shao-Jiang Wang, and Wang-Wei Yu.
\newblock {Chameleon dark energy can resolve the Hubble tension}.
\newblock {\em Phys. Rev. D}, 103:L121302, Jun 2021.

\bibitem{2021APh...13102605D}
Eleonora {Di Valentino}, Luis~A. {Anchordoqui}, {\"O}zg{\"u}r {Akarsu}, Yacine
  {Ali-Haimoud}, Luca {Amendola}, Nikki {Arendse}, Marika {Asgari}, Mario
  {Ballardini}, Spyros {Basilakos}, Elia {Battistelli}, Micol {Benetti}, Simon
  {Birrer}, Fran{\c{c}}ois~R. {Bouchet}, Marco {Bruni}, Erminia {Calabrese},
  David {Camarena}, Salvatore {Capozziello}, Angela {Chen}, Jens {Chluba},
  Anton {Chudaykin}, Eoin~{\'O}. {Colg{\'a}in}, Francis-Yan {Cyr-Racine}, Paolo
  {de Bernardis}, Javier {de Cruz P{\'e}rez}, Jacques {Delabrouille},
  Jo~{Dunkley}, Celia {Escamilla-Rivera}, Agn{\`e}s {Fert{\'e}}, Fabio
  {Finelli}, Wendy {Freedman}, Noemi {Frusciante}, Elena {Giusarma}, Adri{\`a}
  {G{\'o}mez-Valent}, Julien {Guy}, Will {Handley}, Ian {Harrison}, Luke
  {Hart}, Alan {Heavens}, Hendrik {Hildebrandt}, Daniel {Holz}, Dragan
  {Huterer}, Mikhail~M. {Ivanov}, Shahab {Joudaki}, Marc {Kamionkowski}, Tanvi
  {Karwal}, Lloyd {Knox}, Suresh {Kumar}, Luca {Lamagna}, Julien {Lesgourgues},
  Matteo {Lucca}, Valerio {Marra}, Silvia {Masi}, Sabino {Matarrese}, Arindam
  {Mazumdar}, Alessandro {Melchiorri}, Olga {Mena}, Laura {Mersini-Houghton},
  Vivian {Miranda}, Cristian {Moreno-Pulido}, David~F. {Mota}, Jessica {Muir},
  Ankan {Mukherjee}, Florian {Niedermann}, Alessio {Notari}, Rafael~C. {Nunes},
  Francesco {Pace}, Andronikos {Paliathanasis}, Antonella {Palmese}, Supriya
  {Pan}, Daniela {Paoletti}, Valeria {Pettorino}, Francesco {Piacentini},
  Vivian {Poulin}, Marco {Raveri}, Adam~G. {Riess}, Vincenzo {Salzano},
  Emmanuel~N. {Saridakis}, Anjan~A. {Sen}, Arman {Shafieloo}, Anowar~J.
  {Shajib}, Joseph {Silk}, Alessandra {Silvestri}, Martin~S. {Sloth},
  Tristan~L. {Smith}, Joan {Sol{\`a} Peracaula}, Carsten {van de Bruck}, Licia
  {Verde}, Luca {Visinelli}, Benjamin~D. {Wandelt}, Deng {Wang}, Jian-Min
  {Wang}, Anil~K. {Yadav}, and Weiqiang {Yang}.
\newblock {Cosmology Intertwined II: The hubble constant tension}.
\newblock {\em Astroparticle Physics}, 131:102605, September 2021.

\end{thebibliography}
\end{document}